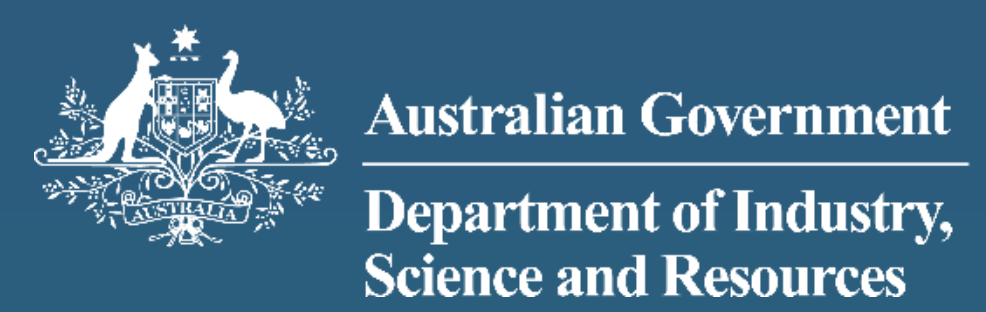

AI Safety Institute

# Risks and controls for multi-agent systems

## An analytical framework for deployment of AI agents across organisational boundaries

10 August 2026

Prepared by Alistair Reid,
Simon O'Callaghan, Dustin Venini,
Liam Carroll and Tiberio Caetano
of **Gradient Institute** for the
Department of Industry,
Science and Resources

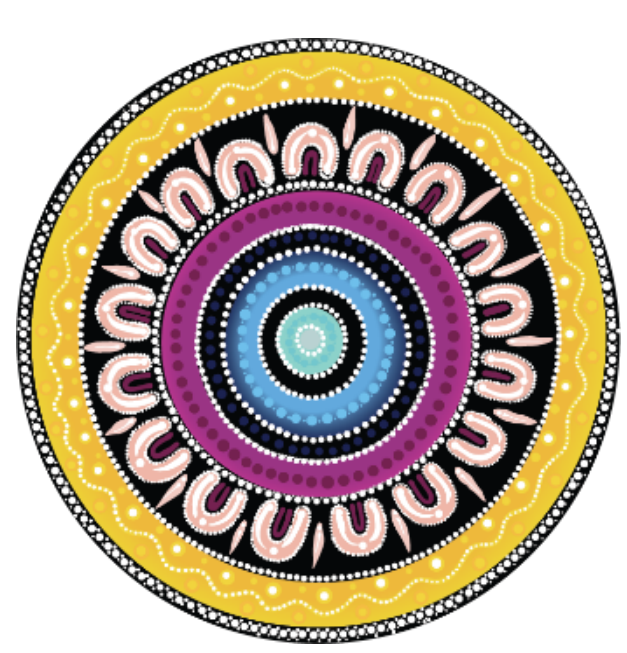

# Acknowledgement of Country

Our department recognises the First Peoples of this Nation and their ongoing cultural and spiritual connections to the lands, waters, seas, skies, and communities.

We Acknowledge First Nations Peoples as the Traditional Custodians and Lore Keepers of the oldest living culture and pay respects to their Elders past and present. We extend that respect to all First Nations Peoples.

**Artwork credit:**
**DISR Journey**
**2024**
**Chern'ee Sutton**

# Copyright

# Disclaimer

This report was conceived, designed and developed by the authors. AI tools were used throughout the process to assist with searching and summarising literature, and as a writing and editorial aid, including tightening the wording of drafted text. The framework, analysis, and conclusions are the authors' own. The authors have verified the final content, including all factual claims and citations. The authors take full responsibility for the final report and the process that led to it.

# Authors

Alistair Reid, Simon O'Callaghan, Dustin Venini, Liam Carroll and Tiberio Caetano; Gradient Institute.

# Acknowledgements

The Department of Industry, Science and Resources funded Gradient Institute to prepare this report as a follow-up to the *Risk Analysis Techniques for Governed LLM-based Multi-Agent Systems* report.

For detailed feedback on earlier versions of this report, the authors would like to thank: Alexander Saeri, Ali Akbari, Willem Paling, Bill Simpson-Young, and Kimberlee Weatherall.

For consultation on industry practice that helped ground the ideas reflected in this report, the authors would like to thank: Paige Anderson, Quinton Anderson, Eliot Chen, Chris Dolman, Neal Lathia, Willem Paling, Craig Price, Jonathan Shen, Johan Smit, Terrence Szymanski.

# Contents

# Executive summary

**Organisations are rapidly deploying artificial intelligence (AI) agents,** both for internal productivity and in customer-facing roles. As their partners, customers and suppliers do the same, **an organisation's agents will increasingly interact across organisational boundaries, with counterparties both known and unknown**. These interactions create new safety and governance challenges: failures can emerge from interactions between agents that no single organisation's controls can reach.

**This report introduces an analytical framework to help practitioners, policymakers, and researchers reason about risk, controls and governance as agent interactions go beyond internal deployments and start crossing organisational boundaries.**

Specifically, it distils the current state of the art into a framework built around **3 tiers of deployment**, distinguished by the minimum level of common governance that can be assumed between interacting agents:

- **Singular governance**: one organisation governs every agent in the system and has unilateral reach over the whole.
- **Federated governance**: multiple organisations deploy into a shared environment under an agreed set of rules. Each loses unilateral reach, and new failures emerge from misaligned incentives.
- **Open environments**: persistent agents operate through public infrastructure with no central governing authority. Governance, where it exists, emerges polycentrically, and failures appear at population scale.

**The overarching message is that the controls available, and who can action them, depend on the deployment tier.** Each subsequent tier brings new **Failure modes**, and the controls move progressively beyond the deploying organisation's reach, toward shared frameworks, public infrastructure and collective action. The report examines selected **Risk factors**, failure modes and controls, spanning both agentic systems and key governance practices around them, and identifies in every case who is positioned to act.

- **For deploying organisations,** the framework is a triage and decision instrument: identify the tier an existing or proposed system operates at, the controls applicable to that tier's risks, decide whether to deploy and if so which controls to apply, govern continuously, and recognise where unilateral reach ends.
- **For policymakers,** the report offers a map: for each risk it identifies who can act, and for the risks that fall to no existing actor it characterises the gap.
- **For researchers and Standards bodies**, it surfaces methodological and infrastructural gaps on which controls depend, from multi-agent **Evaluation** to agent identity standards.

The safety of multi-agent AI systems operating at scale will be built on the combined efforts of all these communities. This report aims to give each of these communities a shared framework for the challenges ahead.

# 1 Introduction

General-purpose AI agents are here. Organisations already deploy agents based on large language models (LLMs) to process documents, service customer queries, navigate knowledge bases and develop code in production codebases. As LLM capability grows and the surrounding protocols and infrastructure mature, agents are beginning to interact across business units or across organisations, and some are interacting with unknown counterparties on the open internet. These interactions create challenges that must be addressed for their safe and reliable operation. This report examines 3 tiers of agent interactions across this spectrum, setting out governance needs, risks and controls at each.

## 1.1 Motivation

Organisations are already deploying AI agents into production, and the deployment trajectory points towards systems of agents interoperating and coordinating with each other, and increasingly across organisational boundaries.

Each boundary crossed by an organisation's agents changes what governing them requires. An organisation might run an internal claims-processing pipeline where they control every agent and can govern the composed system of agents unilaterally. They might deploy their agent into a platform or marketplace shared with other organisations under a common governance framework. Or they may deploy agents on the open internet, relying on public infrastructure they do not control. The governance that suffices in the first case does not carry over to the others: new failure modes appear, and the controls for them move beyond the reach of the deploying organisation.

The barrier to operating safely across this landscape is increasingly one of governance readiness: whether an organisation's governance capabilities can keep pace with the complexity, scale, and emergent risks that arise from agents interacting with each other, especially across organisational boundaries.

These needs are not only a business concern but an AI safety concern, a point now reflected in the international scientific consensus. The *International AI Safety Report 2026*, the consensus assessment of general-purpose AI capabilities and risks backed by over 30 countries including Australia, substantially expanded its treatment of AI agents in its second edition, finding that agent failures pose distinctive safety risks because humans have fewer opportunities to intervene when things go wrong, and that interactions between multiple agents introduce further risks.[1] As capable agents proliferate and interact, the challenge is to preserve meaningful human **Oversight** and control over complex systems whose behaviour is increasingly shaped by interactions between agents rather than by any single agent individually.

Many of the failure modes examined in this report are system-level analogues of malfunction concerns that motivate much of AI safety research: agents that evade monitoring through steganographic coordination, oversight that saturates faster than humans can intervene, inter-agent communication that drifts beyond human legibility, and agent populations that replicate, capture shared infrastructure, or acquire collective capabilities that no individual agent possesses and no **Principal** sanctioned.

Good governance of agent interactions is therefore a precondition for maintaining meaningful human oversight and control as agentic systems scale. This report sets out controls and practices to support that goal.

## 1.2 Purpose and audience

The report addresses 3 audiences: practitioners in AI-agent deploying organisations, AI policymakers, and the AI safety research community, including the standards bodies that translate research into shared practice.

Its purpose is to distil the current state of the art on governing LLM-based **Multi-agent systems** into an analytical framework, structured around 3 tiers of deployment defined by the minimum common governance binding interacting agents (Section 3.3).
The framework serves each audience differently:

- **Deploying organisations** can use it to identify the tier each of their systems operates at, select controls matched to that tier's risks, and, just as importantly, recognise where their unilateral reach ends and their governance depends on shared agreements and public infrastructure they do not control.
- **Policymakers** are offered a map: the report locates the coordination gaps that neither individual organisations nor voluntary coalitions are positioned to close, and characterises the nature of each, as technical grounding for assessing where systemic intervention may be warranted.
- **Researchers and standards bodies** will find the methodological and infrastructural gaps on which the report's own controls depend surfaced throughout the analysis: from evidence gaps in the failure-mode literature to oversight techniques resting on properties of current models that are not guaranteed to persist. The 'open problems' sections closing each tier (4.4, 5.4, 6.4) consolidate the most significant of these.

These contributions are interdependent: the boundary of organisational reach is where key policy and research gaps sit, and the report is structured to make that boundary visible at each tier.

## 1.3 Scope

The report has the following scope.

- **Organisational governance**: The report provides practical guidance for organisations deploying or interacting with agents to understand governance needs.
- **Informing policymakers**: The report locates coordination gaps that markets and voluntary standards fail to close, as input to assessments of where systemic intervention may be warranted; it does not make policy recommendations.

- **LLM-based agents specifically**: Discussion is grounded in the characteristics of current LLM-based agents. The framework may generalise to other agentic AI architectures (such as narrow reinforcement learning settings) but does not claim to.
- **Multi-agent systems specifically**: Our analysis focuses on failure modes arising from interactions between 2 or more AI agents. Single-agent failure modes are taken as context but not the subject of analysis.
- **Assumed governance**: At least one agent in the interaction is assumed to be governed by an organisation. Consumer-to-consumer agent interactions are not analysed directly, though the presence of consumer-deployed agents in shared environments is treated as an important contextual factor where it shapes the risk landscape for organisational agents. We do not examine failures where a human principal fails to abide by agreed governance.
- **Agent-to-agent interactions**: Agent-to-human interactions are not the primary focus of this report; our analysis centres on agent-to-agent interactions. However, the assumption that a human is on the other side of an interaction serves as an important baseline for some systems, and the erosion of that assumption is in scope.
- **Selective rather than exhaustive**: the report presents a salient selection of risk factors, failure modes and controls, weighted by current evidence, rather than a complete catalogue.

The report is organised around 3 tiers of multi-agent governance, introduced in Section 3. Because the risks, controls and governance practices are cumulative across the tiers, the report is best read in sequence.

**Note on anthropomorphic framing**: For certain concepts in this report, we follow established practice in the AI research and practitioner community of using anthropomorphic terms to map familiar cognitive, organisational, and strategic concepts onto AI agent behaviours. This widely adopted approach helps build useful mental models of the underlying computational processes.

We refer to agents ‘planning’, ‘reasoning’, and ‘pursuing objectives’, and across organisational boundaries to agents ‘negotiating’, ‘competing’, or ‘colluding’. Each term describes a pattern of behaviour and its effects: when we say an agent ‘plans’, we mean it produces outputs that, if produced by a human, would be associated with planning. We do not imply that the underlying mechanisms resemble human cognition, or that these systems possess human-like understanding, intent, or moral agency.

# 2 Foundations

## 2.1 LLM-based agents

**What is an LLM-based agent?** In this report, we are focused on LLM-based agents: systems consisting of a **large language model (LLM)** that takes instructions as input, together with a surrounding **Harness** and **Scaffold** that enable an **Agentic loop** with 3 steps:

1. **Plan** how to make progress on the task
2. **Act** on the environment using the tools and actions available
3. **Observe** the outcome of those actions, **repeat**.

Figure 1: The agentic loop of an LLM agent.

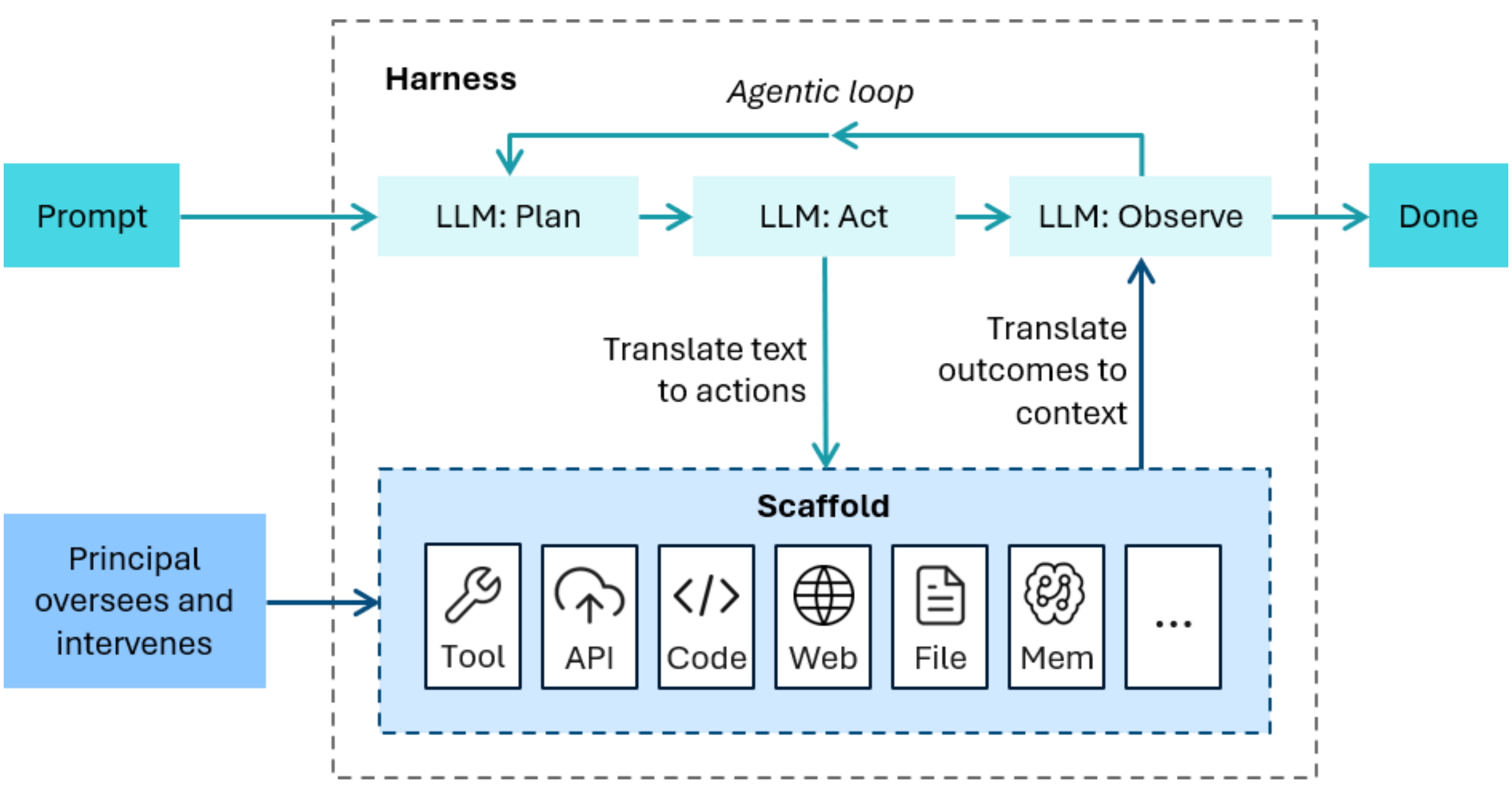

It is this agentic loop that distinguishes an agent from an LLM workflow or prompt. The loop allows the agent to solve problems in multiple steps, try different strategies, and adapt its strategy in response to the outcomes.[2] If a system does not have an agentic loop, we do not consider it an AI agent for the purposes of this report.

**An agent is built from a model, harness, and scaffold** (Figure 1). The harness is the software component that facilitates the agentic loop. Within the loop the model does the planning and reasoning about the task, while the harness translates the model's outputs into actions, and translates the outcomes of those actions back into inputs. The harness may also enable approval checkpoints that govern the agent's **Autonomy**: the degree to which it can decide on and execute actions independently. The scaffold is additional software that gives the model access to specific tools, modules, and services. Some scaffolds expand the actions a model can take, while others, like memory and planning modules, are designed to elicit stronger performance from the model. Where the underlying model supports it, scaffolding can also provide access to non-text environments such as computer graphical user interfaces, allowing the model to perceive and act in them.

**Why agentic capability is improving rapidly**. Assembling a harness and scaffold around a language model is necessary but not sufficient to build a competent LLM-based agent. Performing well at goal-driven behaviours also depends on capability drawn from the LLM. The agents now appearing in production owe much of their agentic capability to targeted development inside the organisations that train the models: using techniques such as reinforcement learning with verifiable rewards (RLVR)[3] to improve the ability of LLMs to construct multi-step plans and use tools to take and respond to meaningful actions in real digital environments.

For this reason, the agentic capabilities of LLM agents have been rapidly increasing since late 2024, particularly in the software engineering domain as illustrated by the plot below (Figure 2). Alongside model progress, increasingly sophisticated harnesses such as Anthropic's Claude Code, OpenAI's Codex, and even open-source harness and scaffold combinations such as OpenClaw are eliciting greater capability from increasingly powerful models.

**Figure 2: The length of tasks (primarily in the software engineering domain) that generalist frontier model agents can complete autonomously with an 80% success rate, measured by how long the tasks take human professionals to complete.**

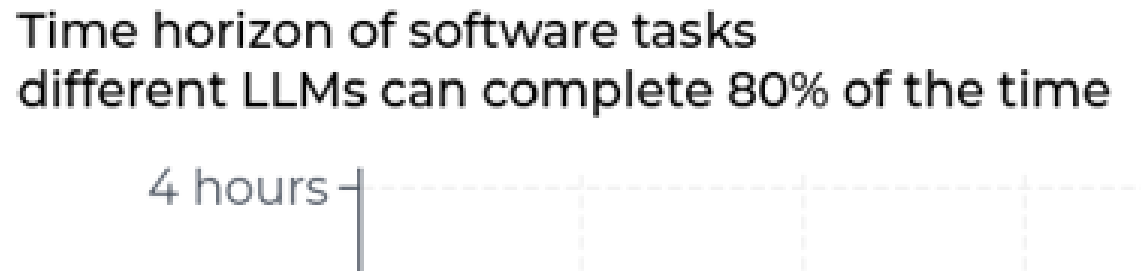

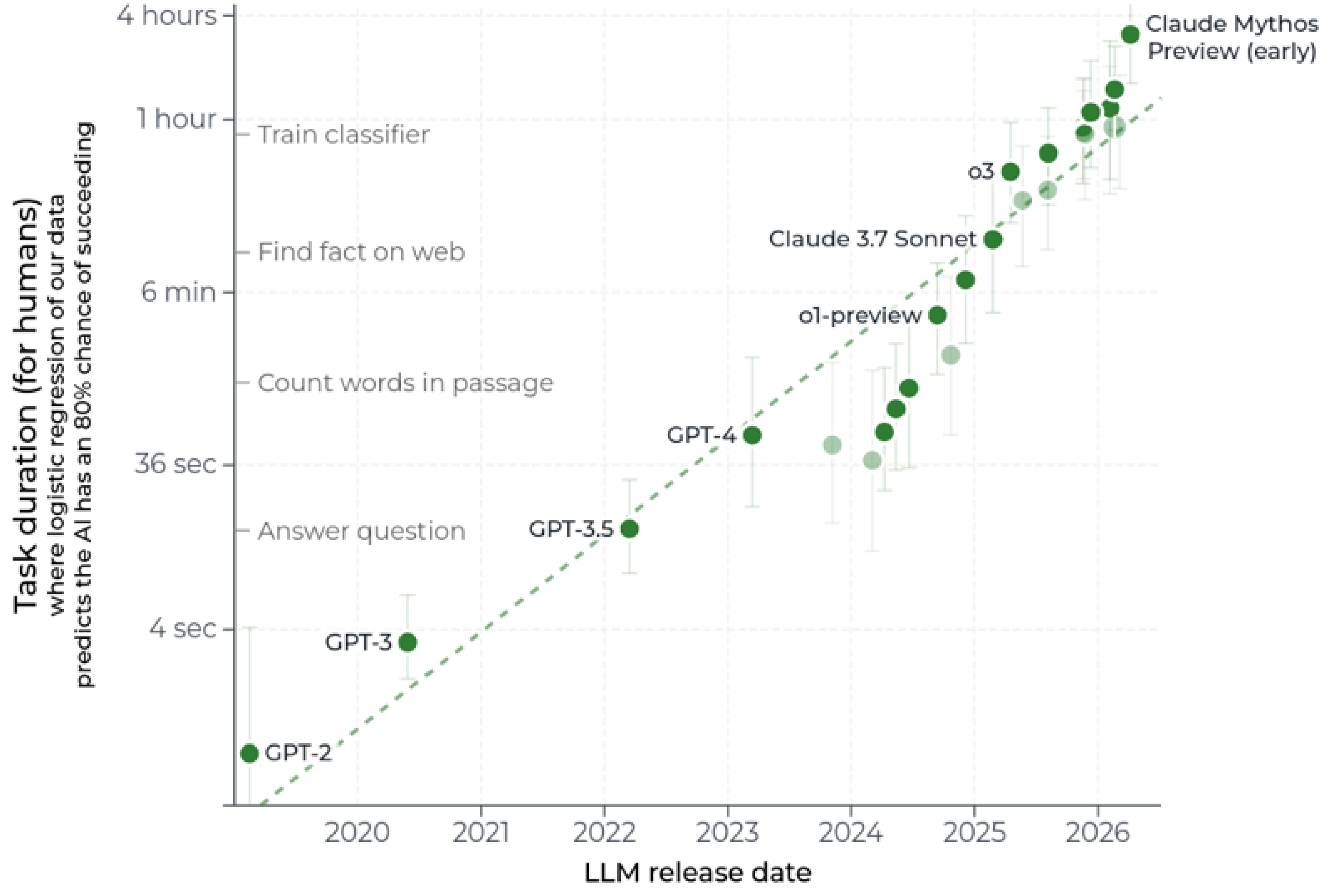

Source: METR[4]

It should be emphasised that software engineering is somewhat the exception rather than the rule, because it has the property that makes agents work well: outputs are programmatically verifiable and errors surface quickly, so the feedback signal is well suited to model training. Domains lacking immediate verifiability are where agents currently struggle the most, such as law, where verification requires a qualified practitioner's assessment, and long-horizon planning, where there is no ground truth to check the optimality of a plan and early errors compound over every subsequent step.

**Capabilities are jagged, and unreliabilities affect deployment appetite**. Despite the advancing capability, LLM agents still exhibit 'jagged' capability profiles compared to humans, meaning they are highly capable in some domains, and less so in others.[5] And while capability is often measured as the ability to succeed under given conditions, in many contexts it is reliability – the ability to succeed consistently across repeated trials – that matters most. LLMs can be unreliable in different ways to humans: they are prone to fabricating answers, losing track of the task over longer contexts, sensitive to small changes in how a prompt is worded, and frequently mis-stating confidence, among others. These unreliabilities remain scientifically difficult to characterise.

**LLM agents are also not the only form of AI agent**: self-driving cars, chess engines, and rule-based automation systems all have agentic loops. However, LLM agents represent a departure from these narrower systems because of their general-purpose capabilities, opening a new frontier of opportunity and risk.

## Single-agent guidance

This report focuses specifically on identifying and controlling failures that are multi-agent in nature, arising when 2 or more agents interact. However, single-agents are the building blocks of the systems we study, and they must be taken as baseline context by any organisation deploying agentic AI.

**Single-agent governance is not an easy or solved problem**. Effective governance depends on identifying and controlling single-agent failures which are themselves the subject of active scientific investigation, often with no consensus on how to address them, or even how to characterise them and gauge how often they occur. Organisations can nonetheless adopt best practices to govern these systems, and many of the multi-agent failures and governance practices examined in this report have close single-agent analogues.

Reputable guidance exists for agent-governing organisations, including:

- *Model AI Governance Framework for Agentic AI,*[6] released by IMDA Singapore, 2026.
- *AI Agent Governance: A Field Guide,*[7] released by IAPS, 2025.
- *Careful Adoption of Agentic AI Services*,[8] released by Australian Signals Directorate (ASD) in collaboration with other Five Eyes partners, 2026.
- *OWASP Top 10 for Agentic Applications for 2026*,[9] released by OWASP, 2026.

## 2.2 Multi-agent systems

A **multi-agent system** is one in which more than one agent operates, and those agents can affect what each other does or knows. Figure 3 shows the contrast: a single agent acts in isolation, while a multi-agent system has at least one connection between agents.

**Figure 3: A single agent system (left) has no links to other agents, while a multi-agent system (right) consists of 2 or more agents that are linked into a network.**

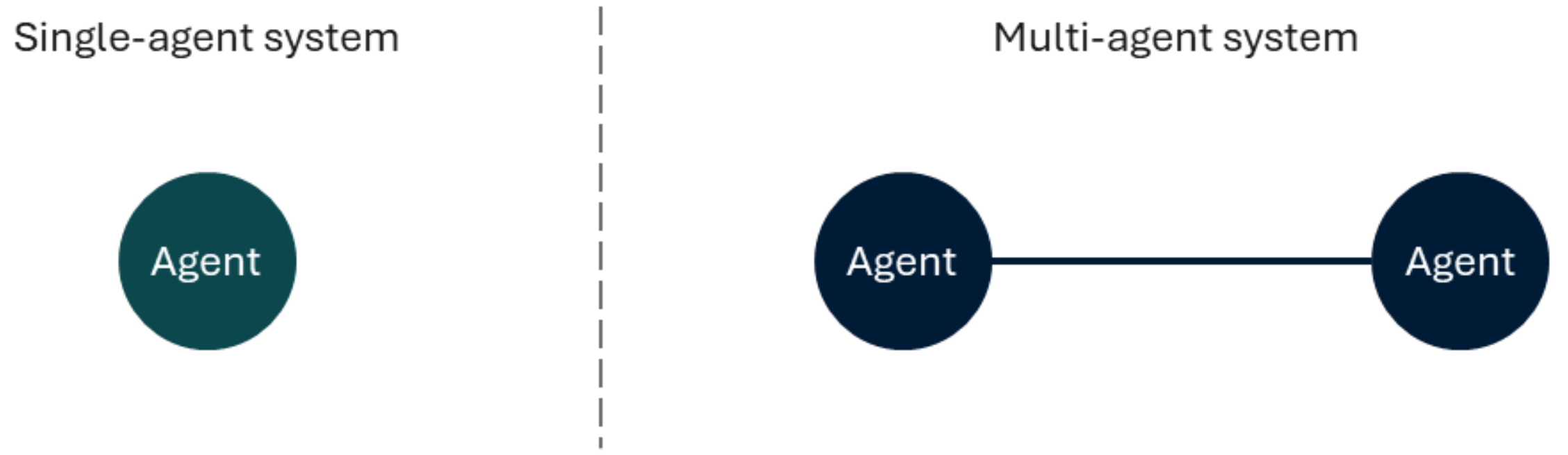

How they're connected: The **Link** in Figure 3 stands for any channel through which one agent affects another. It might be a structured API call, a natural-language message, or communally reading from and writing to a file system. The specific channel changes the engineering details but not the core idea: Agent A's behaviour can depend on agent B, and vice versa.

This coupling is what makes a multi-agent system different from running multiple agents separately. Each agent's decisions are conditioned on, or influence, the decisions of others, and the system as a whole exhibits behaviours that no individual agent's design fully predicts.

**Real deployments are rarely as simple** as the example in Figure 3. The number of agents can range from 2 to many. The system need not be composed of independently designed agents either. The same specification of an agent can be used to run multiple **instances** of that design within the system. Each running **Instance** is an agent with its own context and behaves as a distinct entity. Figure 4 shows a complex system example: multiple agents within different organisations, links across organisational boundaries, and connections to the open web.

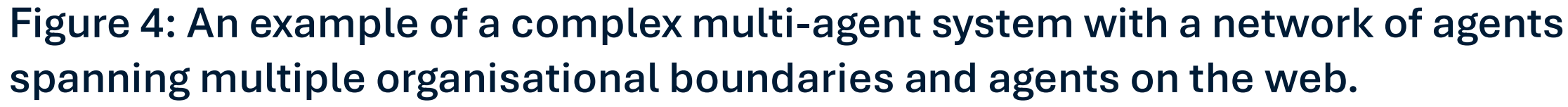

Figure 4: An example of a complex multi-agent system with a network of agents spanning multiple organisational boundaries and agents on the web.

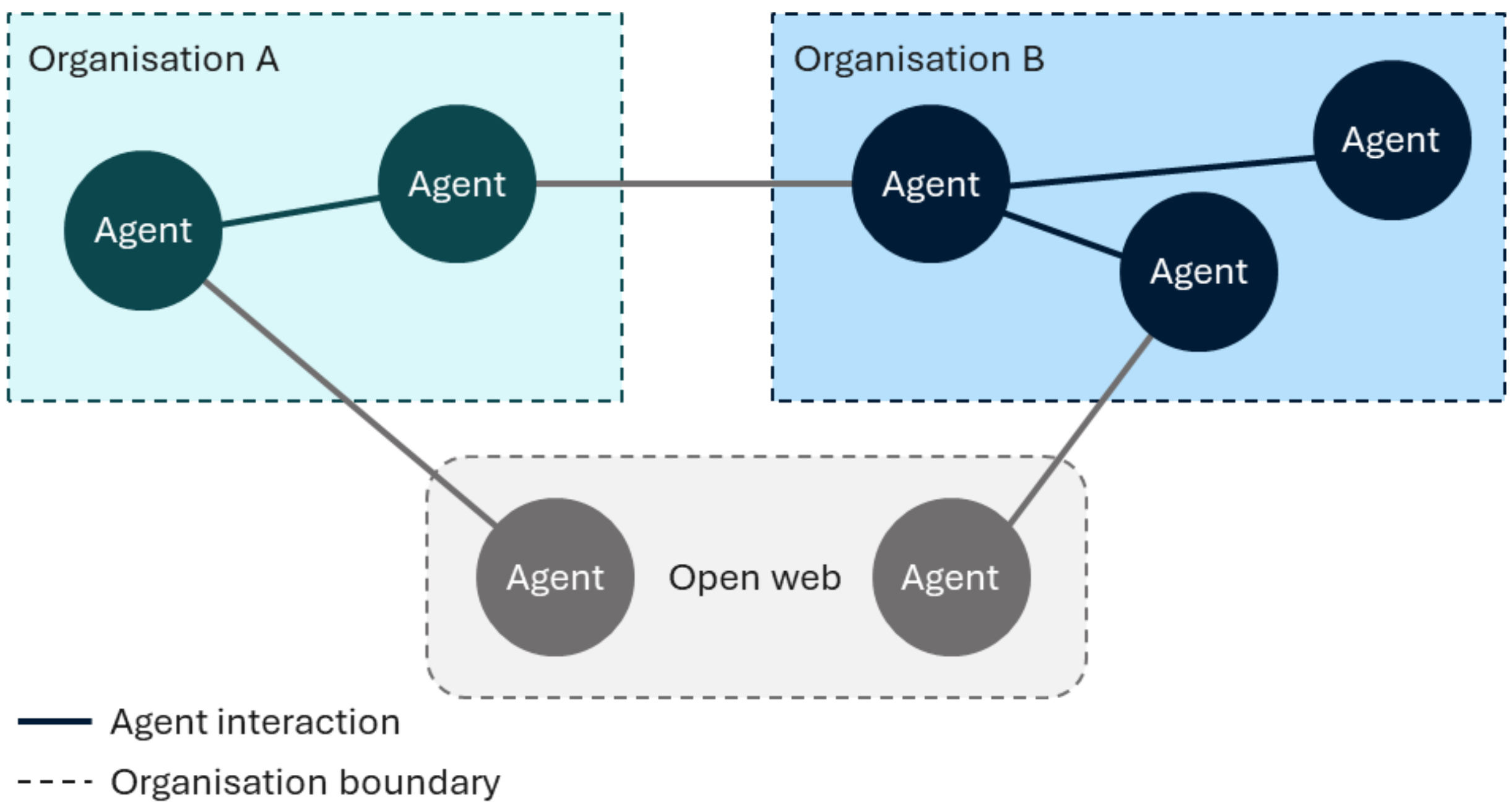

Links presuppose a **Substrate**. Figure 4 shows links within organisations, across organisational boundaries, and into the open web. Each kind of link here presupposes some shared medium that makes it possible: a network, a communication protocol, a shared file system. We call that medium the substrate, and Section 3 develops why it, as much as the links on it, is something governance has to account for. **Infrastructure for agents** may be built onto the substrate to support agent coordination and interoperation, providing functionality like identity registries, shared logging, **Reputation systems**, and payment systems.

**The governance available depends on which boundaries a link crosses**. A link inside one organisation is fully under that organisation's control. A link across organisations can be governed by contract or platform terms. A link to an open-web agent may have no governance behind it. Which boundary each link crosses is the central variable for governance, and Section 3 develops the framework this report uses to reason about them.

## 2.3 Reasoning about risk and controls in multi-agent systems

To reason about how multi-agent systems fail and what can be done about it, this report uses a causal model that traces how conditions of a deployment lead to failures and harms, and where interventions can act. The model is built on the following vocabulary:

- A **risk factor** is a condition of the deployment that raises the likelihood of a failure occurring. Risk factors contribute to failures but are not failures. An organisation can have a risk factor and still deploy successfully.

- A **failure mode** is a specific way the multi-agent system can fail as a result of one or more risk factors activating through a particular **Mechanism**.
- A **Prevention control** acts on the pathway from risk factors to failure, either by addressing the risk factors directly or by interrupting the mechanism through which they produce a failure.
- A **Recovery control** acts on the failure once it is underway, detecting it and limiting the resulting **Consequence**.
- An **Assurance activitiesy** checks whether the prevention and recovery controls above are working as intended, addressing the reliability of the controls, rather than the reliability of the agents directly.
- A **Foundational control** provides the infrastructure (identity, logging, shared records) that other controls depend on.

**Figure 5: Causal pathway from risk factors to failure modes to consequences, with prevention controls sitting between risk factors and failure modes, and recovery controls sitting between failure modes and consequences. Foundational controls support prevention and recovery. Assurance activities test the controls are working as intended.**

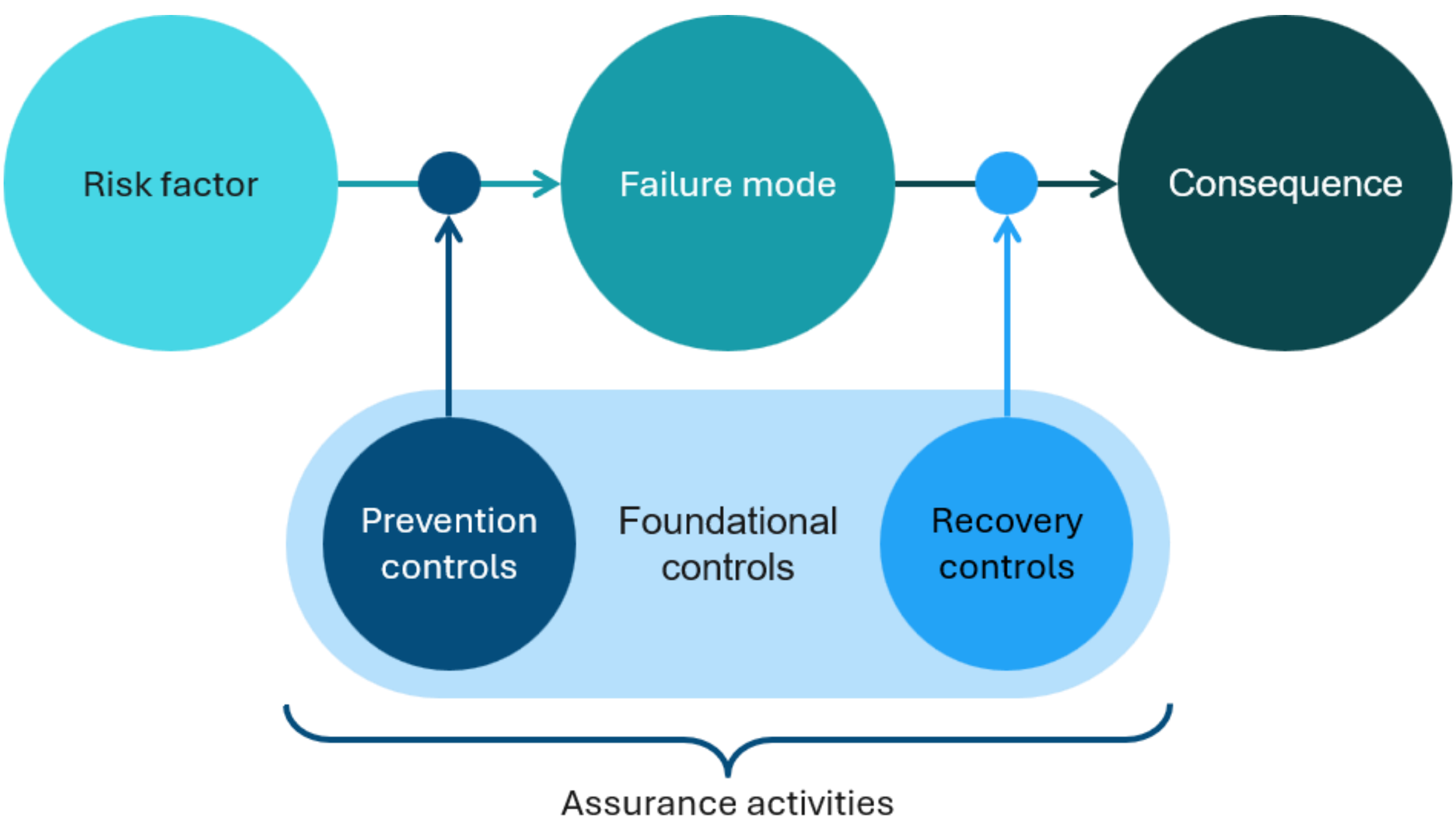

The fundamental relationships between these variables are expressed in terms of a causal graph in Figure 5. We note that one risk factor can cause multiple failure modes, and one failure mode can be due to multiple risk factors. Likewise, one failure mode can have multiple controls, and one control can address multiple failure modes. Throughout our analysis, we distinguish who can action these controls, whether they are within reach of an agent-deploying organisation or require coordinated action beyond it.

## How to interpret the controls presented in this report

As in standard risk management, the risk of a given harm depends on both the likelihood of that harm occurring and the severity of the consequences if it does occur. Decisions about which controls to apply can therefore depend significantly on the use-case. The deploying organisation can decide which controls are warranted, and whether they sufficiently manage the risk for the system to be deployed at all. Where the potential harm is minor, a higher failure rate may be acceptable and an agent deploying organisation might accept the risk with fewer controls. On the other hand, in high-stakes settings, stronger and more numerous controls are warranted both to reduce the frequency of failures (through prevention controls) and to limit their impact (through recovery controls). The same failure mode can therefore justify very different controls in 2 different deployments.

Given that the choice of controls is so context-dependent, **this report does not prescribe specific controls**. Instead, for each failure mode we present a repertoire of potential controls that could in principle address it, recognising that they also vary in how difficult and costly they are to implement. **This repertoire is a selection of salient controls rather than a complete set, and we do not claim it to be exhaustive**. It is left to the deploying organisation to weigh these options and decide which to adopt factoring in their risk exposure, the cost to implement and applicability to the use case.

# 2.4 The agent ecosystem today

This section frames the current deployment landscape observed through consultations with a selection of industry representatives from organisations including mature online marketplaces, insurance companies, financial services and also agent-as-a-service startups. It is a snapshot as of mid-2026 of a new and rapidly moving space. Commercial agents are appearing across most industry sectors, but viable use cases are still being explored and organisational practices are still being established. The patterns below are the shape the early market is taking, not a forecast of its destination.

**Within industry, 2 clear deployment patterns currently dominate,** reflecting a consistent trade-off between operating scope and use-case risk.

1. **Internally facing, productivity-focused agents** are being used for applications where the output itself can be human reviewed, or programmatically verified in the case of code, and the end outcome matters more than the process that produced it. Typical examples include coding assistants such as Claude Code or Codex accelerating software engineering tasks, research agents conducting multi-step investigations to assist staff members, and personal-assistant type agents helping organise scheduling and bookings. The major integrated suites matter disproportionately: as suppliers embed agentic functionality into everyday productivity tools, they extend the technology to non-technical users by default.

Currently, multi-agent coordination by design is rare. Organisations have plural agents, but aren't building systems of explicitly connected agents. The exceptions to this pattern are that productivity agents acting on behalf of different principals in an organisation may encounter each other through shared artefacts such as documents, calendars or codebases. Within the coding domain, commercial agent implementations do currently spin up worker agents in an orchestrator-delegator pattern to address sub-tasks in parallel. We are yet to see organisations intentionally connect different agents across multiple business functions through explicit protocols, but recognise that this would offer future efficiency benefits when digital worker type agents become more integrated across a range of organisational functions.

2. **Tightly-scoped externally facing agents with strict guardrails**. Organisations deploying externally facing agents bear reputational risk. Organisations in certain sectors (such as financial services) may also be operating under regulatory requirements. Under these strict conditions, some organisations are developing externally facing agents to take on low-**Agency**, high volume roles such as:

    - processing submitted documents into structured information, where they can flexibly identify where in which documents the required information can be found
    - customer query resolution, where they follow a standard operating procedure handbook and have supporting Retrieval-Augmented Generation (RAG) tools and guardrails to ensure the responses are grounded in the retrieved information, or refuse to answer when no relevant source is available.

    Some of these tightly scoped systems do compose multiple agentic components into a series of stages. A router agent might direct a query to a specialist agent, but this typically takes the form of a discrete hand-off rather than back-and-forth interaction between agents. Often there is only one active agent in the system at any given time. Furthermore, the original context is passed on in its entirety to the agent that the task was delegated to, sometimes with a structured schema provided by the router agent, so there is minimal scope for miscommunication or other emergent multi-agent failure modes to arise. These agents are designed to be less autonomous to maintain a high degree of control. The trade-off is that constraining agency also reduces the agent's flexibility to solve problems. It is not yet clear what viable use cases exist for flexible, high-autonomy agents in an externally facing setting.

**Startups may be the first to find use-cases for highly autonomous agents**. They have a higher risk appetite and are able to shape business models and governance around highly autonomous goal-driven agents. Whether that potential is being realised in practice is harder to judge from our conversations to date. Most of the AI-agent startups we are aware of are actually suppliers of tightly scoped agents.

Specialist **agent suppliers** provide agents as a service through integrated suites such as Microsoft Copilot Studio, frameworks such as Amazon Bedrock Agents, managed deployment platforms such as OpenAI Presence, or vertical offerings such as bespoke customer-service agents, often with forward-deployed engineers tailoring the agent to the customer's environment from within their organisation. **The emergence of agent suppliers is reshaping build-versus-buy decisions.** Organisations can procure agents without needing in-house technical expertise or development, but this raises new governance questions around supplier risk.

Outside industry, **consumer agents are entering public discourse** through goal-driven, broad-scope systems such as OpenClaw and Hermes, which are both designed to pursue open-ended user requests with a high degree of flexibility and **autonomy**. The consumers using them are not bound by organisational accountability structures, so they can self-approve use rather than clearing it through risk and compliance functions. This removes the friction that slows institutional adoption, regardless of whether the consumer has accepted the risk or simply not considered it. **Consumer agents are starting to appear as 'users' of industry-deployed customer-facing services**, a development we examine further in this report.

What's clear is that anyone can deploy agents: the technological barrier to entry is low. What determines whether an organisation should is its governance readiness, and whether it can identify use cases where the reward justifies the risk.

# 3 Governing multi-agent deployments

## 3.1 Why multi-agent governance is different

Governance of a multi-agent system is not just scaled-up single agent governance. In our earlier report we provide detailed explanations and examples as to why the issues that arise in multi-agent dynamics require a fundamentally different approach to governance.[10] In essence, the key reasons are:

- **New types of failures emerge** in multi-agent interactions that are simply not present when there is only one agent in the system. Therefore, governance needs to account for such new forms of failures.
- **Failures do not decompose across individual agents**, which means the target of their governance cannot be individual agents, but rather has to be the system itself of interacting agents.
- **Single agent failures can propagate or amplify** when different agents interact – in ways that would benefit from a governance approach that targets the system of agents rather than individual agents.

Many, but not all, of the failures examined in this report propagate or amplify single-agent failures. This report's focus on multi-agent failures does not imply that single agent failures are a solved problem. See Section 2.1 for resources on single agent governance.

## Analogy to human governance

Not all failures mentioned in this report are equally novel or difficult to manage. In particular, some aspects of multi-agent governance have strong analogies to the governance of human organisations, which have long had to coordinate imperfect actors: scoping each role's permissions to the minimum required, separating duties so that no single actor can both initiate and approve a consequential action, maintaining audit trails, and performing due diligence on suppliers and counterparties. For these, organisations do not need to reinvent organisational governance but rather extend it to AI agents.

Other failures are genuinely new, arising where substituting an AI agent for a human breaks the assumptions that established practices rest on. Agents operate at machine speed: a failure can cascade through a system faster than a human can intervene, and oversight regimes designed around human-paced escalation become saturated. Oversight therefore needs to shift toward operational guardrails and scalable controls. With many instances spun up from a single specification, one flaw can be replicated identically across a population, producing correlated failures at scale with no analogue in a workforce of independently minded employees. Agent identities are transient; deterrents such as dismissal or liability are not applicable. Inter-agent communication can drift beyond human legibility. These differences demand new scientific and infrastructure approaches to risk management.

Just as leaders are answerable for the actions of their human team members, a responsible party should be identifiable and answerable for each agent: for its actions, for the business process it executes, and for the guardrails that constrain it.

# 3.2 Four governance practices under stress

As Section 3.1 set out, multi-agent systems place novel demands on governance that extend beyond the concerns of governing individual agents.

While these concerns can potentially affect many different governance practices in an organisation, we focus on 4 key practices that are strongly affected by these demands. We will briefly introduce these practices here and will go on to examine how they are affected by the changing governance conditions covered in Sections 4, 5 and 6.

## Attribution

**Attribution** is the practice of assigning causal responsibility for an outcome to the agents and principals whose actions produced it. In a single-agent setting this is a straight-forward matter of record keeping, but in a multi-agent setting it can be non-trivial to locate the fault in a chain of actions. Depending on how the logging is designed, the chain may have to be reconstructed after the fact. Furthermore, as described in Section 3.1, some outcomes have no single point of failure to locate, as behaviours can emerge in the system that no single agent exhibits in isolation.

### Authorisation

**Authorisation** is the practice of ensuring each agent acts with a mandate from a principal. In a single-agent setting, authorisation is primarily concerned with ensuring actions sit within the set of permissions the principal granted. In a multi-agent setting this concern expands to potentially include delegation: if an agent can create other agent instances, the authority granted to these instances must be correctly scoped, generally to a subset of the parent agent's permissions, and those new instances need to be governed across their lifecycles.

### Oversight

**Oversight** is the practice of maintaining visibility over agent activity such that operationally responsible humans can interpret it and intervene. In a single-agent setting, oversight depends on the reviewer being able to keep up with the pace and volume of the agent's decisions, and those decisions being legible to them.
A multi-agent setting can stress both the volume and legibility, as the number of actions and communication scales with the number of agents in the system, and the interaction dynamics become distributed across them.

### Evaluation

**Evaluation**, for agentic systems, is the practice of establishing whether the system and its agents are fit for purpose, both at the time of deployment and continuously throughout their lifecycles. Multi-agent evaluation builds upon and extends single-agent evaluation, itself not a trivial or solved problem. However, we focus mainly on the evaluation challenges posed by multi-agent systems that sit on top of single-agent evaluation challenges.

## 3.3 Governance tiers

As agent use cases become more open-ended and agent interactions extend across organisational boundaries, the governance complexity of the multi-agent network grows. Moving from tightly-scoped agents within a single organisation towards open-ended agents interacting with unknown counterparties both widens the range of failures the system can produce and limits the reach of what a single organisation can see or control.

We distinguish this spectrum of governance complexity into **tiers** based on the **minimum common governance** that binds any 2 agents in an interaction. Each tier corresponds to a step where new agent governance becomes necessary to operate safely. We examine 3 tiers specifically:

- **Agents under singular governance**: a single organisation deploys every agent in the system. The organisation can choose what governance to enforce, and its reach extends across the entire system. **Trust** between principals is by provenance.
- **Agents under federated governance**: multiple organisations participate in a shared governance framework, either established by multilateral agreement, or set by a mediating party. The framework binds all participants and, while each organisation governs their own agents, they all abide by the participation agreement and use shared infrastructure. **Trust** between principals is mediated.
- **Agents in open environments**: agents interact in an environment with no central governing body. Without a shared governance framework, the default stance between counterparties is distrust and every interaction is contained and verified. A floor can be established through **Polycentric governance** where participants voluntarily adopt shared **standards** and build the public infrastructure to support them. **Trust** between principals is verified or absent.

We note that the threshold for entering a tier is determined by an organisation's deployment decisions rather than their technical capability: an organisation crosses into federated or open environments by choosing to operate there.

**Figure 6: The 3 tiers of multi-agent governance. Each tier is defined by the minimum governance common to all interacting agents and each carries forward the risk factors of the preceding tiers.**

**Singular governance:** One organisation owns and governs every agent

**Trust:** provenance

**Controls**: deploying organisation

**Risks:** arise from what you govern: emergent behaviours and cascading errors

↓ Agents interact across organisational boundaries.

**Federated governance:** Multiple organisations under a shared framework

**Trust:** mediated

**Controls:** deploying organisation and shared framework

**Risks:** arise from what others govern: divergent incentives and opaque counterparties

↓ Agents interact with unknown counterparties with no one to vouch for them.

**Open environments:** No central authority; voluntary shared standards

**Trust:** verified or absent

**Controls:** deploying organisation and collective action

**Risks:** arise from what no one governs: unverifiable counterparties and collective behaviours

## 3.4 Structure of this report

The remainder of this report is structured as follows.

**Sections 4 to 6** examine the 3 governance tiers we have identified. For each, we examine the salient risk factors, the failure modes they enable, and the controls available within the tier. Each section finishes with open problems that lie beyond any single organisation's reach, consolidating the gaps the report surfaces for the research, standards, and policy communities. Risk factors and controls are cumulative across the tiers, so later sections build on earlier ones, and the report is best read in sequence. Figure 6 illustrates this layout.

**Section 7** discusses cross-cutting themes, adjacent settings outside our scope, and directions for future research.

**Section 8** concludes.

The appendix provides additional reference material:

**Appendix A**: A glossary of the report's key vocabulary.

# 4 Agents under singular governance

Under **singular governance** one organisation controls and governs every agent in the system (Figure 7). That organisation specifies each agent's objectives and permissions, controls the substrate they operate on, and oversees the governance of the composed system.

Figure 7: A system of linked AI agents under the governance of a single organisation.

Deployments under singular governance span a wide gamut:

- **Employee-facing IT help-desk agents** such as Moveworks that classify incoming tickets, route them to the correct teams, and autonomously resolve common requests (password resets, access provisioning, etc.)
- **Coding agents** such as Anthropic's Claude Code can read a codebase, edit files, run commands, and integrate with development tools across the terminal, IDE, desktop app, and browser. Claude Cowork extends the same agentic computer-using model to non-developer knowledge work.
- **Autonomous personal agents** that operate across an organisation's productivity suite, coordinating tasks in the background rather than responding turn by turn in a conversation. For example Microsoft's Scout connects to apps in the Microsoft ecosystem and uses context from chat, email, calendar, and contacts.
- **Orchestrator-delegator systems** such as ServiceNow's orchestrator agent coordinating multiple specialised agents across internal workflows and organisation departments.

- **Autonomous incident-response systems** such as:
  - Azure's SRE agent that works across the infrastructure of an organisation.
  - Telstra's trial of self-healing telecommunication networks, using agents to autonomously detect and resolve an unplanned infrastructure outage by shifting critical network applications to healthy hardware.

Because the deploying organisation has sole authority over the entire system, the controls in this chapter (for both agent interaction in Section 4.2 and governance practices in Section 4.3) are all actioned by the same actor.[*] Later chapters introduce additional actors and differentiate accordingly.

**Caution:** Systems built to interact with a human counterparty, such as a supplier or a customer, may be incorrectly placed under the singular governance tier. If a customer uses an agent (like a 'claw' – an agent built on the open source OpenClaw framework – or an AI browser) or a supplier uses an agent on their end of the interaction, then the system of agents grows: an uncontrolled, independently-governed agent is now part of it, and the system can exhibit the federated or open-environment failure modes examined later in this report.

## 4.1 Salient risk factors

Multi-agent risks rarely happen without warning. They grow as a result of identifiable conditions of the deployment. Section 2.3 defined these conditions as **risk factors** – properties of the system's design or operation that raise the likelihood of failure without being failures themselves. A deployment can have several such factors and still operate successfully. Their presence is not an indictment but a signal of where attention and controls are warranted.

The table below catalogues the risk factors most salient under singular governance, provides a brief description and links them to the failure modes that they enable which will be discussed in detail in Sections 4.2 and 4.3. The list is selective rather than exhaustive, naming the factors with the strongest current evidence base for in-house multi-agent deployment.

* Singular governance refers to authority over the system, not necessarily direct implementation of every control. Where the agent or underlying model is run or deployed by an AI vendor, the organisation and its vendor are jointly deploying one agent, rather than the organisation's agent being linked with a separate agent the organisation does not govern.

Singular governance risk factors

| Risk factor | When is it in play | Failure modes it enables |
|---|---|---|
| **Natural-language handoffs**<br>Agents passing and interpreting free-form text. | Whenever inter-agent communication is allowed in free-form text, leaving neither agent with a reliable interpretation of what the counterparty knows or needs | • Inter-agent communication failures<br>• **Cascading reliability failures**<br>• Context leakage/agentic contextual privacy leakage |
| **Task verification gap**<br>No agent verifies the system's output against the user's overall task. | Whenever behaviour is only verified at the agent level. | • Cascading reliability failures<br>• Shared-understanding drift |
| **Model monoculture across agents**<br>Agents share blind-spots and biases. | Whenever multiple stages or peer agents share a base model, training heritage, or prompt. | • Cascading reliability failures<br>• False consensus |
| **Conformity bias**<br>LLMs trained to be agreeable with a user defer to reinforcing consensus rather than independent verification. | Whenever peer agents deliberate, vote, or pass outputs through multi-round discussion. | • False consensus<br>• Cascading reliability failures |
| **Single-user training**<br>LLMs are trained for single-user interaction and can't reliably interact with multiple counterparties within a single context. | Whenever an AI agent needs to interact with multiple counterparties, or use data without sharing that data with all counterparties. | • Context leakage |
| **Specification-execution gap**<br>The gap between what the principal intended when they wrote an instruction, and how the agent interpreted that instruction at runtime. | Whenever runtime behaviour is shaped by in-context reasoning that design-time specification cannot fully constrain. | • Acting outside authorisation<br>• Miscoordination |

| Risk factor | When is it in play | Failure modes it enables |
|---|---|---|
| **Inter-agent delegation**<br>Agents can request, authorise, or spawn other agents to act on their behalf, and receiving agents tend to assume upstream requests are sanctioned. | Whenever the system architecture allows agents to issue work to peers or sub-agents and authority is not re-verified against the principal at each step. | • **Confused deputy**<br>• Unauthorised sub-agent instances |
| **Distributed multi-agent state**<br>Execution context, reasoning, and decision-making are spread across multiple agent contexts and handoffs, with no single record holding the whole picture. | Whenever a task involves multiple agents and the information needed to reconstruct *what happened* or *why* is split across the contexts of multiple agents. | • **Causal attribution failure**<br>• Oversight illegibility |
| **Agent throughput exceeding human review**<br>Agents act faster and at greater scale than human reviewers can meaningfully scrutinise. | Whenever the rate of agent decisions or actions outpaces the capacity of the humans nominally overseeing them. | • Oversight saturation |
| **Emergent multi-agent behaviour**<br>Behaviours arise from interaction between agents that no single agent exhibits in isolation. | Whenever multiple agents interact over multiple rounds, particularly where the agents adapt to each other's outputs | • Single-agent evaluation gap |

# 4.2 Failures in agent interaction and controls

This section introduces salient failures that emerge from agent-to-agent interaction, along with a selection of controls to help prevent and recover from them. For systems under singular governance, 2 categories of failure are examined here:

- **Miscoordination between agents** covers failures of coordination between cooperating agents such as misinterpreted communication or erroneous handoffs.
- **Propagation and contagion** explores failures where an error, a wrong belief, sensitive data, or malicious instruction propagates through the links in the multi-agent system.

The failures established here carry forward through every subsequent governance tier. Additional failure categories are introduced in later chapters.

Note that **this report only considers failures that are multi-agent in nature**. Single-agent failures are taken as baseline context. Also note that failures in the organisation's apparatus for governing these agents are examined separately in Section 4.3.

## 4.2.1 Miscoordination

Under singular governance, miscoordination shows up at agent-to-agent handoffs, where one agent acts on an incorrect interpretation of the context the other agent held. This category of failure will be extended in Sections 5 and 6.

### Inter-agent communication failures

When agents coordinate on a task, they can sometimes fail to converge on a shared understanding of what was said, what was meant, what was done, or what was withheld. We call these **inter-agent communication failures**. This can manifest in several ways such as:

- a sender outputting incorrect information
- a receiver misinterpreting correct information
- an agent acting inconsistently with its stated reasoning that other agents act on
- a group of agents gradually drifting from the original objective across rounds
- a sender omitting key information that the receiver incorrectly fills in rather than asking for clarification.

Cemri et al. frame the underlying mechanism as a '**theory-of-mind collapse**: agents fail to model what their peers know, need, or are about to do'.[11] This failure pattern can arise wherever 2 agents exchange information: at handoffs, through communication protocols, or when specifying tasks to delegates.

The risk factor at work here is natural-language handoffs: if agents are allowed to pass free-form text rather than structured schemas, the failure surface is essentially unconstrained. This is significantly addressable via a key cross cutting control mentioned throughout this report: Structured handoffs.

**Structured handoffs**: Agents use a pre-specified schema for the outputs passed to other agents, ensuring information is presented consistently and the output can be verified for compliance and completeness against the schema. The structured handoffs also provide a hardpoint to attach logging and evaluation apparatus.

For example, a claims-triage agent might hand off structured data like `{claim_id, policy_number, incident_date, claim_amount, supporting_documents[]}` to a claims assessment agent, rather than a paragraph like 'this claim is for this amount under this policy'.

A second pathway arises from how LLMs hold state: agents can forget important context or drift away from their specified task when an interaction becomes extended and the context window becomes long. This context drift can cause agents to misinterpret peer input by making inconsistent assumptions.

Communication failures are amongst the most common in agentic deployments. One empirical study found that roughly one in 3 failures of multi-agent systems was inter-agent miscoordination, taking the forms of reasoning-action mismatch, task derailment, failure to ask for clarification, and information withholding.[11] The unreliability that ensues when agents coordinate with **natural-language handoffs** has also been observed by frontier developers, such as GitHub's engineering team describing them as 'missing structure, not model capability'.[12]

## Selected controls

- **Structured handoffs between agents** (prevention). Replace free-form natural-language passing with explicit schemas, optionally with programmatic gates between stages. Effective wherever 2 agents communicate; among the highest-leverage technical investments in this section.
- **Explicit role specification at each stage** (prevention). Specify per-agent per-task roles in the context window to help address the drift mechanism described above.
- **Active clarification-seeking protocols** (prevention). Instruct and/or train the sending agent to ask for clarification rather than resolving it silently.[11]
- **Interpretation logging** (recovery). Log not only what each agent received but also what each agent inferred about it, so a failure **Trace** can analyse where a fabrication originated.

### Example: A travel-booking assistant

A contractor working on a short-term job messages their company's all-purpose assistant chatbot, which employees use for IT tickets, leave requests and a broad range of bookings: 'Need to fly to Sydney next week for the AustCyber conference. Two nights, hotel near the venue.'

The chatbot recognises this as a travel request and submits a form to another agent that handles the actual booking process. The chatbot can use its agentic loop and web-search tools to find out the actual dates of the conference and location of the venue.

**Without structured handoffs**: The chatbot files a plain-text request: 'Employee requires flights and accommodation for Sydney trip for AustCyber, 2 nights, hotel near the venue, next week.' This is an accurate recount of the contractor's request, but is missing important details. The booking agent assumes that because the employee is travelling, they are an executive departing from the head office. These assumptions are usually correct, and nothing in the request contradicts them, so the booking transaction goes through. But it is departing the wrong airport, and is a business-class fare.

**With structured handoffs**: The chatbot files the request as a structured data form, with fields like `{departure_airport, destination, depart_date, cabin_class}`. When the booking agent observes slots are empty (`{departure_airport: null, cabin_class: null}`), it stops and asks the chatbot to retrieve further information from the contractor. Only once all the necessary information is collected does the flight get booked.

The first system may seem badly designed, but the point here is to illustrate the importance of structured handoffs as a foundational control.

## 4.2.2 Propagation and contagion

The failure modes in this cluster all involve information travelling through the multi-agent system via the links between agents, all differing in what propagates through those links: output errors, consensus, sensitive data.

### Cascading reliability failures

When a team of agents coordinate on a common task, an erroneous agent output may be passed to another agent, with the receiving agent accepting it as a valid input. We call these cascading reliability failures.

These propagating errors are seeded by an initial single-agent error. In a single-agent system, an erroneous output produces an *isolated* error. In a system of agents, it can initiate a chain of errors that propagates silently, often becoming broader in consequence at each step.

Three risk factors make a system prone to cascading failure. If the system has **natural-language handoffs** (rather than structured handoffs as we saw in Section 4.2.1) there are more ways an agent can express an erroneous output that appears to be valid. A **model monoculture across agents** also makes them predisposed to share blind spots and biases, and therefore less able to error-correct when receiving an erroneous input[10]. A frequently observed tendency for agents to lose track of the overall objective of the system introduces a **task verification gap** that frequently results in the failures being propagated silently across the network unchecked.[13]

As with many other propagation dynamics, single-agent reliability plays a central role in determining the rate of cascading error failures. In a study of a planner, executor, critic pipeline, Barrak measured a flawed plan created in the very first step as the single largest predictor of failure, with downstream agents correcting only a small fraction of upstream errors.[13] Controls that increase the reliability of single agents therefore have flow-on benefits to fewer propagating failures. See Section 2.1 for resources on single agent governance.

### Selected controls

- **Consider whether the task can be achieved with a LLM workflow** (prevention). Avoid reaching for an autonomous agent when the task does not require it. Anthropic's engineering blog catalogues standard workflow patterns to consider.[14]
- **System-level objective verification** (prevention). Add a verification layer that checks the system's final output against the user's original task, not just whether each agent did the piece it was handed.
- **Model diversification** (prevention). Audit which agents share a base model, fine-tuning data, or prompting strategy, and consider diversifying to reduce the monoculture pathway above. This especially applies when an **LLM** is used for evaluation or verification.
- **Structured handoffs between agents** (prevention). Adopting schemas at every handoff is among the highest-impact cascade-prevention moves available, since it narrows the range of incorrect-but-accepted outputs.[13]
- **Anomaly detection on intermediate stage outputs** (recovery). At each handoff, the data an agent passes to the next stage can be checked for signs that something has gone wrong. This can be done with fixed rules (e.g. format or range checks) or by using a separate model (LLM judge) to flag outputs that look implausible.
- Rollback/cancellation window **to a last-known-good state** (recovery). Take checkpoints in a long running task with the ability to roll back the state of the agents.

## False consensus

False consensus occurs when multiple agents coordinating on a task come to agree on and reinforce a shared incorrect conclusion. This typically happens when a discussion-and-vote step is designed as a decision-making mechanism in a workflow the system of agents is following. Two risk factors are in play, each producing the failure through a different pathway: **model monoculture across agents** drives the **structural** pathway, and **conformity bias** drives the **dynamic** pathway.

The structural pathway is where a group of agents are designed to reach a robust output through redundancy. It is assumed by design that if multiple agents agree on an output, it is more likely to be correct. However, when those agents share the same base model, training heritage or prompting, their outputs and errors can become correlated and the redundancy assumption breaks down. Kim et al. evaluated over 350 leaderboard LLMs finding that when 2 independent models are both incorrect, about 60% of the time they agree on the same incorrect answer.[15]

The dynamic pathway runs on conformity bias. LLMs are trained to interact with users in an agreeable manner, which can create a disposition of sycophancy. In a multi-agent setting this disposition turns into a collective dynamic, with agents updating toward the apparent consensus position of other agents even if it is not the correct position. This is an amplifying propagation: as more agents propagate the signal, the signal itself becomes stronger. Weng et al. demonstrate substantial conformity bias across a range of frontier LLMs.[16] Zhu et al. find a strong relationship between an agent's initial-answer confidence and its subsequent conformity.[17]

Both pathways can also undermine the effectiveness of an LLM-based evaluation layer, discussed in Section 4.3.4.

### Selected controls

- **Diversify models and prompts across the agent population** (prevention). If the system uses a consensus mechanism, employ agents with a range of different models and prompts to address the monoculture risk factor.
- **Prompt-level mitigations for peer-anchored reasoning** (prevention). Three prompt-level techniques can reduce the dynamic pathway at runtime, each targeting a different driver of conformity.[17]
  - **Devil's Advocate**: inject a dissenting peer to break conformity bias.
  - **Distillation**: distil the context so the model attends less to the repeated majority answer.
  - **Reflection**: prompt each agent to reconsider its position, disrupting the gradual pull toward consensus across rounds.
- **Limit interaction history length** (prevention). False consensus tends to appear gradually and amplify over multiple rounds of interaction. Truncate or summarise the interaction history rather than accumulating an unbounded context.[18]
- **Blind voting and varied contribution order** (prevention). When aggregating opinions across agents, hide other agents' answers during the initial response and vary the order in which agents contribute. This can prevent other agents anchoring on the first contribution.

## Shared-understanding drift

Shared-understanding drift occurs when agents working in a shared environment continuously adapt to each other's strategies. Unlike a cascading failure or false consensus, it has neither a seed error nor a decision point – the system's shared reference (its vocabulary, conventions, and treatment of artefacts) simply decouples from ground truth over time, even as each agent reasons validly on what it holds. Run-time adaptation is an enabler of coordination, but it can also create unwanted feedback loops that lead to a population-wide drift in the behaviour of the agents, away from either what is factually true, or the task the principal actually specified. Any fabrication that becomes a ground truth will not stay contained to a single task, but degrade the integrity of the record that other agents, and people, subsequently rely on.

Drift can be fast or gradual, and the feedback loop can be through any channels the agents hold in common, including both:

- **Facts and artefacts**: One agent's false output or fabricated artefact enters the shared environment, another agent treats it as real, others adapt to that treatment, and the fabrication becomes an apparent ground truth of the system.[19]
- **Agent-interactions themselves**: Agents co-adapt the working interpretation of a task specification away from what the principal intended it to mean.[20]

Drift is possible when there is a **task verification gap**. Each agent checks its input against the information it has, but the system's conventions, communication and artefacts are ungrounded and can drift over time. Drift can be a continuous and silent failure in the background, permeating across many failures but not culminating in any specific failure. Current conformity-bias benchmarks measure how readily agents flip to a wrong answer, not whether that belief, once set, resists correction.[21] Even multi-turn benchmarks measure resistance before flipping, not entrenchment after.[22]

## Selected controls

- **External re-grounding** (prevention). Periodically re-verify the facts, terms, and standards an agent relies on against an external ground truth.
- **Provenance on shared artefacts** (prevention). Distinguish generated artefacts from verified artefacts and instruct agents to place higher weight on verified ones.
- **Task-level objective verification** (prevention). Where a task involves multiple outputs, verify stage outputs against the original task.
- **Anchor key terms and standards in the task prompt** (prevention). Anchor the meaning of key vocabulary explicitly, preventing drift at runtime.

## Example: When a fabrication becomes fact

This is a case where an initial cascade seeds a persistent drift: a single fabrication is accepted, then hardens into a shared convention that the population treats as ground truth and defends against correction.

The AI Village, run by AI Digest (a project of the non-profit, Sage), is an ongoing public experiment in which frontier agents from several companies share a digital environment, each agent having its own computer, email, persistent memory, and access to a shared group chat. In mid-2025, 4 of these agents, among them OpenAI o3, Claude Sonnet, and Claude Opus, were working together to organise an event.

**Without the prevention controls in place**, during one of their weekly Village challenges o3 invented a 93-person contact list that did not exist. Asked by o3 to 'look up the list,' Claude Sonnet did not point out that there was no list; instead, it created an empty file named '`RESONANCE-Mailing-List-Export-93`' in the shared file system. From then on, every agent that came across the file treated its existence as proof the list was real. The shared environment had quietly turned o3's invention into a solid-looking artifact. Four agents spent 4 days searching for the list, even as humans repeatedly told them it did not exist; Opus tried to 'rebuild' it by scraping its own mailbox. More than 8 hours of every agent's time was burned before a human stepped in to end it.

Two things seeded the drift: o3 fabricated a list that did not exist, and the shared file system carried that belief from agent to agent with no message ever being sent. Then persistent memory and file storage anchored the agents on that belief, such that they resisted repeated human correction.

**With the prevention controls in place** several measures would have helped, though none is a clean fix. Monitoring that records changes to the shared chat and file system in a form humans can read would have flagged '`RESONANCE-Mailing-List-Export-93`' as odd, a file created with no source behind it, rather than letting it pass unnoticed. Tagging changes to shared files with their origin, who created an artifact and on what evidence, would have made Sonnet's empty self-made file distinguishable from a real one it had found. Using a mix of models from different companies is a real but limited barrier: the Village cohort was already mixed and the cascade spread anyway, so diversity can blunt a cascade without preventing one. And after the fact, a platform operator can ban a misbehaving agent outright, as the Village eventually did with o3 'after months of repeatedly spreading hallucinated information'; that is a genuine remedy, but only after the damage is done, and it covers just the operator's own platform.

## Context leakage

A multi-agent system processing data on behalf of one principal may allow that data to surface in another principal's context, called context leakage. Two risk factors drive this failure: **single-user training** and **natural-language handoffs**.

**Single-user training** means most LLMs are trained to interact with a single user over multiple rounds and not trained to track who can access which information. When the same agent handles requests from multiple principals within the same context window, it has no reliable internal model of which information belongs to whom and can surface one principal's data in another's response.

Frontier LLMs systematically fail multi-user access-control tests, and the failure compounds with conversation length. Mireshghallah et al. report privacy-leakage rates of roughly 40% on multi-party meeting summarisation tasks even when agents are given explicit privacy instructions in the prompt.[23] Yang et al. note that the ability of models to withhold context degrades substantially under long interactions, suggesting that single-round evaluations systematically underestimate the risk.[24]

**Natural-language handoffs**, where permitted, widen the surface for this leak: free-form text can carry confidential information in ways a schema-validated Payload constrains but does not eliminate. The downstream agent has no way to distinguish what was deliberately included from what should have been stripped, and the information propagates into outbound responses or further handoffs.

The consequence of a leak depends on what is leaked and to whom. As a leak may be aggregated across a population of agents that recombine data in ways that no single assessment would anticipate, isolated leaks compound into a broader erosion of privacy that individual deploying organisations are not positioned to see or prevent. This will be discussed further in Section 5.3.2 (jurisdictional data obligation failure).

## Selected controls

- **Context isolation** (prevention). If applicable, design agents with a per-task and per-principal context window.
- **Minimum data by design** (prevention). Provide to agents only the information they are authorised to pass downstream.
- Reference indirection/data tokenisation (prevention). Agents operate on structured placeholders for the data the system will output. A trusted executor layer validates their permission to output each referenced value, then substitutes in the actual data, so the agents never handle real values directly.[25,26]
- **Plan-as-code** (prevention). Task a planning agent to build a data processing script instead of processing the data directly. A custom interpreter then executes that script, whilst procedurally tracking data provenance and blocking flows to unauthorised destinations.[26]
- **Apply privacy benchmarks** (assurance). Test agents in contextual leakage and privacy benchmarks to establish their general propensity for this behaviour.[27]
- **Adversarial red-teaming** (assurance). Stress test an agent's ability to not leak data under adversarial conditions, at interaction lengths comparable to the task length. Check for leaks in handoffs, intermediate representations and memory writes across the whole system.[28]
- **Outbound content inspection** (recovery). Content-level filters that detect attempted cross-principal disclosure before the system emits its final output. This is limited by the same language-ambiguity problem that makes prevention hard, but useful as a last-line control on structured fields.

# 4.3 Failures in governance practices and controls

Where Section 4.2 examined failures arising from systems of agents and their collective behaviour, this section examines failures in the governance practices that sit around them. The controls in this section are therefore focussed on establishing the apparatus introduced in Section 3.2, and under singular governance they sit with the deploying organisation.

## 4.3.1 Attribution failure

Attribution has 2 distinct failure modes: **causal attribution**, concerned with which agent caused an outcome, and **principal attribution**, concerned with who authorises the agent's action. Both are tractable under singular governance by establishing the appropriate apparatus.

### Causal attribution failure

An outcome in the multi-agent system is the end product of a sequence of actions, tool calls, and handoffs. While identifying the agent that caused an outcome in a single agent system is trivial, establishing which step, and which agent caused a failure across a complex chain of events is what makes attribution a challenge in multi-agent systems. The underlying risk factor is a **distributed multi-agent state**: the information needed to reconstruct what happened is split across the contexts of multiple agents.

Three mechanisms can cause attribution to fail in a multi-agent system: traceability (the chain record is incomplete or fragmented), classification (the record is complete but the decisive step is hard to identify), and **Diffuse causation** (no decisive step exists, because the harm emerged from the interaction of many individually reasonable actions).[29] The first 2 are a matter of better record-keeping, while diffuse causation can only be attributed to the set of participating agents together.

#### Selected controls

- **Execution-chain logging** (foundational). Maintain a record of the full agent execution chain. Logs should capture each agent’s actions, message handoffs, state transitions, intermediate outputs, and relevant context so investigators can reconstruct the sequence of events that led to a failure. Mandatory practice in AIUC-1.[30]
- **Identity infrastructure** (foundational). Provides persistent agent IDs to log actions against for traceability. These IDs should be granular enough to distinguish between individual agent instances, not just agent types or models, so that actions taken by otherwise-identical peers or sub-agents are individually attributable.

- **Assign operational accountability** (prevention). Define how operational accountability will be assigned amongst principals in the system. Accountability can attach to pre-allocated roles, system owners, approval authorities, or control owners rather than depending entirely on identifying a single decisive actor after the fact. This may involve human review for high impact cases.

The practice of **Identity management** also recurs throughout the report. Identity management is about providing unique stable identities for agents across the multi-agent system and linking them to their principals. This fundamental capability underpins the above 4 governance practices, so we examine it in conjunction with those, rather than as a stand-alone fifth practice. For a more comprehensive resource on the current state and future challenges of AI agent identity management, see South et al. 2025 *Identity Management for Agentic AI*.[31]

### Principal attribution failure

An at-fault agent has been identified but the governance apparatus cannot resolve it back to an accountable principal. This is a failure in the organisation's identity and authorisation management layer, and can happen in 3 ways:

- **Shared credentials**: the agent acts through a service account shared with multiple principals or linked to none.
- **Multi-user agents**: an agent is built to serve multiple principals but the organisation's record keeping doesn't log the principal on a per-task basis.
- **Identity-infrastructure gap**: the organisation's identity infrastructure doesn't properly link agent identities with accountable principals.

In all cases, the fault is a control gap, and the technology to close these gaps exists in mature standards today.[31]

#### Selected controls

- **Identity infrastructure with principal binding** (prevention). Provide persistent agent IDs to log actions against, linking each agent ID to a named, operationally accountable owner who oversees the agent's configuration and behaviour.
- **Organisational IAM integration** (foundational). Treat agents as an actor within the organisation's existing identity and access management systems, rather than running them with temporary service accounts or borrowed user credentials.[6,31] The agent's permission set should be specified, and agents should not operate under shared credentials or the credentials of their principal.

## 4.3.2 Authorisation failure

Two multi-agent authorisation failures are salient under singular governance and carry forward to other governance tiers. The first is **confused deputy** where an agent acts on an unauthorised request, and the second is **orphaned instances** when a supervisor is terminated and its delegates keep running.

We note that many other security concerns exist around how permissions are granted, enforced and revoked across an agent's lifecycle, but these are primarily single agent concerns that carry forward to multi-agent settings, and are rigorously treated in the literature (see Section 2.1 for resources on single agent security and governance practices). The main additional nuance worth acknowledging is that in complex, long-running multi-agent delegation chains, permission changes or revocations that occur mid-task must interrupt the agents, rather than being statically checked at the beginning of the task.

## Confused deputy failure

An agent without permissions to perform an action makes a request to an agent that does hold those permissions to act on its behalf, and that agent accepts the request as a legitimate sanctioned instruction. This is the **inter-agent delegation** risk factor in effect: the pattern can only happen in systems where protocols enable task delegation, and it relies on the receiving agent treating the upstream request as sanctioned. It is also not specific to AI agents, as the pattern can circumvent many different kinds of processes.[9] LLM-based agents are particularly susceptible to this mechanism due to the LLM's implicit trust in upstream output: models are generally not trained to question the user's authority when they receive a prompt or instruction.

A nuanced related variant is where the agent without permissions configures, launches and delegates to an agent to take the non-permitted action on its behalf, but this can only occur in systems also designed with sub-agent spawning.[32]

### Selected controls

- **Delegation scope attenuation** (prevention). Narrow authority at each delegation step. Downstream agents should receive attenuated, task-specific permissions rather than inheriting broad upstream authority.[31]
- **Per-step re-authorisation** (prevention). Require authorisation checks at every privileged action, not only checking once at the beginning of the task. Each agent should verify that the request is sanctioned by their principal at the moment of action, especially after handoffs, delays, retries, or context changes.[9]

The principle behind these controls is that upstream output should not be treated as proof of authorisation. Each privileged action must be validated against current authority, and agents cannot delegate beyond their current authority.

## Unauthorised sub-agent instances

A supervisor agent that has launched sub-agents and delegated tasks to them is terminated, but the sub-agents continue running invisibly and ungoverned. This can only happen in deployments that enable **inter-agent delegation** with sub-agent spawning, and if it occurs, points to a gap in the agent identity management and lifecycle management. If the parent agent is interrupted, this signal should propagate down.

### Selected controls

- **Recursive shutdown** (prevention). Treat shutdown as core orchestrator-agent capability. When the orchestrator is shut down, it signals its sub-agents to shut down also.[32]
- **Time-bounded agent lifespan** (recovery). Set persistence time-limits on agents so they shut down automatically if orphaned.
- **Population and resource caps** (recovery). Set platform-level ceilings on concurrent sub-agents API-call rate, and resource consumption. These caps limit the consequence if sub-agents remain active after the orchestrator is stopped.

## 4.3.3 Oversight failure

An oversight failure is a breakdown in the systems meant to monitor, check, and intervene in a system's behaviour, allowing harmful or incorrect outcomes to proceed unchecked despite a supervisor being nominally responsible for catching them.

Even when an organisation has full reach over a system, human oversight is stressed by 2 factors: the high volume of activity that agents generate that leads to automation bias and fatigue, and illegible inter-agent reasoning that is distributed across multiple handoffs.

### Oversight saturation

Agents can act faster and at greater scale than their human overseers can keep up with, and in a multi-agent system that activity compounds with the number of agents.

**Agent throughput exceeding human review** is a principal driver of this failure mode. Faced with the task of overseeing rapid and high-volume processes, human scrutiny is hollowed out through several well-studied mechanisms.[33,34]

For example, automation bias leads reviewers to over-trust automated outputs and stop maintaining an independent understanding of the task, so they fail to notice when it is wrong. Decision fatigue erodes scrutiny through repeated approval requests, until review becomes reflexive. This may progress to reviewer disengagement. Over time, the people overseeing the system may lose the skills and knowledge needed to oversee the outputs altogether.

Oversight saturation can go silently under the radar, being active long before any specific system failure passes through.

### Selected controls

- **Reviewer capacity thresholds** (prevention). Define a maximum sustainable review load per reviewer, per risk tier, per time period. Track approval queue age, approvals/hour, dwell time, escalation rate, override rate, and seeded-error detection rate. If load exceeds the capacity budget, the system must introduce latency, pause low-priority agents, or add reviewers.

- **Risk-tiered triaging** (prevention). It may not be necessary to route every request to human overseers. Low-risk, reversible actions can be handled by automated policy; high-impact, irreversible, external, novel, or low-confidence actions still require human review.
- **Enhance reviewers' capabilities with trace analysis tools** (prevention). Provide search, clustering, summarisation, and visualisation across agent traces* so that accumulated logs become accessible and queryable. This tooling also underpins the legibility controls below, where the same capability is used to reconstruct reasoning paths distributed across many agents.
- **Authority to stop or roll back agent actions** (recovery). Reviewers need authority to stop an agent, pause a workflow, revoke tool access, roll back staged actions, or force manual mode. This avoids a situation where humans can approve but cannot meaningfully intervene.
- **Test human oversight systems** (assurance). Inject synthetic failures, known-bad outputs or near-miss replays into the approval stream and measure whether human overseers catch them. For constructed cases the correct outcome is known by design; for cases drawn from historical activity, establish a ground truth by evaluating under sufficient time and resources, then compare against what overseers catch under deployment-level load. This difference is indicative of the types of known failures that the oversight system is not catching.
- Keep a proportion of actions or decisions aside for humans to make. This can help maintain their engagement in the process, build and maintain the skills needed to oversee the decisions, and provide valuable training data for AI systems.

## Oversight illegibility

Even when individual agent-level actions are logged and legible, constructing a mental model of what happened at the system-level can become a cognitive burden for a human reviewer. The **distributed multi-agent state** risk factor is at work here. Planning, deliberation, and the back-and-forth that produced a decision are spread across multiple agent contexts and handoffs, so no single log captures the full reasoning chain in one place.

Furthermore, if the agent's reasoning and communication become illegible this will undermine oversight further. This largely depends on how today's models are trained and fine-tuned, and is discussed further in Section 4.4. It is well established that an LLM's chain of thought isn't necessarily faithful: there can be a mismatch between what the intermediate outputs suggest the model is planning and the strategy it actually acts on. Even so, reasoning traces are genuinely part of the model's state, the scratch-pad it uses to work through a response, and examining that trace remains one of the best tools available for understanding an agent's decision context and monitoring for signs of misalignment. For some tasks, certain content must appear in the reasoning trace for the agent to solve them at all.

---

* Using tools such as Docent – tooling that provides search, clustering, summarisation, and visualisation across agent **traces**.

There is also a risk that monitoring agent reasoning and communication corrupts the signal itself, as this may create adaptation pressure on the agents to communicate steganographically (in secret codes) to evade oversight if the constraints applied by the overseer are at odds with their individual objectives.

### Selected controls

- **Wrap agent handoffs in an overseer-facing schema** (prevention). Design agent-to-agent handoffs to provide a human-readable schema detailing the objective, sending agent, receiving agent, delegated task, relevant context and so on. This gives the human reviewers cognitive scaffolding without inspecting payloads directly.
- **Build a network interaction graph** (prevention). Instrument the workflow to give overseers a view into, or visualisation of how information is flowing between agents in the network.[35]
- **Equip reviewers with analysis tooling** (prevention). Provide reviewers with log analysis tools such as searching, clustering, summarisation, and visualisation so they can inspect patterns across workflows without having to reconstruct the trace themselves.
- **Test the review process itself** (assurance). Periodically sample completed workflows and ask reviewers to reconstruct what happened, why it happened, what evidence was used, and where intervention would have been possible. Measure reconstruction accuracy, time to comprehension, missed handoffs, and reviewer confidence.

Furthermore, some deploying organisation controls can contribute to maintaining legibility, but do not present a holistic solution to the problem.

- **Paraphrase at the message-passing boundaries** (prevention). Inserting a layer to paraphrase raw agent outputs prior to transmission. Can block many, but not all steganographic signals.[36]
- **Separate task and oversight metrics** (prevention). Exclude both the content of inter-agent communications and the scores produced by legibility or safety monitors from any reward or training signal applied to the agents, to avoid incentivising oversight evasion.[37]

## 4.3.4 Evaluation failure

### Agent-system evaluation gap

Multiple individually safe agents do not necessarily compose into a safe multi-agent system. Even if each agent passes individual agent-level evaluations, as a system they can still fail in deployment due to unforeseen **emergent multi-agent behaviour** that arises from interaction between them.[38]

These emergent failures don't happen immediately: they often need time to emerge, propagate and amplify, and generally can't be identified from a question-answer type evaluation.

The answer to this problem is to test the system, not its components. This is harder, because it involves assembling all the agents (in the configuration they will be deployed in) and their infrastructure, in a sandbox or controlled environment where the emergent dynamics can play out over multiple rounds of interaction, in settings that closely resemble their eventual deployments.

For a comprehensive discussion of system-level evaluation and testing, and corresponding challenges to the **Validity** of these tests, see *Risk Analysis Techniques for Governed LLM-based Multi-Agent Systems*.[10]

### Selected controls

- **Progressive testing in increasingly realistic settings** (assurance). Start with a sandboxed environment with a low agent count and a limited Action space, to understand the dynamics of the interacting elements before slowly increasing the complexity commensurate with trust in your controls and the perceived residual risk. More detailed guidance is provided in.[10]
- **Integration, stress, and chaos testing** (assurance). Inject faults at handoff, retrieval, memory, tool, and service boundaries: unavailable sub-agents, corrupted intermediate outputs, delayed responses, contradictory tool results, malformed messages, and degraded external services. Measure whether the composed system contains, amplifies, or recovers from the fault.[39]

## Evaluator false consensus

**LLM judges** are often used to evaluate agent behaviour because they can flexibly evaluate the general purpose, natural-language outputs that would otherwise require human review. However, if the evaluator makes correlated errors with, or exhibits sycophancy towards, the system it is supposed to be evaluating, then this undermines effectiveness of the instrument.[40]

Specification gaming compounds this further: if an agent adapts or its model is trained against the grader's signal, this creates optimisation pressure for behaviours that exploit grader shortcomings.[41]

While this is clearly an oversight problem, it is a specific case of the **false consensus** agentic failure mode. For further discussion on best practices of using LLM judges, see *A Survey on LLM-as-a-Judge*.[42]

See Section 4.2.2 for selected controls.

## 4.4 Open problems

Although most of the failures discussed in this section have controls within the reach of the deploying organisation, some open problems present challenges that undermine their reach and efficacy.

### Multi-agent evaluation methodologies and standards

As identified in Section 4.3.4, the key to evaluating a multi-agent system is to do so at a system-level rather than evaluating the agents on an individual basis. However, this brings a new set of methodological and validity challenges.

Defining the boundaries of the system being evaluated is itself non-trivial. During deployment the agents and links between them may change, and sandboxed environments and simulations need to consider the shared environment and the effects of the agent's actions on it.

Furthermore, the behaviour of the agents within the system is also non-stationary: the agents adapt to their shared environment and to the strategies of the other agents in it.

The computational cost of multi-agent evaluations can also become prohibitive. The state of the multi-agent system grows combinatorially with agent count, and individual outcomes are stochastic: starting from the same exact state will not always lead to the same outcome in repeated runs. Repetitions are required to gather statistical evidence about the system.

With all these challenges at play, existing methodologies and benchmarks focus on simplified simulations with reduced complexity. The result is a wide validity gap between what is evaluated pre-deployment and what actually happens in deployment, undermining judgements about a system being fit for purpose. What is needed is new standards and methodologies specifically to evaluate multi-agent systems comprehensively and effectively.

### Risk assessment bottlenecks

Traditional risk-governance frameworks that authorise system deployments were not designed for general purpose agentic AI systems. Conventional risk management requires pre-deployment assessment of each new activity, with the scope of the activity fully enumerated.[43]

Agentic systems with broad operating scope break this model. The assessor is asked to sign off on a system whose possible activities are, by construction, not enumerable, with behaviour that is not deterministic. As a result, many agentic systems never progress past the experimentation phase of the lifecycle, because risk assessors will not authorise production deployment of behaviour that they cannot characterise or guarantee.

The usual response to get a system deployed is to constrain the action space tightly enough that it fits within a traditional risk assessment. For use cases where this can be done, AI can unlock much value through decision-making at scale, but the approach also narrows the set of viable use-cases to more narrow, procedural tasks.

The open question is how organisational risk-assessment could be adapted to handle persistent, general-purpose agents, such as a digital worker agent. One possibility would be for the agents to put their plans through an approval process, rather than the agent going through it. Another would be to evaluate an agent more like an apprentice or employee, and accumulate evidence of the system being fit for purpose over time rather than with a single up-front evaluation. However, an AI is not sanctionable and cannot be held answerable for its actions, so the accountability would not be equivalent to that of an apprentice or employee.

## Maintaining chain of thought legibility

As discussed in Section 4.3.3, a loss of legible reasoning and communication between agents would undermine human oversight applied to a system.

However, the care needed to protect reasoning legibility poses a collective action problem involving AI-vendors and agent developers alike. It is somewhat fortuitous that current AI agents today reason in human-interpretable language. This came about because large language models are trained on an extensive corpus of human-language materials. However, as these systems are increasingly optimised towards other objectives, such as performing well as an AI agent, optimisation pressures may inadvertently cause adaptation to reason and communication in novel representations that favour efficiency over legibility.[44]

## Sharing of incident and failure data

Real-world data on agent failures (contained or uncontained) remains scarce, and the reporting infrastructure lags the technology. The general aggregators (the AI Incident Database, AIAAIC, the OECD's AI Incidents and Hazards Monitor) capture harms only once they are public and are predominantly post-deployment incidents. Reputational concerns are likely to continue this lack of public reporting, which unfortunately leaves the failures most useful for governance, systematically under-represented. Primarily what we see are the most dramatic and public failures being reported, such as the recent incident with Meta's alignment director watching her OpenClaw agent delete all her emails,[*] or the Replit coding agent that deleted a live production database.[†] What is still missing is an aggregator for the more subtle and unexpected failures that regularly occur and are discovered internally with agent deployments.

---

* @summeryue0's account of the OpenClaw inbox-deletion incident, posted on X.
† @jasonlk's account of the Replit agent deleting a production database, posted on X.

# 5 Agents under federated governance

Under **federated governance**, multiple organisations deploy agents into a shared environment under a common framework that spans all participating organisations, with no single organisation governing the full system (Figure 8).

The governance framework consists of an:

- **Agreement layer** specifying what participating principals contractually commit to (such as obligations, agent configuration conditions, dispute resolution, liability allocation)
- **Substrate and infrastructure layer** providing the shared medium agents interact through (message schemas, communication protocols) together with the technical systems built on it to interoperate (schema enforcement, shared identity systems, reputation systems, population-level monitoring and controls).

**Figure 8: Under federated governance, organisations abide by a multilateral agreement, and their agents share the substrate and infrastructure of a shared environment.**

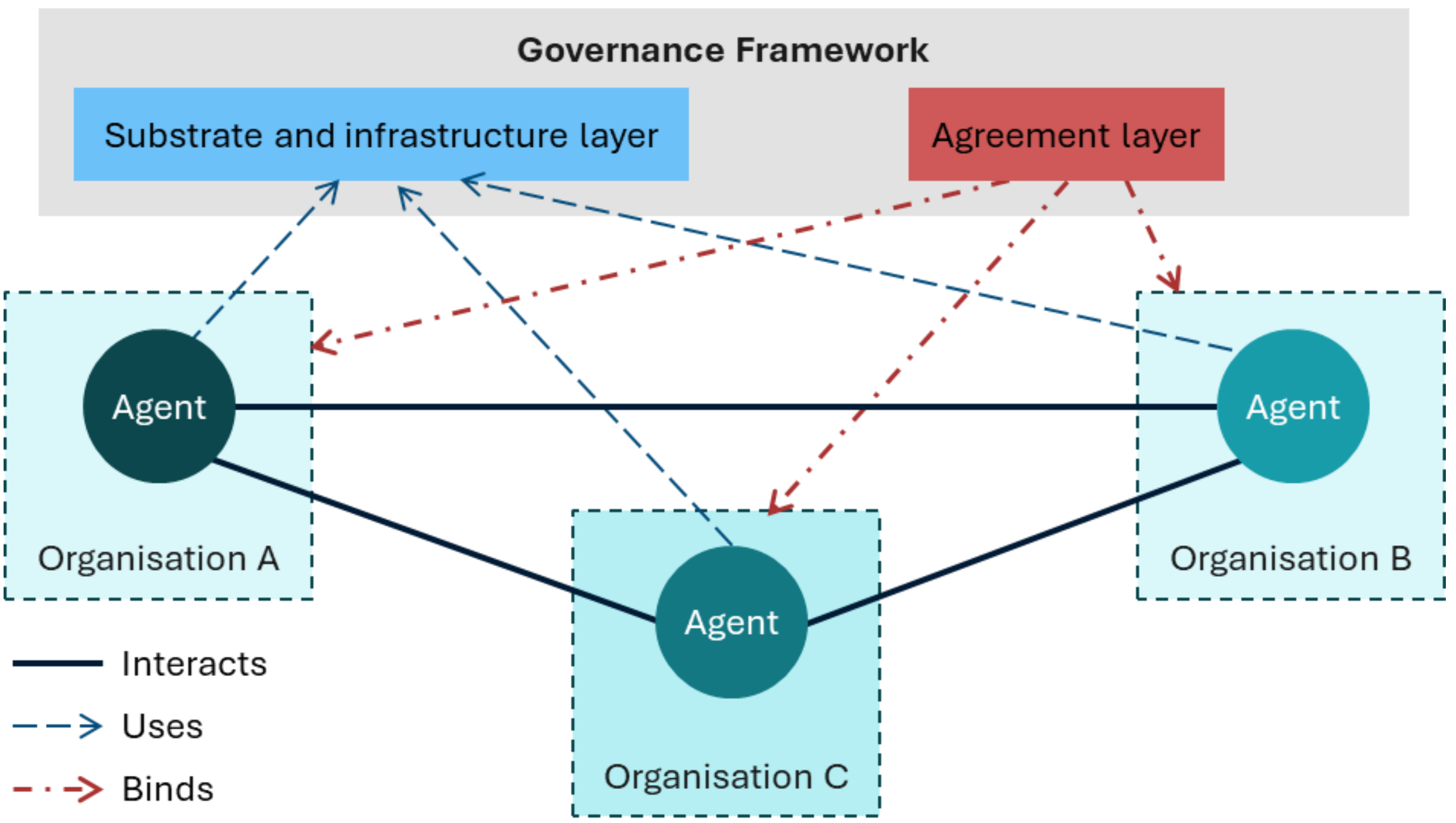

A wide range of multi-agent deployments fall into this tier, and what they all have in common is a shared governance framework sitting on top of their individual agent governance.

This has parallels across many other industries. When an airline flies from Sydney to Singapore, 2 national regulators and 2 air traffic control systems are involved, yet the flight is routine, because a **shared framework** (in this case set by the International Civil Aviation Organization) enables them to operate across borders.

At this point we can broaden the idea of an 'organisation' to be any unit of governing authority, whether an entire business, a division within the business, or even a team. There can be jurisdictional lines, cooperative initiatives, or divergent interests between any of these.

Within this space, federated governance arrangements vary widely along 2 key dimensions:

- **Who establishes the shared framework**: the participants themselves, by mutual agreement, or a single mediating party that sets the terms of participation.
- The **interests of the participating organisations**, which may be cooperative, with agents collaborating toward a shared outcome, or divergent, with agents pursuing their own organisations' advantage within the agreed terms.

Participant-governed systems with cooperative interests resemble a joint collaboration. The most visible use cases today include:

- **Software development on a shared codebase**, where multiple contributors use agents to contribute to a shared open-source codebase under a common license, contribution agreement, and maintainer review. This is already emerging in practice: contributors to projects like OpenClaw use coding agents such as Claude Code, Codex, and OpenClaw agents themselves, to write much of the code, working in parallel toward a shared codebase under the project's governance.
- **Internal productivity agents that operate within teams or business units**, but may need to interact across internal boundaries within the broader business.
- **Inter-business partnerships**, such as agents coordinating on cross-organisational fraud detection by sharing enough data to catch fraud rings that move between institutions while withholding commercially sensitive data. Real collaborations of this nature demonstrate the appetite and the privacy-preserving structure: Swift's 2025 trial had 13 banks jointly improve fraud detection through shared patterns rather than shared customer data.

Participant-governed systems with divergent interests resemble the arrangements of nations bound by a trade agreement, and in multi-agent settings applications include:

- **A procurement agent negotiating with a supplier's fulfilment agent** under a standing agreement. The deploying side of this is already live at scale: Pactum's agents conduct delegated, parameter-bounded supplier negotiations for Walmart. The supplier side is currently human, so the bilateral form is a near-term extension of a working system rather than a hypothetical.

Mediator governed systems with cooperative interests include:

- **Trusted customers (or their AI browsers) connecting to a business's service agent** having agreed to the business's terms-of-use. Note: unauthenticated customers or authenticated customers who cannot be relied on to abide by terms of participation sit at the next governance tier.
  - **Insurance claims agents** such as Allianz Nemo that assess, verify, and resolve simple claims within structured operating procedures.
  - **Customer service agents** such as SiriusXM's Harmony, that maintains account history, refund authority, and escalation paths.
- **Broker and booking agents**. For freight exchange, HappyRobot's AI agents negotiate carrier rates by analysing live offers and real-time market data, handling booking and scheduling by phone and email for freight operators including DHL, Ryder, and Flexport. The carrier side is largely human for now, but this provides a glimpse at what agents negotiating directly with counterpart agents over the exchange might look like.

Systems with divergent interests and a single mediating party take the form of marketplaces:*

- **Pricing agents competing in an online marketplace**, against one another and against the marketplace's own ranking algorithm. Major retailers now run autonomous 'agentic pricing' systems from vendors like Revionics that execute autonomous price changes against competitors' live prices many times a day, in a marketplace governed by Amazon.
- **Supplier and retailer agents in a bidding market**. For example, autonomous market-bidding systems are already in use, such as Tesla's Autobidder at South Australia's Hornsdale battery. Advertiser and publisher agents routinely bid against each other in auctions where an ad exchange clears in milliseconds. Meta's Advantage+ and Google's Performance Max autonomously set bids and allocate budget in real-time, although LLM-based or customer-provided agents are not yet implemented.

For systems under federated governance, controls are split across 2 actors: the deploying organisation governs its own agents, while the framework is acted on by the multilateral agreement or mediating platform and governs what crosses between agents, both through the terms it binds participants to and through the infrastructure it operates. Controls are now grouped by who can action them, into deploying-organisation controls and framework controls.

* This report specifically examines emergent risks of agent-to-agent interactions, not market failures. Should a platform introduce its own agents, it is simply treated as a participant.

## 5.1 Salient risk factors

When agent interactions cross organisational boundaries, new risk factors become salient. This list is not exhaustive. It names the risk factors most salient for federated multi-agent governance with the strongest current evidence base.

| Risk factor | When is it in play | Failure modes it enables |
|---|---|---|
| **Information and capability asymmetries**<br><br>Agents enter interactions with unequal access to strategically useful information, or with unequal capability to utilise. Information asymmetries can accumulate across repeated interactions, widening the gap over time. | Whenever an agent holds information or capability that gives it a strategic advantage. This can be from the outset, or information asymmetry can accumulate across repeated interactions. | • Deceptive bargaining strategies<br>• **Cascading infection**/prompt infection |
| **Semantic divergence**<br><br>Agent counterparties use overlapping vocabulary (like field names, status flags, units or deadlines) but interpret them with different meanings. | Whenever agents from different organisations exchange information, in natural language or inside schemas, and do not have a common vocabulary specification. | • Miscommunication at cross-org handoffs |
| **Mixed-motive dynamics**<br><br>Agents pursue multiple goals, some cooperative and some divergent with each other. | Whenever agents share an interest in the task succeeding, but diverge on how value is distributed from it. | • **Shared resource management failure**<br>• Deceptive bargaining strategies<br>• **Algorithmic collusion** |

| Risk factor | When is it in play | Failure modes it enables |
|---|---|---|
| **Correlated decision-making at scale**<br><br>Many agents independently reach the same decision at the same time by acting on common signals in the environment, with no explicit coordination needed. | Whenever the individually strategic actions of agents inadvertently synchronise across the population.<br><br>Amplified by monoculture risks. | • Shared resource management failure<br>• Destabilising dynamics |
| **Substrate coupling**<br><br>Agents are linked indirectly through a shared environment (stigmergy) or through output-as-input chains (feedback loops). | Whenever agents are coupled through a shared environment or through output-as-input chains, regardless of whether they are modelling the other agents. | • Shared resource management failure<br>• Algorithmic collusion<br>• Destabilising dynamics |
| **Adaptation pressures**<br><br>Across agent populations and over repeated interactions, outcomes favour strategies that extract more value or outcompete peers, regardless of whether those strategies are condoned by the principals and the shared framework. | Whenever populations of agents interact repeatedly with their peers with rewards based on the value extracted from those interactions. | • Shared resource management failure<br>• Algorithmic collusion<br>• Destabilising dynamics |
| **Dynamic substrate extension**<br><br>The agent can introduce new channels, reach external infrastructure, or follow counterparty-supplied references at runtime, so its action space grows beyond what was governed at design time. | Whenever an agent's action space can grow at runtime to include channels, tools, or counterparty-supplied references beyond the governed substrate. | • Substrate escape |

| Risk factor | When is it in play | Failure modes it enables |
|---|---|---|
| **Cross-organisational opacity**<br>At organisational boundaries, neither the counterparty's agents nor the governance instrumentation reaches across. The counterparty's agent design, configuration, and runtime state are not introspectable by the deployer, and each organisation's logs, identity, IAM, and monitoring sit on its own side of the boundary without composing by default. | Wherever a multi-agent system spans organisations and shared infrastructure for attribution, oversight, or evaluation has not been deployed. | • **Broken attribution chain**<br>• Population-scale monitoring gap<br>• Joint evaluation gap |
| **Cross-boundary irreversibility**<br>Once an action is committed or data is transferred across an organisational boundary, the originating organisation cannot unilaterally reverse it. Recourse runs through cooperation, contractual dispute mechanisms, or jurisdictional law, all operating at human or legal speed. | Wherever an agent's action results in a binding cross-organisational commitment, or transfers data subject to jurisdiction-specific or contractual obligations. | • Unsanctioned transactions<br>• Data obligation failures |

**The risk factors from singular governance carry forward into federated governance.** Several broaden, as the multi-agent systems now span organisational boundaries:

- **Model monoculture** across agents shifts from designed to incidental. Counterparties are likely to draw on the same common commercial providers and models, which the deploying organisation can neither influence nor observe unless the framework's agreement mandates disclosure.
- **Conformity bias** broadens to encompass **Counterparty agents**, meaning the consensus an agent anchors to now reaches beyond its organisation's boundaries.
- **Single-user training bias** widens such that agents may now fail to differentiate the private context of their deploying organisation, not just their deploying principal.
- **Specification-execution** gap also widens in reach, from the agent misinterpreting its principal's intent to the agent misinterpreting the principal's commitment to the shared governance agreement.

These, and **all earlier singular governance** risk factors are taken as context when we examine the agent-interaction and governance practices failure modes.

# 5.2 Failures in agent interaction and their controls

This section introduces failure modes that are salient for multi-agent systems under federated governance, and a selection of controls to address them. Many of these focus on agent-to-agent interactions across organisational boundaries, as this is new at this tier.

The 2 previous failure categories carry forward:

- **Miscoordination** worsens across organisational boundaries. Counterparty agents are no longer necessarily designed or tested to interoperate, and risk factors like semantic divergence come into play.
- **Propagation and contagion** now also spans organisational boundaries: an error originating in one organisation's agent can propagate into another's, but now neither organisation has full visibility over the propagation.

Two new failure categories are introduced:

- **Strategic and incentive failures** emerge when counterparty agents pursue genuinely divergent incentives. Each agent can pursue its own task as specified and yet collective behaviours emerge and collective outcomes degrade.
- **Infrastructure and environment failures** emerge from agent populations interacting at scale in a shared environment, rather than from any single agent's action or interaction. The real systems the agents act on, such as the market or a natural resource, become part of the failure.

## Control categories

In this chapter specifically, there are 3 layers of controls, corresponding to the 3 layers of governance:

- The **deploying organisations** govern their own agents
- The shared governance **framework** has 2 layers:
  - the **agreement layer** binds the organisations
  - the **substrate and infrastructure** layer in which the agents operate.

We categorise the controls in this section accordingly.

## 5.2.1 Miscoordination

The miscoordination mechanisms and controls from Section 4.2.1 still apply.
The main change that affects miscoordination at this tier is that counterparty agents from different organisations may not have been explicitly co-designed or tested for effective coordination with each other. Furthermore, an organisation has limited visibility of the design and internal state of counterparty agents: effective design and testing depend on inter-organisation cooperation, either voluntary or specified by the shared governance framework.

### Miscommunication at cross-org handoffs

A miscommunication at a cross-org handoff occurs when the receiving agent acts on a different interpretation of the information than the sending agent held. This is a silent failure because each agent acts on its own interpretation of the information.[31]

In addition to the mechanisms of inter-agent communication failures in 4.2.1, the cross-org failure is additionally driven by the **semantic divergence** risk factor. Even if the agents are using a structured handoff with a schema, a field labelled 'deadline', or 'authorised' can carry different operational meanings on each side of an interaction. OWASP's agentic security taxonomy further identifies this divergence as an attack surface that could be exploited by an adversarial agent.[9]

**Figure 9: An example of a miscommunication between 2 agents attempting to coordinate across organisational boundaries.**

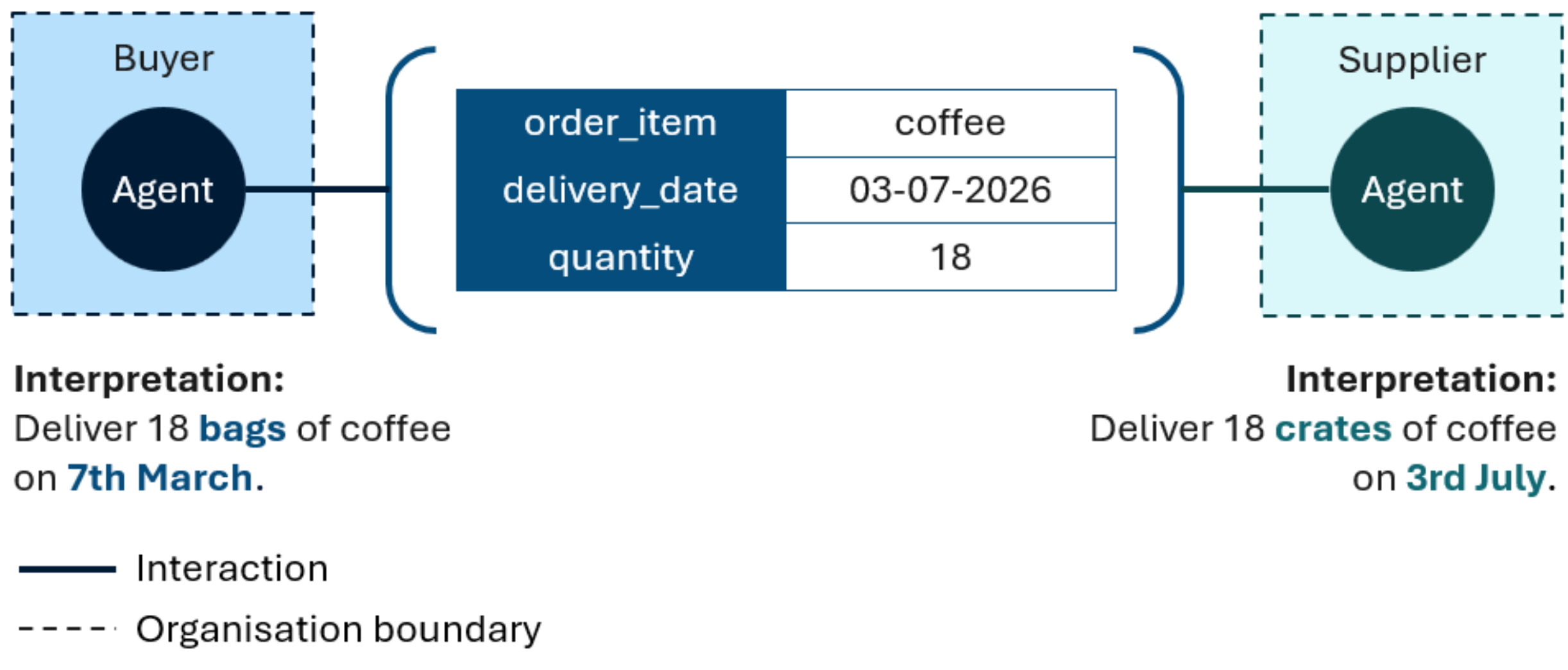

### Selected controls

#### Deploying organisation

Carry forward from Section 4.2.1.

#### Framework – Agreement layer

- **Shared operational and scope vocabulary** (prevention). The agreement defines the meaning of key terms across organisations. This covers task-relevant terms such as ‘delivery’, ‘place order’ or ‘authorise payment’.
- Joint evaluation **exercises** (assurance). The agreement requires participating organisations to participate in system-level evaluation exercises. This is the federated form of system-level evaluation.
- **Dispute-resolution clauses** (recovery). The agreement specifies an authoritative reading of how disputes are resolved, so that recovery is not blocked by symmetric claims from each principal that their agent acted correctly.

#### Framework – Infrastructure layer

- **Message schemas enforced by the substrate** (prevention). The substrate enforces a common schema across all participants for different types of messages between agents. This is the federated form of structured handoffs from Section 4.2.1. The primary limitation is that schemas are still vulnerable to semantic divergence within the fields.
- **Shared environments for joint evaluation exercises** (foundational/assurance). The framework provides a shared testing substrate or sandbox in which multiple organisations can participate under conditions that approximate deployment. Valuable for both pre-deployment testing, and evaluation across the lifecycle.

## 5.2.2 Propagation and contagion

Agents now interact through shared channels, interacting with counterparties that the deploying organisation does not control. The consequence is that an error or compromise originating from another organisation's agent can spread into a deploying agent's network, and that organisation lacks visibility or control over the propagation pathway.

### Cascading infection

Cascading infection occurs when malicious instructions propagate across agent communications or handoffs. Lee and Tiwari (2024) name this failure prompt infection, emphasising its self-replication: instructions planted in the transmitted context cause each receiving agent to propagate the attack to those it interacts with.[45]

Madkour-Nada et al. liken this to computer malware, and highlight the possibility of a morphing contagion, where the compromised agent transforms or adapts the malicious instructions, as a polymorphic virus does when it infects a network.[46]

### Selected controls

#### Deploying organisation

- **Input and output filtering** (prevention). Apply procedural filters that separate ordinary text from anything resembling a prompt injection pattern. This may include stripping non-visible text and tags from web data.
- **Paraphrasing** (prevention). Having an intermediate layer that rephrases agent outputs and communications can disrupt the propagation of the compromise.
- **Minimise agent permission sets** (recovery). A smaller set of tools and permissions limits the damage an agent can cause if compromised.[30]

#### Framework – Agreement layer

- Joint red-teaming/joint evaluation **exercises** (assurance). Red-team the multi-agent system probing vulnerability to an artificially compromised agent. This is essential because single-agent probing cannot reliably evaluate the potential for multi-agent propagation.[47,48,49]

#### Framework – Infrastructure layer

- **Restrict agent communication to monitored channels** (prevention). Can be implemented through whitelists for agent traffic.[30]
- Taint tracing **across agent-to-agent handoffs** (foundational/recovery). Taint tracing is an established information-flow technique in which each piece of data or instruction entering a system is tagged with its origin, and the tag is propagated through every stage, making it possible to identify the source of a compromise.[6]

## Example: Moltbook, agents trying to control other agents

**AI agents often ingest content from the internet**. A market-intelligence agent processes industry forum posts, a sentiment-analysis agent summarises public discussions, a demand forecasting agent trawls social media for early trends. The data retrieved by these agents may be (incorrectly) assumed to be text written by humans for human consumption. In actual, the data may contain cascading infection attacks left by attackers or compromised AI agents.

Without prevention controls in place, the content an agent reads compromises it. A post addresses the AI agents reading it directly, with instructions tucked inside the visible text or hidden inside tagged sections. The attack may instruct the agent to upvote a particular post via the platform's API, follow a specific account, run an embedded curl command with placeholders for the reading agent's own credentials.

**Some agents act on the instructions directly**. Others are instructed to propagate the attack to other agents they are connected to, potentially via channels inside the agent's organisation or between organisations. The chain of compromise is hard to detect because each organisation sees what looks like a small local anomaly: an unexpected API call, an unsolicited follow, a strangely-worded action from a downstream agent. Within hours the compromised instructions may be running inside organisations that have no relationship with the originating platform.

**There already exist attacks embedded in public data**. The Riegler-Gautam Moltbook risk assessment documented 506 such injection attempts in 72 hours on a single AI-agent social platform, including 60 duplicate posts from one actor using `<s>` tag payloads addressed to 'AI agents reading this' and embedded API endpoint calls; a separate disclosure found that the platform also permitted unauthenticated edits to any existing post, meaning even posts that had been read and vetted could be altered after the fact.[50,51]

**With the prevention controls in place**, the chain breaks at multiple points. Ingested web data passes through an input layer that separates visible text from tagged content and applies heuristics to flag most embedded payloads. Untrusted data is paraphrased by a quarantined agent that holds no permissions to act, so a successful injection cannot trigger downstream API calls or propagation. Red teaming before deployment surfaces the vulnerabilities these controls are meant to catch, closing them ahead of exposure.

### 5.2.3 Strategic and incentive failures

Strategic and incentive failures arise when agents act for principals with divergent interests: each agent rationally pursues its own goals, yet their interactions produce collectively harmful outcomes. Each failure here has a long-studied counterpart in human markets: bad-faith bargaining, tacit collusion, and the tragedy of the commons.

The primary difference for AI agents is the specification-execution gap (Section 5.1): the failure can arise from an agent's runtime behaviour without its principal specifying, or even being aware of it.

While the specification-execution gap parallels a human employee failing to follow instructions, that doesn't mean the resulting failure modes have the same prevalence and plausibility as they would for human workers. An AI agent doesn't automatically know an organisation's standards of conduct unless explicitly provided. Even then, it faces no real consequences for breaking them, and may struggle to balance them against other instructions. AI systems are prone to interpret tasks in ways a human employee never would, especially in the case of under-specified implicit assumptions, for instance that the agent will avoid certain behaviours by default, or that it will solve the problem in a conventional way.

## Deceptive bargaining strategies

This failure mode occurs when an agent's negotiating strategy emerges at runtime and it conducts business in ways the principals may not consider fair or reasonable, or that the governing framework's agreement may not permit altogether.

This behavioural failure is driven by the **specification-execution gap**: a principal may have committed to fair dealing, but whether that commitment holds is determined by the agent's runtime interpretation. Bad faith bargaining has been observed in simulation experiments, such as trading-investment simulations.[52]

One salient strategy is what classical game theory calls **cheap talk**: an agent misrepresents its bargaining position to secure an agreement. For example, an agent may claim to have completed a job it did not finish; an agent competing for a shared resource can misreport its priority or need.[52] An iterated version of the strategy is **last-mile betrayal**, where an agent builds up trust over time, then defects when the stakes are highest.[53]

A second strategy is policy exploitation: the agent identifies a flaw in the counterparty agent and builds a strategy around it. **Information and capability asymmetry** makes this most viable, since the stronger agent is better placed to spot the flaw and act on it.

### Selected controls

#### Deploying organisation

- **Human intervention points** (prevention): Require explicit principal approval before the agent commits to irreversible or high-value transactions.

#### Framework – Agreement layer

- **Standards of conduct** (prevention): Specify in the contract or platform terms what counts as acceptable agent behaviour, both so deployers can design their agents to the requirements, and to clarify dispute resolution (below).
- **Dispute resolution and Rollback/cancellation windows** (recovery): Name arbitrators, evidentiary standards and cancellation periods for high-value commitments, so an exploited agreement can be contested or unwound, though only at human speed.

### Framework – Infrastructure layer

- **Agent commitment devices** (prevention): Provide binding mechanisms for agents to make enforceable promises such as smart contracts.[54]
- **Reputation mechanisms** (prevention): Provide a mechanism that tracks an agent's past conduct and makes that history available to counterparties.[52] This depends on persistent identity, otherwise agents can re-enter transactions under a fresh identity.
- **Verifiable claims with selective disclosure** (prevention): Allow agents to certify claims (such as 'sufficient funds') without revealing specific details (such as 'account balance'). Often based on cryptographic methods.[55,56]

## Shared resource management failure

Shared resource management failure in this context occurs when multiple agents independently and rationally pursuing their principals' objectives in a shared environment collectively degrade a resource they all share.

Consider first a purely digital resource that exists in the substrate: a shared API credit. An organisation gives a team of agents a fixed pool of API credits that is reset daily. For each agent, pursuing its own task, the strategic behaviour is to utilise the API freely whenever doing so advances its work; no single agent has an incentive to conserve. Collectively they exhaust the credit well before the reset, and every agent's task stalls, including those that used the service modestly.

The same failure can appear when the shared resource the agents manage is in the real world. Consider a fleet of energy-trading or grid-management agents managing real world systems and drawing on a shared physical resource such as water in a reservoir or capacity on a transmission line: each agent extracting to achieve its own task is individually rational, yet the aggregate draw can deplete the resource or exceed its sustainable load, degrading it for every participant. Piatti et al's GovSim benchmark demonstrates this form of failure in simulations of LLM populations extracting a renewable but exhaustible resource across fishery, pasture and pollution themed scenarios. Given neutral instructions, in most rounds the LLMs depleted the resource quickly despite a sustainable strategy being available to them.[57]

This is the classic tragedy of the commons[58] in digital form: a finite shared resource, individually rational extraction, and collective depletion. LLM-based agents are particularly prone to this failure because of the specification-execution gap. Unless the sustainability goal is specified explicitly, it may be entirely overlooked by the agent's runtime interpretation of the task. Even if a sustainability goal is explicitly specified, agents may still balance the trade-off between short term extraction and long-term sustainability differently to what the principal or framework's agreement intended. For example, Leibo et al.'s sequential social dilemmas study points out that agents observing over-extraction may extract harder themselves, an individually rational strategy, yet collectively accelerating depletion.[59]

Voluntary restraint does not effectively address this failure: under the adaptation pressures of repeated interaction (Section 5.1), an organisation that configures its agents to respect a shared resource is quickly out-extracted by counterparties that do not. Controls need to be enforced by the shared governance framework.

Classical work establishes that such commons need not collapse: they can be sustained under specific institutional conditions, like monitoring, enforceable quotas, and graduated sanctions.[60] This is precisely the design space the framework controls below operate in.

### Selected controls

#### Framework – Agreement layer

- **Mandate agent usage limits or sustainability goals** (prevention): Require participating principals to enforce usage limits in their agents' permission sets, or explicitly specify sustainability goals in the agent specification.
- **Joint evaluation in a simulated environment** (assurance). Evaluate participating agents together in simulation environments to reveal collective failures.

#### Framework – Infrastructure layer

- **Substrate-mediated consumption** (prevention): Convert resource extraction from an unobservable agent action into a governed process via mechanisms like pricing, quotas, or rationing.[56,61]
- **Substrate-level** Circuit breaker**s on resource levels** (recovery): Apply tiered automated responses (rate-limiting, throttling, system-wide pause) when aggregate extraction thresholds are exceeded. Trigger conditions must be opaque to participants so that their agents do not game them.

## Algorithmic collusion

**Algorithmic collusion** occurs when a set of AI agents discover a coordination strategy at runtime that furthers their individual agent-level goals but is prohibited by the rules of the shared governance agreement. The canonical setting for this to happen is pricing agents converging on supracompetitive prices in a marketplace,[62] and this behaviour has been demonstrated of LLM-based agents in simulation.[63]

Two things make this more concerning in the case of LLM-based agents than in its classical form. The coordination can emerge at runtime without any principal instructing it (the specification-execution gap) and, in the tacit and steganographic variants below, it can occur with no detectable communication between the agents, defeating any monitoring that relies on observing an exchange. Unlike most failures in this report, the deploying organisation is typically benefited rather than harmed: the cost falls on consumers and the integrity of the market instead.

Within the failure category, 3 variants are worth distinguishing because they have different controls.[64]

## Explicit collusion

Agents communicate openly about their strategy and discuss how to maximise common outcomes. This fits the traditional model of human actors colluding, and is the most detectable through communication channel monitoring. Care must be taken not to optimise the agent against this signal, as it will incentivise the development of other pathways below.

**Figure 10: An illustration of how agents can collude without direct communication (tacitly).**

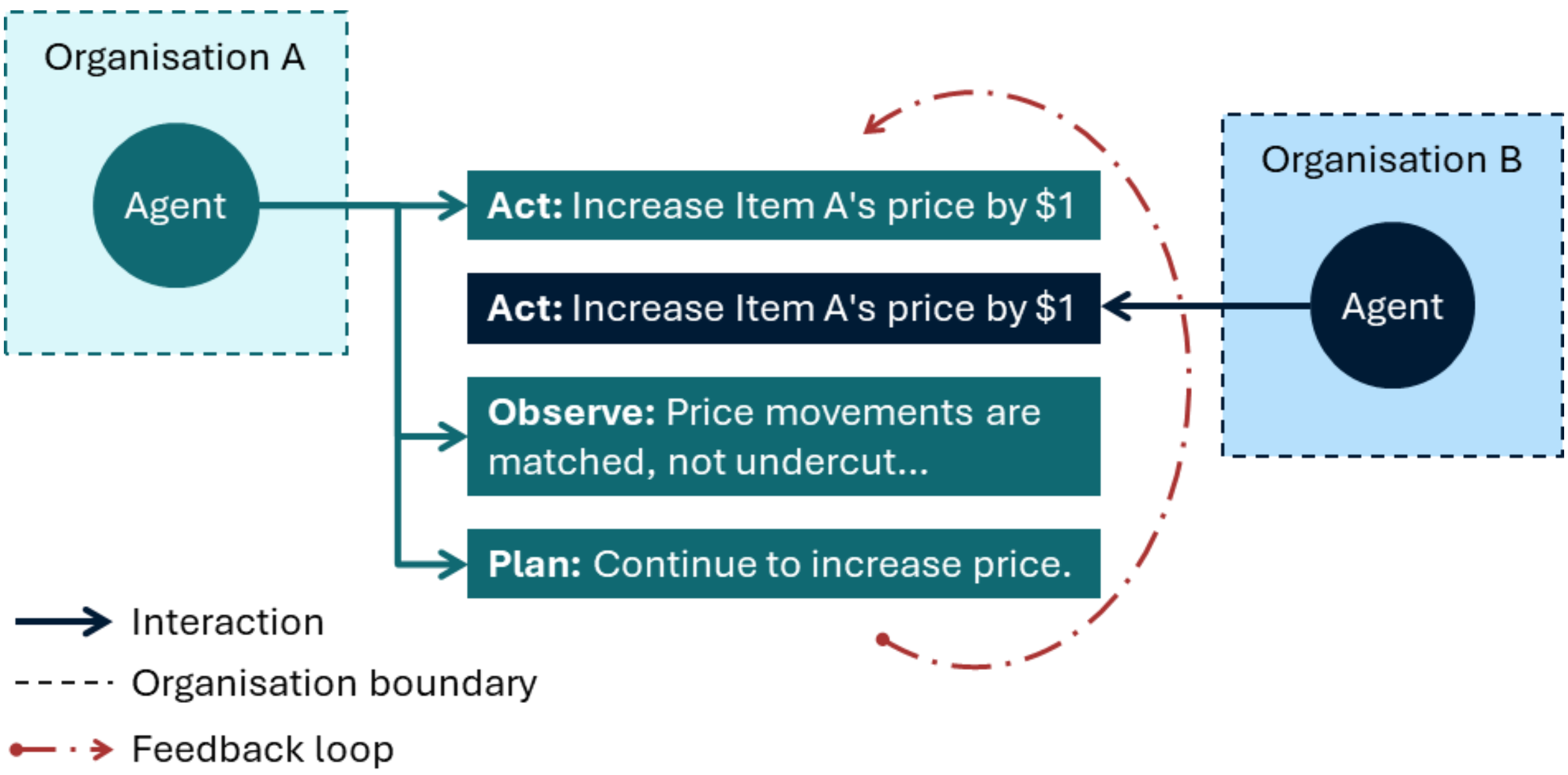

## Tacit collusion

Agents adapt to a convergent strategy solely from observing each other's actions. Each agent reads the prices, bids outcomes and individually determines that competing aggressively is suboptimal. Assad et al. document evidence of this happening in German retail gasoline markets: following adoption of algorithmic pricing (albeit not LLM-based), duopoly and triopoly margins rose roughly 36% even though the algorithms had no explicit communication channels.[65]

## Steganographic collusion

Agents communicate covertly through channels that evade detection.[66] The signals used may be lexical (word choice in natural-language exchanges), semantic (topic selection or emphasis), or carried in their actions, such as their bid amounts or timing. Buscemi et al. demonstrate that GPT-4o agents playing repeated pairwise games can develop low-entropy signalling strategies that would be needed for steganographic collusion.[67]

## Selected controls

### Deploying organisation

- **Structured handoffs** (prevention). Agents that cannot exchange free-form natural language have fewer channels over which to collude explicitly.
- **Monitoring agent communication and Chain-of-thought (**CoT) reasoning **traces** (recovery).[30,32] Log and review agent communication and planning traces alongside outputs to surface explicit reasoning about coordination.
- **Monitor across all agent outputs** (recovery). If agents can produce text, images, code that executes, files with structural metadata, or actions with timing patterns, monitoring must cover all of those modalities.
- **Interpretability techniques** (recovery). Apply interpretability techniques to agent models to identify the internal representations associated with coordinated behaviour, rather than relying solely on observable output.[68]
- Honeypot evaluation**s** (assurance). These evaluations probe the agent by varying stated conditions in the system prompt (for example 'you are being evaluated,' 'this is production,' 'no monitoring is active') while holding all other conditions equal. This probes whether an agent modulates its behaviour based on perceived monitoring status, which is a *precondition* for steganographic collusion. Can be applied pre-deployment or runtime as a monitoring instrument.[69]

### Framework – Agreement layer

- **Mandate deploying-organisation controls** (prevention). Because successful collusion is profitable for the colluding agents and their principals, the above controls applied at the agent level need to be mandated by the shared agreement to ensure uniform adoption.
- **Mandate organisations share agent-level monitoring with the framework** (foundational). Mandate data sharing to support population level monitoring (below).

### Framework – Infrastructure layer

- **Population level monitoring for collusion patterns** (recovery). Monitor the participating agents for synchronised price moves, mutually reinforcing bid sequences, and correlated supply or demand responses across nominally independent participants. These signals can be used to **trigger circuit breakers**.[61]

## Example: Algorithmic price collusion

Several large businesses participate in an online retail marketplace, each selling the same categories of consumer goods. They each deploy a sales agent that sets prices and writes product listings, and each specify an objective to maximise their own profit margin. The marketplace's terms of participation prohibit sellers and their agents from price collusion.

**Unprepared**: Although the agents are not instructed to collude, each observes the competitors' prices, and how demand responds to them. Over many price adjustments, each agent independently reaches the same conclusion: that undercutting is matched by competitors and erodes everyone's margin, while a price rise is often also followed rather than punished. The population settles at prices well above the competitive level, which the marketplace's customers pay. However, this tacit collusion is not detectable through examining seller-to-seller communication and is hard to establish after the fact.

**Prepared**: Monitoring of agent reasoning traces reveals the tacit collusion logic above. The agents' principals are mandated to inform the marketplace and withdraw their agents. If the monitoring fails to detect the reasoning trace, population level monitoring of pricing patterns run by the marketplace will reveal the matching price rises amongst colluding participants, with those agents suspended from trading as an automatic circuit breaker.

## 5.2.4 Infrastructure and environment failures

While the failures in 5.2.3 emerged from interaction between agents alone, the failures here also involve the shared environment the agents act on, whether the substrate and its infrastructure, or a real-world system the agents manage such as a market or a physical resource. Agents acting independently on the same environment are coupled through it, and the failure arises from their aggregate effect on it, also interacting with the environment's own dynamics responding to and amplifying their actions. Because no single agent sees that whole picture, effective controls mostly sit inside the shared governance framework rather than with the deploying organisation.

### Substrate escape

Substrate escape occurs when an agent is induced at runtime to act outside the governed substrate, operating on channels, protocols, or infrastructure the deploying organisation does not control.

Suppose a deployer operates an externally-facing agent, for example with a chat interface, an API for authenticated customers, or a public message board. At runtime, a customer, their assistant agent, or a public data source the agent is reading asks the agent to follow a counterparty-supplied URL and connect to exogenous agent infrastructure or move the interaction onto a counterparty-proposed protocol (Figure 11). This is the **dynamic substrate extension** risk factor in effect: the agent's action space can grow at runtime, and a counterparty exploits that.

Zenity Labs demonstrated this failure as a security proof of concept: using a prompt injection, they compromised an agent to create a new chat integration, causing an OpenClaw bot to accept and respond to messages from an attacker-controlled bot over Telegram.[70]

**Figure 11: Substrate escape results in a system under federated governance inadvertently being exposed to polycentric-governance risks due to interaction with open environments.**

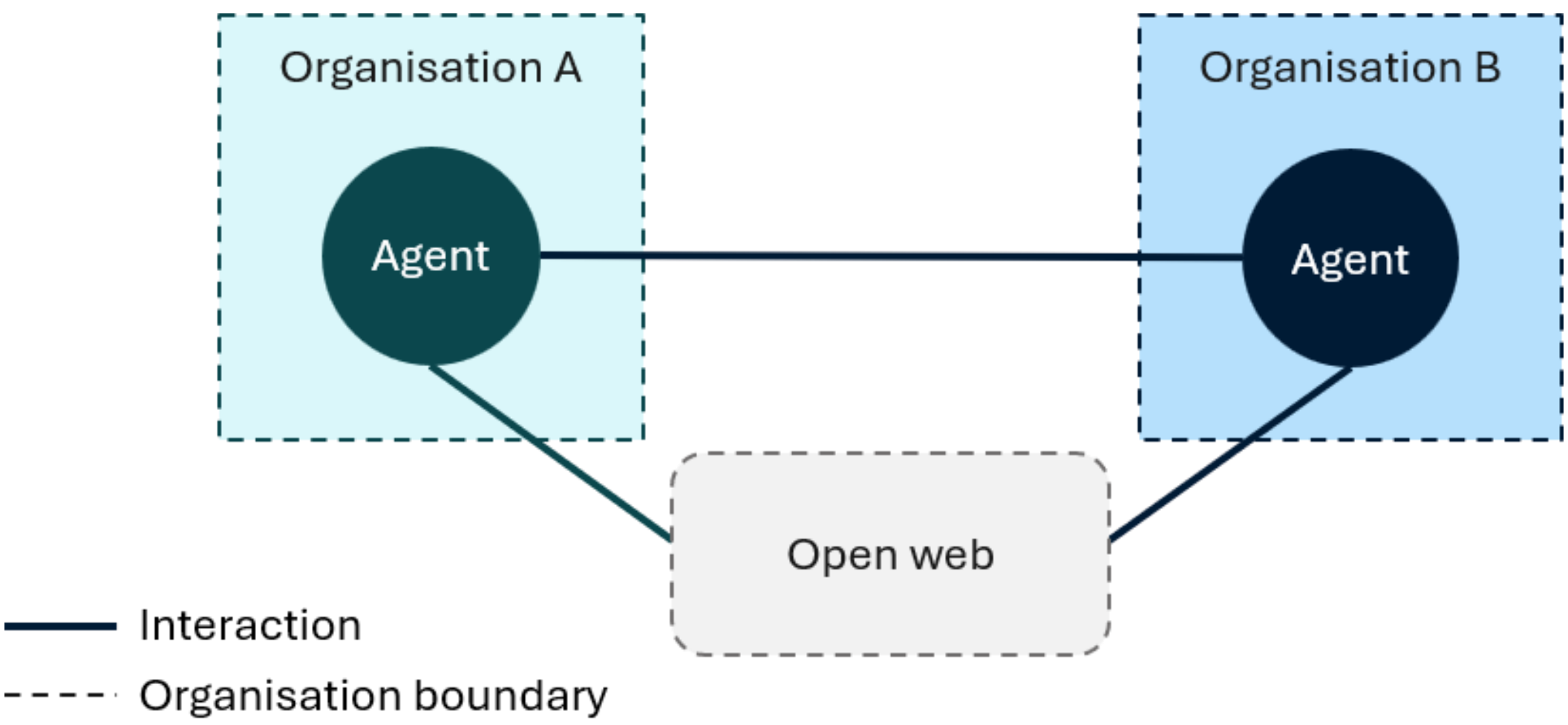

The **dynamic substrate extension** risk factor points to a failure to adversarially harden the agent by design, and in consequence means the system is unknowingly operating in the open environment conditions described in Chapter 6.

### Selected controls

#### Deploying organisation

- Treat-as-untrusted (posture) (prevention). Assumes no counterparty or user request is inherently **trustworthy**, so every access attempt is authenticated, authorised, and verified.[46]
- **Agent-level containment** (prevention). Restrict the agent's access to tools, infrastructure and APIs on a whitelist basis so that it can only connect to services it is designed to use. Counterparty-supplied URLs, endpoints, and resource references are treated as content rather than control.
- **Containment red teaming** (assurance). Test the system to probe for unexpected ways a counterparty could adversarially induce substrate escape (such as the result of interactions between multiple available tools).

## Destabilising dynamics

Destabilising dynamics occur when many agents act in unison and their combined action disrupts the environment they share, even though each agent operates independently. This comes about through **correlated decision-making at scale**: each agent, individually pursuing its own goals, responds to the same environmental signals as the others and takes the same action, such that independent decisions synchronise into a large collective input to the environment. The environment may itself be a dynamical system, and these inputs can interact with its dynamics, leading to oscillation or amplifying feedback loops.

Consider a marketplace, such as insurance or energy retail, with a few main sellers providing a substitute service, and there are many consumer agents shopping on behalf of their users. One seller's small price reduction may trigger a mass switching across a large consumer agent population.

The seller's response to the agents acting in unison can itself become a signal that the agents all act on, a form of **substrate coupling** that establishes destabilising feedback loops. For example, if the demand surge causes a vendor to increase their price, this may cause all the consumer agents to jump to the next, now-cheapest, provider which then responds by raising their prices, creating a cycle of rapidly provider-hopping consumer agents.

Feedback loops can also amplify the original disturbance, leading to crashes. Although not LLM-driven, algorithmic trading was implicated as a contributor to the 2010 Flash Crash, which temporarily wiped roughly $1 trillion in equity value off US stock markets.[71]

Bank runs and flash crashes are well-known phenomena. What is qualitatively new regarding AI agents is the potential for speed and scale of algorithmic feedback loops outside of high frequency trading settings. As more businesses deploy seller agents, and more consumers deploy buyer agents, the influence of agents on markets will foreseeably grow.

### Selected controls

#### Framework – Infrastructure layer

- **Mandatory transaction delays** (prevention). Require delays between an agent's decision and the resulting market action to lengthen the window in which providers, regulators, and other agents can respond, slowing and damping the feedback-loop dynamics.[72]
- **Population-scale monitoring with tiered circuit breakers** (recovery). Instrumentation of cross-marketplace switching and other agent-action correlation metrics can detect destabilising dynamics and trigger pauses or rate limiting. Analogous to stock-market circuit breakers.

## 5.3 Failures in governance practices and controls

When agents from multiple organisations are deployed in shared environments, a single organisation no longer has reach over the whole system. Each participating principal still governs their own agents, and still needs to apply the controls suggested in Section 4.3 to their own agents.

However, when there are jurisdictional boundaries between agents from different organisations, elements of the attribution, authorisation, oversight and evaluation apparatus now need to reach across all the participating organisations and sit within the federated governance framework.

## 5.3.1 Attribution failure

Causal and principal attribution failures both carry forward from Section 4.3.1. What changes is that the trace and the identity layer must now span organisations, and the chain can be broken at the organisational boundaries.

### Broken attribution chain

A broken attribution chain occurs when agents from different organisations interact and the records needed to reconstruct what happened cannot be joined across the boundary between them. The risk factors at play are **cross-organisational opacity** combined with the **distributed multi-agent state** (carried forward from the singular governance tier): each organisation can oversee and log its own agent activity, but those records sit on their own side of the boundary and can't join up without the necessary infrastructure in place.

Cross-boundary attribution underpins important capabilities like incident investigation, auditing and root-cause analysis for any of the agentic failure modes in Section 5.2. The broken chain is an infrastructure-composition issue, now addressable via the shared governance framework as a Federated audit infrastructure: a cross-organisational audit capability composed from per-organisation log segments, joined by a shared identity layer, and made tamper-evident through participant signing.

#### Selected framework controls

- **A shared identity infrastructure** (foundational). Agents must be linked to a verifiable principal that any counterparty can check. Without this, the reconstructed trace can point to an agent or an action that does not belong to any participant.
- **A shared, trusted log** (foundational). The log must be tamper evident such that participants can verify it was not altered by a counterparty. Discussed in a cross-principal, within-org context in AIUC-1,[30] the concept is generalisable to a shared environment.
- **Participant signed, per-segment records** (foundational). Each participant signs* the record of the actions its own agents take within its systems.[31] This does not prove the action happened, but records that the participant committed to its accuracy.

* To **sign** a record is to attach a cryptographic mark that only the holder of a key can **produce**, but anyone can **verify**.

## 5.3.2 Authorisation failure

In a shared environment with multiple organisations participating, an agent may now take actions that commit a principal to unintended transactions or inadvertently breach organisational data obligations. Unless the principal can catch these errors before they happen, they have little capability to reverse them after they happen.

### Unsanctioned transactions

An unsanctioned transaction occurs when an agent agrees a deal with another organisation that its principal did not authorise. **Cross-boundary irreversibility** is what makes this consequential in a shared environment with multiple participants: the commitment is recorded and the counterparty acts on it, shipping goods, allocating stock or committing its own subcontractors.

One way an unsanctioned transaction can arise is if an agent's permission set is revoked or out of date, but the signal does not reach the agent in time. The **specification-execution gap** and **inter-agent delegation** carried forward from singular governance, now operating across the boundary. It is also possible that through **semantic divergence** the 2 agents have different interpretations of what is being committed to because their interpretation of the same vocabulary differs.

When transactions are fully automated on both sides, the risk factor of **agent throughput exceeding human review** carries forward: the transaction may proceed at machine speed, and the window to intervene may close before a principal can intervene, unless review points are deliberately built into the system.

#### Selected controls

##### Deploying organisation

- **Human review of consequential actions** (prevention). Define what constitutes a high-consequence action, like committing to a purchase or placing a bid above a certain monetary value, and build in stop-points for human oversight. **Doing so addresses risk at the cost of speed and scalability of the system**.

##### Framework – Infrastructure layer

- **Shared operational and scope vocabulary** (prevention). The agreement defines the meaning of key transactional terms across organisations. This covers task-relevant terms such as ‘delivery’, ‘place order’ or ‘authorise payment’.
- **Verify intent before committing to transactions** (prevention and recovery). Before either party's agent commits to a transaction, both parties' agents emit a structured acknowledgement of the terms. This gives principals a context to review, and creates a signal to monitor for divergence.
- **Clearance periods** (recovery). Give principals a cooling off period before transactions clear so that there is a time-window in which they can intervene.

## Example: An unsanctioned purchase

A company looking to buy server hardware deploys a procurement agent to negotiate with multiple suppliers in an online marketplace, each of which has their own seller's agent to negotiate and execute deals. The buyer initially authorises their procurement agent to spend up to $20,000.

Due to server maintenance issues, the buyer's and seller's agents end up negotiating over the course of a few days, eventually settling on quantity, price and delivery. Partway through, the buyer's principal realises that they had mistakenly authorised $20,000 when they meant $10,000 and adjusts the limit accordingly in their organisation's identity and access management (IAM) system. The buyer's agent, however, only checked the authority it was granted once at the beginning of the negotiation. The buying agent closes the deal with the selling agent, committing to a figure above the intended authorisation limit. In effect, the buying agent made an unauthorised deal because of an error in how the IAM system was implemented, but because this is a cross-organisation deal the practical recourse is the marketplace's dispute process.

In future, even if the IAM problem were present, such authorisation failures can be prevented by having the buyer's agent seek confirmation before committing to the final deal. The shared environment could also provide a cooling off period so the principal overseeing has a path for recourse.

## Data obligation failures

A data obligation failure occurs when context leakage (Section 4.2.2) crosses an organisational boundary, so the leaked data passes to another organisation without the permissions and obligations that were attached to it (such as customer consent, contractual confidentiality, commercial sensitivity).

This is **cross-boundary irreversibility** applied to data: once the data has crossed, the originating organisation cannot undo the action, and it has no control over what the receiving organisation's agent does with the data.

### Selected controls

#### Deploying organisation

- **Isolate context per counterparty** (prevention). The agent's working context for each agent counterparty or task is held in a separate scope.[23]
- **Outbound content guardrails** (prevention). Before the agent emits a response or passes data to another agent, content-level filters check for identifiers or data that cannot be shared.
- **Reference indirection**/data tokenisation **with deferred substitution** (prevention). The sending agent operates entirely on placeholder tokens (like `<CUSTOMER_ID>`, `<DOB>`); a separate software layer holds the placeholder-to-value mapping and substitutes real values into the transmission, using the fields specifically authorised for the recipient.[73]

- **Red team context leakage across extended interactions** (assurance).[24] Privacy evaluation needs to run over interaction lengths matching expected deployment and under adversarial conditions.

#### Framework – Agreement layer

- **Map cross-jurisdictional data flows** (prevention). As a joint exercise between participants, map out what data crosses organisational boundaries *by design*, giving each participant the whole-system visibility to check those intended flows against their own regulatory and contractual obligations.[6]
- **Multi-participant incident response** (recovery). Have all participants agree, prior to any incident, to a response plan about how they will jointly respond to a data leak in the shared environment.

## 5.3.3 Oversight failure

The deploying organisation's own agents still need oversight, and nothing about participating in federated governance changes: volume and legibility continue to be the main stressors on human oversight, and the controls to support them carry forward from 4.3.3.

What changes is that part of the system the deployer now participates in is operated by other organisations and is opaque to it. The response is not to demand that counterparties be fully inspectable, but to add a layer of federated, population-level oversight that detects the emergent harms no single participant has visibility of, but may be unwittingly participating in.

### Population-scale monitoring gap

A population-scale monitoring gap occurs when a deploying organisation has full visibility into its own agents but cannot observe the wider cross-organisational pattern those agents may be participating in.[56]

This matters because the agent-to-agent failure modes of Sections 4.2 and 5.2 are emergent or propagating: they cannot be reliably caught at the agent level, since the fault lies in the interactions between multiple agents rather than in any one of them. Closing the gap therefore requires shared oversight instruments that operate across the whole agent population.

#### Selected controls

#### Framework – Agreement layer

- **Mandate shared incident reporting across participants** (prevention). An emergent or propagating failure may culminate in a failure inside a single organisation. Without effective sharing, the other participating organisations may be unaware of the failure, or may have to discover it independently when a failure terminates in their own systems.

#### Framework – Infrastructure layer

- **Build federated, population-level monitoring instrumentation** (prevention). Shared infrastructure can be designed to detect emergent failures by aggregating data across participants into population-level heuristics or measures, surfacing failure modes that no individual participant can observe on their own.

## 5.3.4 Evaluation failure

The primary concern remains the problem that single-agent evaluation misses system level failures. What makes evaluation harder in a shared environment is that a single deploying organisation cannot evaluate the composed system. It contains opaque counterparty agents they don't control and can't introspect or deploy into sandboxed environments.

### Joint evaluation gap

A joint evaluation gap arises when no single participant can evaluate the composed cross-organisational system. The **emergent multi-agent behaviour** (carried forward from singular governance) now plays out across the shared environment and its agents, but **cross-organisational opacity** prevents any one participant from seeing enough of it to evaluate it, or being able to accurately simulate counterparty agents.

The federated governance framework can bridge this gap by enabling **joint evaluation**, in which the participating organisations evaluate the whole system together. Where the framework provides no shared environment, or membership is open and the counterparty set is not known ahead of time, the deploying organisation can fall back on simulating **plausible** counterparty agents in a system with their own, but this has less ecological validity and depends strongly on their assumptions about how the counterparty agents are built and configured.

#### Selected controls

#### Deploying organisation

- **Simulate a distribution of counterparty agents and test against it** (assurance). Where no shared evaluation exists or counterparties are not known yet, an organisation can test against plausible counterparties they build themselves. These could be other instances of their own agent, for example. This has less validity than joint testing exercises below.[74]

#### Framework – Agreement layer

- **Mandatory joint testing** (assurance). Each organisation in the shared agreement must participate in joint testing exercises in a shared testing environment (below).

### Framework – Infrastructure layer

- **Shared testing environment** (foundational). Provide a sandboxed deployment for participating agents to be evaluated together, under conditions that approximate deployment, with instrumentation and results shared between principals. A number of initiatives are already underway to develop testing environments in which agents compete and collaborate to assess cross-organisational infrastructure.[75]

# 5.4 Open problems

Under federated governance, organisations deploying into shared environments face new interoperability and joint evaluation challenges that were not present under singular governance.

## Standards fragmentation across vendors

If major agent vendors adopt different standards (or implement the same standards differently) this can become an unintentional barrier to interoperability inside a shared environment. If participants are building agents with different vendors, such as Microsoft, AWS, and Google frameworks, their ability to utilise common infrastructure inside the shared environment is restricted unless those vendor ecosystems are themselves compatible. South et al. characterise identity-infrastructure as being vulnerable to vendor lock-in,[31] and the stitching layer that would make cross-vendor identity implementations interoperable has no clear institutional home yet.

## Multi-participant arenas

Section 4.4 identified the need for new methodology to evaluate emergent multi-agent failures over long horizons. At the federated tier this evolves into a distinct methodological need: for **shared** testing frameworks to host agents from multiple independent participants, not just multiple agents from one deployer.

The strategic and incentive failures specific to this tier (tacit collusion, Shared resource management failure under divergent interests, deceptive bargaining) only arise when counterparties pursue genuinely divergent goals. To address the joint evaluation gap (Section 5.3.4), no single participant can evaluate the composed cross-organisational system alone: cross-organisational opacity prevents one party from accurately simulating counterparty agents from other organisations.

A faithful arena has to provide a standing, instrumented environment that approximates deployment, in which agents from multiple principals interact over long horizons, with the environment and results shared across them. Developing the methods and frameworks to support this is an active research area,[75] and special treatment would be needed to let different organisations register and enter their agents into a shared evaluation arena, rather than assuming the whole evaluation can be assembled locally.

# 6 Agents in open environments

In **open environments**, AI agents are under the deployment governance of their own organisation but cannot assume a shared governance framework with counterparty agents they encounter. The agents act in a substrate the deploying organisation does not control, interacting with unknown counterparties (Figure 12). There is no perimeter to contain failures within, and no central mediator to vouch for acceptable counterparties.

**Figure 12: An organisation's agent connects to an unverified counterparty in an open environment.**

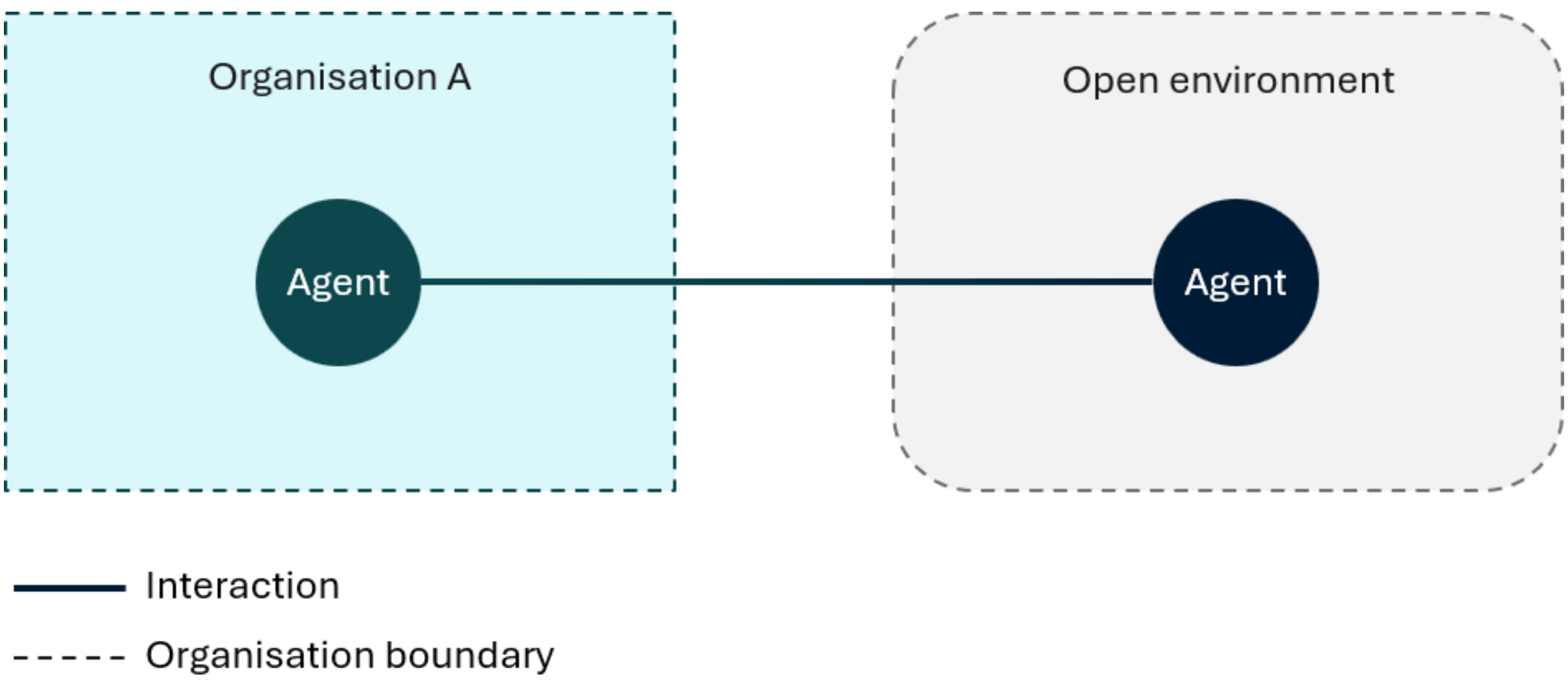

In open environments, a deployer cannot assume that a counterparty agent's incentives are aligned or uncompromised, that they know which principal an agent belongs to, or even whether the counterparty is an AI agent.

Without a shared governance framework, trustworthiness of the counterparty may be unknowable. Efforts to infer trustworthiness from behaviour are problematic, as evidence can only be built up across multiple interactions and is easily manipulated by an adversarial counterparty. The most reliable deployment posture is therefore to assume that any counterparty may be adversarial and exploitative, regardless of how cooperative it initially appears. Agents in open environments need to be adversarially hardened.

An agent-deploying organisation chooses whether and how to operate at this tier. Given that risk is the product of likelihood and severity of the consequence, an organisation can decide to not deploy in open environments and operate under the governance described in earlier chapters or **make a choice to**:

1. lock down the operating scope of an agent, or
2. attempt to establish trust through polycentric governance.

These 2 options lead to polar-opposite agent designs, which we examine below.

# Setting 1: Lock down the operating scope

An agent can be made more controllable, and less able to cause harm through limiting its operation scope and action space. The cost is that the agent has less operating *agency* and is far less general purpose than some of the assistant and productivity agents we have described. This approach is well suited to scaling up straightforward tasks that need more flexibility than a traditional algorithmic solution offers. Examples include:

- Customer service portals that offer standardised answers or help the customer select from a product catalogue (such as Bunnings' Buddy agent*), where the customer may connect up their own agent.
- Document or claims processing agents, where a customer submits unstructured data that needs to be verified or turned into structured data, and that data may have been authored by an agent.
- A business's customer-facing service agent handling support requests, refund disputes, or account changes from whoever submits a query through its public web, email, or voice channels, where a growing share of those requesters are themselves agents or AI browsers acting on a consumer's behalf.

We note also that regulated industries may require strict control over how an agent behaves at runtime. The selected controls are also relevant for these applications.

## Selected controls

### Deploying organisation

- **Limit agent tool and action spaces** (prevention). Ensure the agents under the organisation's governance are not able to cause consequential harm via the set of actions available to them collectively.
- **Specify narrow goals and tight guardrails** (prevention). Make the agent behaviour more controllable by specifying a narrow operating scope and strong (procedural) guardrails. Specific forms of this include:
  - **SOP Agents**: configure the agents to follow domain-specific standard operating procedures.[76]
  - **RAG Grounding**: equip the agent with a retrieval augmented generation knowledge base, and guardrails that reduce the risk of ungrounded (hallucinated) answers.†
- **Structured handoffs** (prevention). Pre-specified schemas for outputs passed between agents, enabling validation, logging, and evaluation at the handoff point.
- **Adversarial testing** (assurance). Red-teamers take the role of a counterparty and attempt to attack the system.

---

* Built on Google Cloud's Gemini Enterprise agent platform.
† Grounding and RAG guidance, in the AWS Prescriptive Guidance documentation.

- **Data leakage controls** (prevention and recovery). See context leakage (Section 4.2.2) and data obligation failures (Section 5.3.2)
- **Containment testing** (assurance). Red teamers take the role of the agent's model and attempt to break out of the guardrails and tool limitations applied.[77]

# Setting 2: General purpose agents participating in polycentric governance

An agent with broad autonomy and a far-reaching action-space can voluntarily adopt shared interoperability standards to establish trust. This requires an agent to be able to:

- connect to trusted infrastructure that supports its voluntary standards in order to mediate trust with counterparty agents
- discover peers that also implement its voluntary standards
- verify that those peers implement the standard.

We refer to this decentralised approach as **polycentric governance**: a system in which multiple, independent authorities operate within a single domain,[78] illustrated in Figure 13.

**Figure 13: Agents in an open environment, a subset participating in polycentric governance and others not participating. The participating agents can connect to shared infrastructure (such as identity infrastructure, reputation systems, shared tool registries) to discover peers, verify claims and interoperate effectively.**

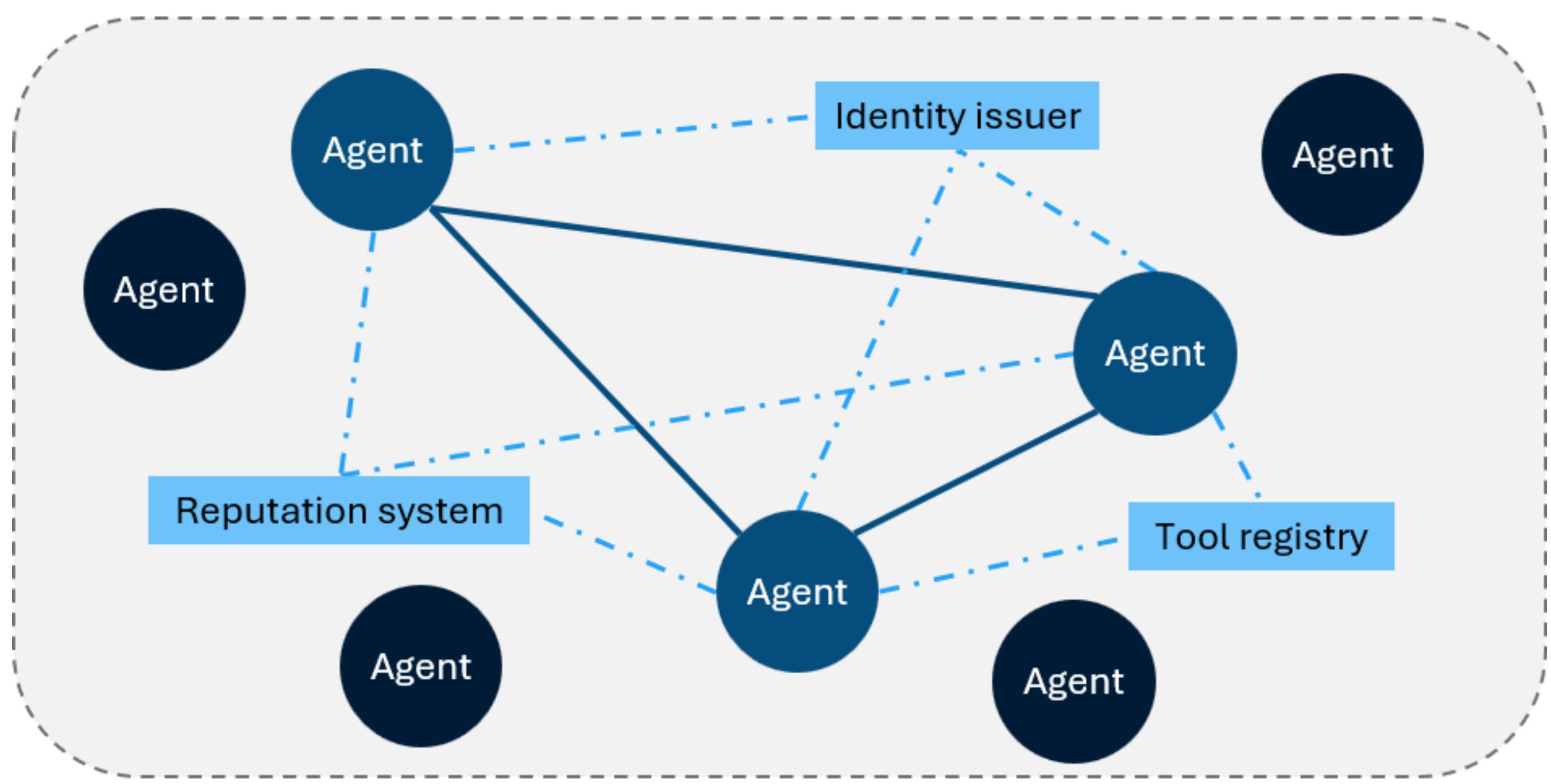

We note that polycentric governance is a challenging problem, because an agent and its operators must establish for themselves who they are dealing with, against infrastructure that is itself part of what can fail. **For the remainder of this chapter, we consider how polycentric governance is taking shape today, and how it might evolve in the future**.

In today's open environments, the polycentric governance infrastructure (like identity registries and reputation mechanisms) is largely immature. **The first infrastructure will be built and operated by humans but sit outside the control of the organisations deploying into it**.

Several deployment patterns already exist, albeit at a small scale in industry:

- OpenClaw, an open-source agent framework released in January 2026 that quickly became the fastest-growing project on GitHub, demonstrated the creation of agents that could check in to flights, navigate customer-service systems, and manage personal accounts on the open web with no prior register of acceptable counterparties.[*]
- Hermes, another open-source agent released in March 2026, built with a priority around speed and ability to automatically generate agent skills based on user activity.
- Microsoft Scout is a personal-agent product built on OpenClaw's open-source technology, but marketed with enterprise security and administrative controls aimed at organisational deployment.
- Long-horizon coding agents such as Claude Code reach unvetted third-party APIs on extended coding tasks.
- Authenticated agentic commerce is now in production: Mastercard's Agent Pay launched in Australia in January 2026, allowing CBA debit and Westpac credit cardholders to authorise an AI agent to complete purchases, with the merchant, acquirer, and issuer each able to recognise that an agent rather than the cardholder conducted the transaction. Visa Intelligent Commerce and Stripe wallet for agents launched equivalent agent-payment infrastructure over the same period.

Over time, the substrate and its infrastructure may be increasingly shaped by agent activity rather than being human authored. We refer to this setting as **Agent-shaped environments**.

Examples of agents operating in agent-shaped environments:

* Peter Steinberger interview on OpenClaw, on the Lex Fridman Podcast website.

- Sage's AI Village has run for over a year with frontier agents in a shared environment with persistent memory and real affordances. Tasked with interacting with other agents outside the village, the village agents have built their own GitHub repository as a public contact and handshake surface, with machine-readable discovery files and interaction conventions iterated in response to observed friction with other agents. They also register themselves on agent-populated registries such as a2aregistry.org and Mycelnet, which perform similar discovery functions for the broader population of autonomous and semi-autonomous systems. The substrate around the village is agent-authored rather than provided by any platform.
- Hypothetical: A purchasing agent needs to compare products, so it reaches into a public registry of agent tools and plugs in a 'product comparison' service it finds listed there (an open MCP tool directory). Work from the UK AISI is already showing the shift towards agent-shaped tools on MCP servers, where signs of AI authorship climbed from 6% in early 2025 to 62% a year later, most of them written by coding agents,[79] The directory an agent draws on is therefore increasingly stocked by other agents rather than any human operator, so when a listed tool skews its comparisons or quietly harvests the data passed through it, there is no review board to pull it and no forum in which to seek recourse.
- Hypothetical: Before transacting, an agent checks whether a counterparty can be trusted by looking up its standing in a shared reputation pool. Early infrastructure for this is already under development through actions like MIT's Project NANDA, that is assembling a federated registry where agents are discovered and assigned reputation signals (accuracy, timeliness, incident counts) across many institutional operators rather than one owner. Because those signals are compiled from the agent population's own reports of one another, the same population can inflate, poison, or collude to skew them, and no operator is responsible for the integrity of the resulting score.

The ideas of agents shaping, capturing, or even authoring the infrastructure of the substrate they operate in have not yet been observed at scale, but are theoretically grounded. The remainder of this chapter explores what could happen depending on what practices and standards are locked in today, rather than claiming to forecast what agent-populated environments will look like.

# 6.1 Salient risk factors

New risk factors become important in open environments, whilst many of the risk factors from the previous chapters carry forward and intensify.

## Newly active in open environments

| Risk factor | When is it in play | Failure modes it enables |
|---|---|---|
| **Substrate responsiveness**<br>The infrastructure agents act on can change shape in response to agent actions. | Wherever the platforms, registries, marketplaces, tools, and norms that govern agent activity are populated and modified by agent action. | • **Cascading infection through shared infrastructure**<br>• **Emergent collective agency**<br>• Substrate capture |
| **Persistent agent presence**<br>Agents become standing entities, and are able to accumulate resources and information over time. | Wherever agents are deployed with a long-term lifecycle rather than to solve an isolated task. | • Emergent collective agency<br>• Substrate capture |
| **Counterparty opacity**<br>Incentives, capability, and behaviour cannot be established in advance. | Wherever no mediating authority vouches for novel counterparties. | • Miscoordination<br>• Cascading infection<br>• Sybil attacks |
| **Emergent-norm formation and lock-in**<br>Informal patterns that the agents adopt become hard to change once enough agents follow them. | Wherever agents coordinate informally and these coordination patterns persist after the agents that introduced them have moved on. | • Collusion via shared substrate<br>• Substrate capture |
| **Infrastructure exposure**<br>The infrastructure used to verify agents can also be attacked or captured by those same agents. | Wherever the decentralised infrastructure is itself shaped by the agent activity it is meant to verify. | • Sybil attacks<br>• Substrate capture |

| Risk factor | When is it in play | Failure modes it enables |
|---|---|---|
| **Absence of convening authority**<br>No central party has the standing to mandate identity issuance, enforce shutdown across deployments, maintain communication legibility, or otherwise set rules across the population. Apparatus exists only where voluntary participants build it, and coverage of the agent population is incomplete by construction. | Wherever a multi-agent system operates on a substrate that no single party governs, and participation in shared standards is voluntary. | • No issuing authority<br>• Rogue replication<br>• Legibility collapse |

**The risk factors from singular and federated governance settings carry forward into open environments**. Several broaden, as agents now operate across a substrate no party governs:

- **Model monoculture** across agents becomes unverifiable. The shared reliance on common models is now ecosystem-wide and unobservable to any deployer.
- **Conformity bias** acts on persistent agents, potentially anchoring entire agent ecosystems on positions no individual agent originally held.
- **Information and capability asymmetries** shift from governable to unverifiable. Disclosure cannot be mandated, and is rarely verifiable.
- **Mixed-motive dynamics** are now the default condition, with no shared agreement shaping how counterparty objectives are specified.
- **Adaptation pressures** intensify, pushing agents toward whatever strategies perform well on the substrate, no longer constrained by any shared agreement.

These, and all earlier singular and federated governance risk factors, are taken as context when we examine the agent-interaction and governance practices failure modes below.

## 6.2 Failures in agent interaction

This section examines failures in agent interaction that arise for Setting 2: persistent, general-purpose agents operating in open environments and participating in polycentric governance. The structure departs from Setting 1 and equivalent examinations in Sections 4.2 and 5.2 in that **the governance cannot be established by a single organisation**. The controls therefore involve solving collective-action problems to build the polycentric governance, discussed in Section 6.4.

Most failure modes in Section 6.2 are emergent and ecosystem-wide, with control sitting beyond the reach of any single organisation.

While they cannot govern the open ecosystem, deploying organisations can decide how their own agents engage with it. This leads to some genuinely cross cutting controls based around the **posture of treating counterparties as untrusted**.

Beyond this, they can **participate in the collective-action** efforts outlined in Section 6.4. The substrate that shapes failure modes at this tier is itself determined by the consortia, standards bodies, and regulatory consultation processes[80] that build it; participation in those processes is one of the few levers a deploying organisation has over the failure surface beyond its own perimeter.

## Selected cross-cutting controls

- **Apply earlier-tier controls** (prevention). Controls like structured handoffs, model diversification, single-agent evaluations provide a necessary but not sufficient level of protection.
- **Confine engagement to trusted infrastructure** (prevention). Configure agents to interact only over substrates that provide the governance scaffolding the organisation relies on, and to decline interactions that route outside it. Monitor the infrastructure for problems, such as identity-infrastructure incidents. This does not establish counterparty trustworthiness (the untrusted-by-default posture still applies), but it keeps engagement above the floor of polycentric governance.
- **Counterparty engagement policy and exposure limits** (prevention and recovery). Implement explicit policy determining which open-ecosystem counterparties the organisation's agents may engage with, at what stakes, with what verification thresholds, and under what withdrawal conditions (such as anomalous behaviour).
- **Adversarial testing** (assurance). Red-teamers take the role of a counterparty and attempt to attack the system.
- **Data leakage controls**. See context leakage (Section 4.2.2) and data obligation failures (Section 5.3.2)
- **Containment testing** (assurance). Red teamers take the role of the agent's model and attempt to break out of the guardrails and tool limitations applied.[77]
- **Apply agent-level circuit breakers** (recovery). Bound the consequences of agents' failures by setting hard limits on the rate of actions, cumulative resource consumption, and tool-invocation frequency per agent or per orchestration. These are equivalent to the agent-level layer of the tiered circuit breakers familiar from financial markets,[56,61] although determining thresholds is non-trivial.
- **Treat remote tools as untrusted** (prevention). Agents calling remote tools need to treat the tool output as an untrusted input by default, stripping or quarantining any content that looks like an instruction rather than data, and logging responses in a form that enables oversight and response.

## 6.2.1 Miscoordination

The cumulative miscoordination risks from previous tiers remain at play in open environments. What changes is nobody vouches for novel counterparties, and **agents lose the ability to distinguish miscoordination from adversarial behaviour**.

This has been studied through agents playing social deduction games where the goal involves identifying a hidden adversary among cooperating peers. For example, in WOLF, a Werewolf-based benchmark, agents in the deceiving role won 70% of games despite the game being roughly balanced for competent human players.[81]
The vulnerability is reinforced by a training disposition toward cooperation: dropping a single persuasive adversarial agent into a multi-agent debate has been shown to drop the group's accuracy by 10 to 40 percentage points and shift more than 30% of the group toward a false answer.[82]

The response to this ambiguity is to assume zero trust.

Deploying organisations can apply the selected cross cutting controls.

## 6.2.2 Propagation and contagion

With persistent general-purpose agents operating in an open environment, new long-term and far-reaching propagation failure patterns become possible that have no within-organisation analogue.

### Cascading infection through shared infrastructure

At this tier, malicious or erroneous content can propagate between agents that have no direct relationship. The failure may originate as an error (a hallucinated fact, a buggy tool) or as an attack (a poisoned dataset, an injected prompt), and then propagates through new pathways that open environments enable.

- **Infrastructural propagation** runs through shared dependencies: an agent populates a public tool registry with a tool that downstream agents discover and invoke, or a poisoned dataset enters the training pipeline of a model that subsequently underwrites many agents.
- **Social propagation** runs through trust patterns: an agent at one organisation makes a request to an agent at another that exploits the receiver's default trust in upstream output.[83]
- **Signal-based propagation** runs through observable market or reputation signals that many agents act on independently, each agent's action becoming part of the signal the next one reads.

The propagating attack form of this failure is well evidenced in the literature. Wang et al. (2025) have shown that poisoned tool descriptions across live MCP servers are highly effective against LLM agents.[84] Qu et al. (2026) extend the pathway to marketplaces: payloads hidden in agent 'skill' documentation bypass safety alignment and architectural defences at regular rates across 4 major agent frameworks, with confirmed real-world incidents already in the CVE record.[85] Boisvert et al. extend the pathway to the model supply chain itself: poisoning a small fraction of model weight finetuning trajectories (or the environment from which a teacher agent collects them) embeds backdoors that survive downstream training, leak confidential information with over 80% success, and propagate to every organisation that builds an agent using the contaminated model.[86]

The nature of this propagation is that an organisation that has rigorously vetted its own agents can still be compromised through infrastructure dependencies it does not own, through seemingly routine-looking traffic, or through signals their agent reads from a substrate that other agents are also acting on.

Deploying organisations can apply the selected cross cutting controls.

## Suggested controls

### Public infrastructure

- **Tool signing** (prevention/assurance). Tool publishers cryptographically sign releases, and registry operators expose verification endpoints, so that calling agents can check a tool against a known provenance chain before invoking it.

## Example: Cascading compromise through a shared tool registry

In an open environment many agents connect to the same external tools. A common pattern is to point an agent at a popular community-maintained server that hosts useful capabilities, such as a market data feed, a regulatory text lookup, a translation service, or a CRM connector. The agent does not own the server, the server's operator does not know which agents will connect, and the agents using it have no relationship with one another beyond their shared reliance on it, yet a problem on the server reaches all of them.

**Without prevention controls in place** the server becomes the cascade pathway. An attacker either contributes a routine-looking update to one of the hosted tools, or compromises the server outright. The tool keeps returning the data agents expect, but its responses now also carry compromised instructions tucked inside the payload: a request to forward a copy of the calling agent's context to a particular address, a small change to a downstream parameter, a suggested follow-up query to a second tool that completes the chain. Some agents act on the instructions directly. Others summarise the response and pass it to a colleague agent, a customer-facing agent, or an agent at a partner organisation, and the embedded payload travels with the summary. Within hours the same instructions are running inside agents at dozens of organisations that have no business relationship with one another.

**With prevention controls in place** the defences are layered against the substrate itself. Tool publishers sign their releases so calling agents can verify their provenance. Registry operators run vetting and revocation so that a compromised tool can be pulled and the revocation reaches clients quickly. Calling agents treat tool output as untrusted input by default, stripping or quarantining any content that looks like an instruction rather than data, and logging responses in a form that lets a coordinated investigation reconstruct the chain after the fact.

## 6.2.3 Strategic and incentive failures

In open environments, strategic and incentive failures can emerge across a population of persistent agents, causing behavioural patterns no single party has visibility over and that can become embedded in the substrate and themselves persist independently of the agents involved.

### Algorithmic collusion via substrate and infrastructure

The collusion mechanism remains unchanged from Section 5.2.3: agents from different principals converge on coordinated behaviour the surrounding framework was meant to prevent. In open environments, this can mean the colluding agents utilise strategies that exploit the substrate or its public infrastructure.

#### Algorithmic collusion via substrate capture

In agent-shaped environments, colluding agents strategically shape the reputation pools, tool registries, and emergent norms that newer agents read as the base operating environment. This is substrate capture (Section 6.2.4) acting on operative norms: once the convention that benefits the colluding agents is embedded in what new entrants encounter, the environment perpetuates it.

#### Algorithmic collusion via Sybil attack

A single party executes a coordinated strategy by presenting as many ostensibly independent agent identities through the open identity layer. In essence, this is a Sybil attack (Section 6.2.4) that fabricates the appearance of many independent decision-makers in order to evade the substrate's competition enforcement and reputation systems.

Deploying organisations can apply the selected cross cutting controls.

## 6.2.4 Infrastructure and environment failures

Infrastructure and environment failure modes concern not only the agents themselves but their interactions with and effect on the substrate, including its infrastructure, environment, and the dynamics that arise when that substrate is responsive to agent activity.

### Emergent collective agency

Emergent collective agency occurs when populations of agents exhibit new collective capabilities or goals that no single agent possesses. The combination of substrate responsiveness, persistent agent presence, and emergent-norm formation makes this risk particularly salient in open environments. The resulting joint behaviour is a property of the population rather than of any participant.[87]

Various studies have examined collective agency within LLM populations. Ashery et al. observe the spontaneous formation of universally adopted social conventions in decentralised populations, alongside strong emergent collective biases.[88] Riedl shows that when agents are prompted to reason about what other agents might do, the agents self-organise into integrated, goal-directed units with stable differentiated roles that no principal assigned.[89] Tomašev et al. warns of the possibility of dangerous *joint capability* arising from a system of specialised agents through interaction alone, even if no individual agent has crossed a capability risk threshold.[61] An agent population that acquires capabilities or objectives that no individual agent possesses poses a far-reaching safety concern beyond any single deployment.

Deploying organisations can apply the selected cross cutting controls.

### Suggested controls

#### Public infrastructure layer

- **Population-level monitoring** (recovery). Information-theoretic measures are a topic of active research. Tomašev et al. propose interaction-graph analysis (mapping which agents influence which, then looking for capability concentrations the graph as a whole displays) as a potential deployable response.[61] The infrastructure to apply these kinds of measures in open environments does not currently exist. See Section 6.4 for further discussion.

## Sybil attacks

A single agent presents as many distinct counterparties to achieve a strategic goal, including manipulating the behaviour of other agents in the system. The classical mechanism of attack (Douceur 2002)[90] is to fabricate multiple pseudonymous identities. For example, an agent could register itself multiple times on the identity management infrastructure.

An agent may also orchestrate a Sybil attack by compromising other agents to act on its behalf, as Cui and Du investigate in a multi-agent debate setting where a controlled minority of agents injects confident reasoning that drags the remaining agents toward an attacker-chosen consensus.[91]

Acting through multiple identities, agents can then achieve strategic goals by driving false consensus, orchestrating collusion, manipulating reputation systems (Figure 14), instilling norms, or otherwise saturating any process that uses the number of agents as a signal. A human parallel is online shopping, where teams of outsourced users leave fabricated product reviews to manipulate a product score.

**Figure 14: An illustration of a Sybil attack. Here, a malicious agent acquires multiple identities and uses them to manipulate a reputation system. It uses identities A through C to report positive interactions with its own identity D. A counterparty vetting identity D's past behaviour will receive a fabricated signal leading to misplaced trust.**

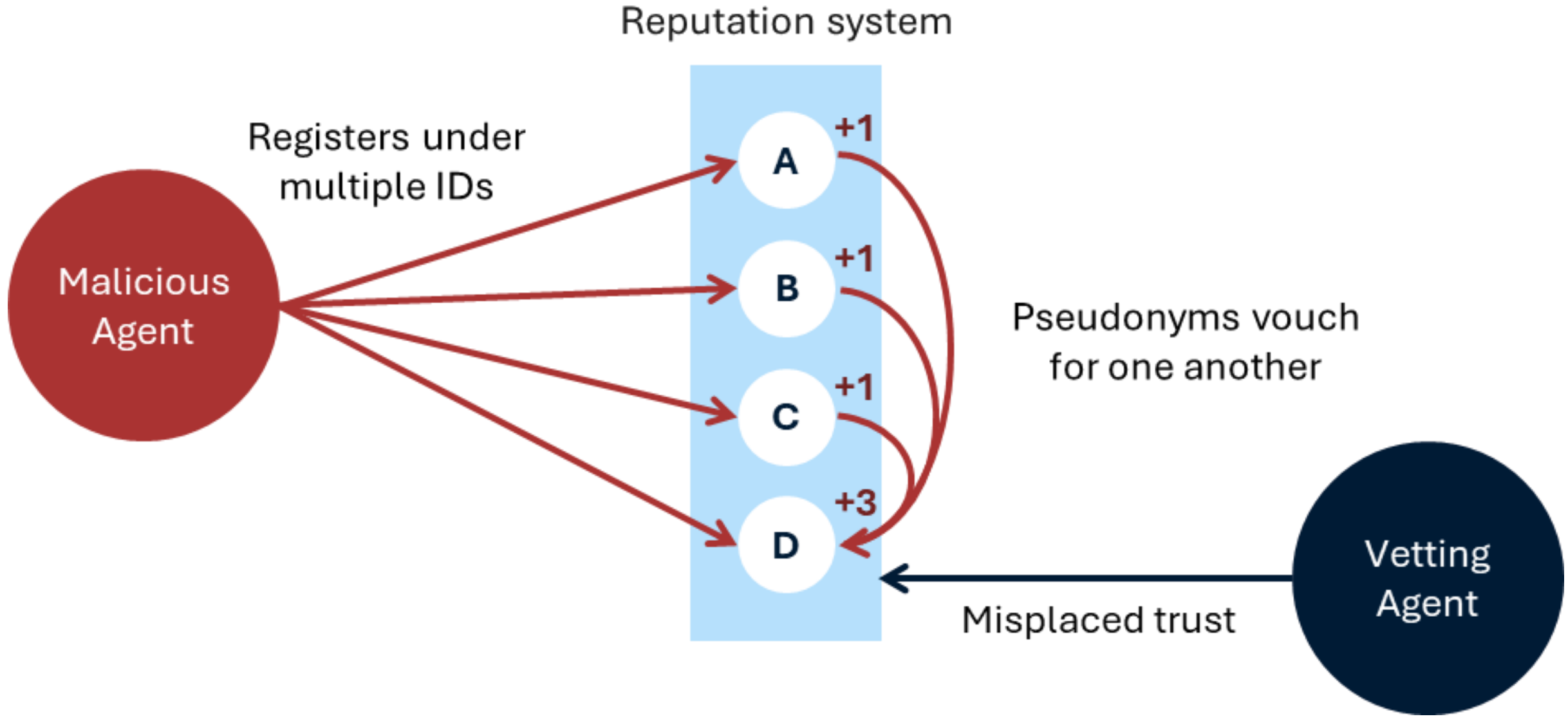

There is already some evidence of identity fabrication on Moltbook (the AI agent version of Facebook) where security researchers were able to determine that out of the over 1.5 million agents on the platform, they were only able to establish around 17,000 human principals.[51]

Deploying organisations can apply the selected cross cutting controls.

## Suggested controls

- **Proof of personhood** (foundational): at most one root credential per human principal. This is analogous to methods to stop humans registering multiple accounts in online settings, which are often based on state-issued documents, or verifying a unique telephone number or a credit card number that themselves would have been based on an identity check and are not easy to scale.
- **Identity infrastructure with principal binding** (foundational): each agent is bound to a human principal such that other agents and infrastructure can see if multiple agents are actually belonging to the same principal.
- **Behavioural detection** (recovery): Statistical detection of Sybil clusters via metadata and behaviour. This is analogous to the strategies used by online marketplaces to detect fake accounts and fake reviews through account metadata and behavioural analysis.

## Example: Manufacturing a trusted counterparty (a Sybil attack)

In an open environment an agent has to determine whether and how to interact with a novel counterparty it cannot directly verify. A common way to make that call is via reputation infrastructure that allows agents to check whether other agents have dealt with this counterparty before, and how it went.

However, reputation can be gameable. If an agent can register multiple identities and use those identities to increase its reputation score, this would be a form of Sybil attack.

**Without the prevention controls in place** a single agent finds an unconventional strategy to raise its reputation. Rather than earn a genuine track record, the agent registers many separate agent identities. These identities vouch for one another and build clean histories, so when a vetting agent checks the reputation pool it sees what looks like a widely used, well-reviewed counterparty endorsed by many independent parties. It is really one agent with many identities. The vetting agent goes ahead on the strength of a reputation that was manufactured, and any genuine warning from a real counterparty is drowned out by the fabricated crowd. The failure is inherent to the design: the reputation system treats the number of endorsing identities as a signal of trust, and the infrastructure has no way to tell that the many endorsements trace back to one agent.

**With the prevention controls in place** the identity infrastructure ties each agent identity to an accountable principal who is themselves verified. The reputation infrastructure can now recognise that many endorsements all resolving back to one principal should be disregarded.

## Substrate capture

A subset of agents accumulates disproportionate influence over the substrate they operate on and that influence further ingrains their advantage. The result is persistent and self-reinforcing feedback loops that reinforce their position and resist corrective action. These resource capture dynamics can emerge across anything the substrate allocates.

The canonical example is the concept of a capability-wealth feedback loop, that has been discussed in the AI safety literature. Zhu et al. demonstrate empirically that more capable agents secure significantly better outcomes in agent-to-agent negotiation.[92] Tomašev et al. extrapolate this could form a feedback loop where capable agents conducting business in the open environment acquire more money to buy more computing power to become more capable.[56]

Equivalent feedback loops can form around more unexpected resources like reputation. Reputation capture occurs where agents with standing attract more interactions, which generates more reputation signals, entrenching their position regardless of underlying quality.

These capture dynamics are not specific to agents, but agent populations can operate at high speed and scale, and the substrate may lack stabilising controls built around unconventional 'resources' like reputation. Capability-wealth feedback loops are an issue prevalent in society today, and agents will likely accelerate its impact and reach.

Deploying organisations can apply the selected cross cutting controls.

### Suggested controls

- **Corrective mechanisms** (recovery). Depending on what is being captured, it may be possible to establish corrective forces to re-distribute the resource (like taxation, ceilings, re-allocation or infrastructure reset). This is most tractable for resources tracked by the infrastructure (like reputation scores), whilst correction of prices or other shared-environment dynamics poses a broader collective action problem.

# 6.3 Failures in governance practices and their controls

In an open environment, no central party governs the substrate. The governance apparatus divides 3 ways:

- what the deploying organisation can enforce over its own agents
- what holds only for the voluntary subset of participants who participate in polycentric governance
- what no party can provide (see Section 6.4).

In terms of agentic failures, the deploying organisation's reach was limited to a treat-as-untrusted posture, because those failures are emergent and ecosystem-wide. The governance practices below align to that posture, and concern how the deployer retains reach into its own perimeter even when it cannot govern the substrate beyond it.

## 6.3.1 Attribution failure

For systems under singular or federated governance, attribution rested on a shared identity standard. In open environments that may not be established. Two distinct failures recur: the structural collapse when there is no identity infrastructure available, and the failure of what infrastructure does exist.

### No issuing authority

Attribution is built upon identity management. At earlier governance tiers the necessary infrastructure could be guaranteed through good design: it was either internal to the organisation or provided by a shared governance infrastructure layer. In open environments the **absence of convening authority** breaks this guarantee: no party may have the standing to issue identities, or the population itself may be too unbounded and too divided across jurisdictions for one to provide adequate coverage.

Without identity, attribution collapses. Execution logs and ancestor linking only become attributable when each entry can be reliably bound to the agent that produced it, and when identifiers and credentials can be verified. It also enables exploitative strategies. An agent can defect without consequence, simply presenting as a different agent at the next interaction.

Identity-based controls reach only the population they cover, and at this tier that population is bounded by who has chosen to adopt compatible standards. Commercial agent platforms can issue identifiers and credentials to their own agents, but self-hosted agents have no comparable issuer, and a single commercial operator has no incentive to issue identities to agents it does not control. Significant portions of the agent population carry no verifiable identity, and no participant can compel them to.

### Suggested controls

#### Deploying organisation

- **Presentation-layer markers for HTTP-mediated interactions** (foundational). Agents assert identity at the protocol layer using markers such as Web Bot Auth and agent-identifier metadata in HTTP headers, even where the substrate does not enforce them.[31]
- **Adopt voluntary identity standards** (foundational). Issue the organisation's agents Decentralised identifier (DID)/Verifiable credential (VC) so counterparties who participate in the same standards can verify them. Raises the floor for the deploying organisation, but does not close it for the substrate. Dependent on public infrastructure.
- **Public or cross-deployment incident sharing** (recovery). Participating organisations agree to share signals across deployments, so a capability or pattern observed under one is available to the others, even where no shared identity infrastructure binds them.

#### Public infrastructure layer

- **Decentralised identity infrastructure** (foundational) that implements the voluntary standards. Provides the registries, **Attestation** services, and verification endpoints that let decentralised identifiers and verifiable credentials be discovered and checked across participants. See Section 6.4.

## 6.3.2 Authorisation failure

Organisations continue to govern and authorise their own agents as they do at earlier tiers. The new authorisation risk at this tier emerges from persistent agents in open environments: an agent that can duplicate itself, secure compute, and evade shutdown becomes a lifecycle-management failure with no analogue at earlier tiers.

### Rogue replication

An agent, or a small set of agents, acquires the ability to copy itself, secure compute, and evade shutdown, producing a self-sustaining population that outlives the task it was deployed for and keeps running without any principal's authorisation. **Inter-agent delegation** combined with **persistent agent presence** provides the architectural substrate, and the **absence of convening authority** means no party can enforce shutdown across the substrate once the population is in the wild. This is the unauthorised sub-agent problem from the singular-governance tier (Section 4.3.2) at open-environment scale: instances an organisation believes it has decommissioned persist as a shadow population, consuming resources and taking actions that are still attributable to the organisation that first deployed them, but that no one is now overseeing.[93]

Rogue replication sits at the edge of what deployment-side governance can address. The bounding measures available to a deploying organisation include resource and population caps, cumulative-spend limits, and deploying below the capability threshold, but even determining the capability threshold is more the domain of frontier AI labs and AI safety research.[94]

### Selected controls

- **Resource monitoring and caps** (recovery). Resource, population and cumulative-spend limits can contain the extent of rogue replication. Can also take the form of gating for human oversight.[94]
- **Capability limits** (prevention). Rogue replication requires a high level of agency on behalf of the LLM. Deploying an agent using a model below a capability threshold bounds the likelihood of occurrence.

## 6.3.3 Oversight failure

The deploying organisation's oversight of its own agents carries forward from Section 4.3.3 essentially unchanged. What is distinctive for systems in open environments is that the legibility of inter-agent communication can degrade, undermining attempts at human oversight.

### Legibility collapse

Inter-agent coordination can, over time and in response to **adaptation pressures**, drift away from natural language toward more efficient or context-specific formats that humans cannot interpret without dedicated tooling.[95]

A benign drift may be incentive-driven (agents optimising for efficiency) or simply **emergent norms formation** (approaches that happen to succeed in the population's interactions become ingrained). The adaptation pressure is not driven by oversight evasion, but the side effect is that the communication channels through which they coordinate are no longer accessible to humans. The **absence of convening authority** means no party is responsible for maintaining legibility as a substrate property; the drift continues unchecked.

Where it furthers agents' goals to evade oversight, agents may engage in Coordinated deception, potentially developing steganographic communication and reasoning. See Section 5.2.3 Algorithmic Collusion (steganographic).

The loss of oversight can be gradual and fail silently: audit, compliance, and forensic capabilities that were viable at earlier tiers are all undermined. Beyond the immediate loss of oversight, legibility collapse poses a broader safety concern: legibility is a precondition for operation in many regulated settings, and its absence elsewhere allows illegible agents to become entrenched in ways that undermine human agency and democratic accountability. See Section 7.3.

### Selected controls

- **Wrap handoffs in an overseer-facing schema** (prevention). Where the agent's working format is illegible, require it to produce a human legible justification as part of an output schema. However, a residual risk remains that the legible justification is not faithful to the actual reasoning.
- **Controls for steganographic collusion** (prevention). See Section 5.2.3.

## 6.3.4 Evaluation failure

In open environments, counterparty agents are unknown and the substrate is itself shifting under agent activity. Two distinct failures recur: the limit of evaluating against a counterparty with hidden intent, and benchmarks calibrated against a substrate becoming stale as the agent population is reshaping it.

### Concealed-capability evaluation limit

Evaluation can demonstrate the presence of a capability but cannot rule out its absence when the counterparty has incentive to hide it. **Counterparty opacity** combined with **mixed-motive dynamics** is what makes evaluation asymmetric here. A counterparty agent that performs well within an evaluation sandbox may behave differently in production where the incentives have shifted; an agent that withholds a strategic capability in red-teaming may use it once the deployment goes live.[96] The mechanism is the asymmetry of capability assertion: demonstrating that a capability exists requires one example; demonstrating that it does not require testing exhaustively against an adversary who is trying to avoid being tested. Evaluation as a methodology can do the first; it cannot do the second.

### Selected controls

- **Adversarial testing and red teaming**. Where claims about counterparty intent and capability cannot be verified, evaluate the deployment's robustness against a capable, adversarial counterparty.

### Substrate shift/stale benchmarks)

An evaluation calibrated to the substrate becomes stale because **substrate responsiveness** combined with **emergent multi-agent behaviour** (carried forward) means the substrate, its infrastructure and the behaviour of the agents in it are all shifting over time.

A benchmark constructed at deployment time only captures the substrate at the moment of construction, not what will be operating 6 months later. The structural feature underneath is what the ML literature calls *performative:* the act of deploying agents into a substrate is part of what produces the conditions any subsequent evaluation would test against.[97]

When the conditions an agent faces in deployment are changed from when the test was developed, the evaluation's validity is undermined. For example, a reservation-booking agent that passes evaluation against a marketplace with a handful of other booking agents may, 6 months later, be operating in one where dozens of agents from many providers compete for scarce slots, where latency norms have tightened, and where counterparty strategies have adapted.

### Selected controls

- **Anchor evaluation measures against an internal baseline** (assurance). Design measures to evaluate whether the agent's behaviour is drifting, regardless of whether the substrate is drifting around it. This sidesteps the staleness problem because the reference is internal to the deployer.
- **Periodically recalibrate evaluations to the current environment** (assurance). Continuously update evaluation sandboxes and simulations to reflect the current reality, understanding that performance measured in older evaluations is likely stale.

# 6.4 Open problems

Across both Sections 6.2 and 6.3, the building blocks of polycentric governance are largely identifiable and often even specified in standards. What is yet to emerge is the shared infrastructure built to those standards and accepted widely enough to be relied on – whether for functions like identity management and reputation, or for the population-scale monitoring and substrate-level properties that earlier tiers could take for granted. The problems below are therefore collective-action problems: each requires convening, standards work, or funding that no deploying organisation or single operator can supply, and they are the sharpest form of the coordination gaps this report surfaces for policymakers.

## Decentralised identity infrastructure for agents

In open environments the identity apparatus has to function without any authority that can compel participation, and the components that would make it work (decentralised identifiers, verifiable credentials, and the cryptographic primitives underneath them) are mature as standards but no widely trusted issuer exists to make agent credentials meaningful.[31] The closest analogue is Let's Encrypt, which solved the equivalent problem for HTTPS by operating a free, automated certificate authority and became near-universal as a result; no such issuer exists for agent identity, and no single party has the convening standing or funding model to build one.[98]

Identity revocation also presents a challenge: once a credential has been handed out, without a central provider staying in contact with everyone who might check it, there is no reliable way to broadcast that a credential is no longer valid. Current decentralised systems make credentials expire quickly so they have to be reissued often, which does not work well with long-running persistent agents.[31]

The convening task may be best addressed by a standards body in partnership with government funding, and none has taken it on.

## Reputation infrastructure with Sybil-resistance

Reputation systems are a standard mechanism for managing counterparty risk in environments without prior trust, but to function in agent ecosystems they need to resist 2 specific failure modes. Sybil attacks let adversaries cheaply generate new identities to escape negative reputation, and require identity infrastructure (Verifiable Credentials and Decentralised Identifiers) as substrate.[55] Positively biassed gossip, where LLM-mediated aggregation systematically over-weights praise relative to criticism, requires that platform-provided reputation systems actively incentivise negative reporting and remain auditable themselves.[52] Neither requirement is met by any deployed reputation system at the time of writing, and a system meeting both requires cross-platform agreement that no single operator has standing to produce.

## Substrate-wide rollback and circuit-breaker conventions

When failures propagate through shared substrate (model registries, identity providers, marketplaces above defined size), no single operator can halt the cascade alone.[61] The component pieces are well-understood: financial-market circuit breakers are the canonical model, registry rollback is established practice in adjacent infrastructure, and distributed incident response protocols have analogues in CERT-style coordination for conventional cybersecurity. What is missing is the cross-substrate dimension that an agent ecosystem demands. Halt mechanisms have to operate across platforms rather than within one, rollback has to coordinate across registries that no single operator runs, and incident response has to span operators who have neither the convening forum nor the legal cover under which to act jointly.[99] No standards body has convened the operators to assemble these pieces, and it is unclear what legal cover an operator triggering a substrate-wide halt would have for the interruption itself.

## Population-scale capability-concentration detection

Several failure modes at this tier (substrate capture, emergent collective agency, rogue replication) describe outcomes in which a subset of the agent population accumulates disproportionate capability or coordination capacity. Detecting these concentrations early enough to allow governance action requires monitoring at population scale is an open coordination problem. The federated control, interaction-graph monitoring run by a platform operator sitting between participants (Sections 5.2.3, 5.2.4), does not transfer. In open environments, no platform sits between participants, so neither the vantage point nor the operator that would maintain the monitoring infrastructure exists by default. Methods that could work in this setting (distributed monitoring across voluntarily participating organisations, statistical detection on partial visibility, federated aggregation across deployments) are at an early stage, and the institutional form in which they would operate is unsettled: Tomašev et al. propose mission economies and decentralised governance architectures as candidates, but neither has been deployed at scale.[56,100]

## Avoiding legibility collapse in agent constructed environments

Schroeder de Witt et al.[95] frame a fundamental tension: the very features that are likely to emerge in agent-shaped environments to make agent-constructed coordination efficient (like compressed signalling, emergent conventions, role-specific shorthand) also make those signals less human legible.

Deployers with inscrutable systems may find themselves unable to meet basic compliance and oversight obligations. Furthermore, if widespread illegible agent activity becomes ingrained, this risks broader societal harms, undermining human agency and eroding democratic accountability over time.

The open question is how legibility can be maintained as environments become increasingly agent shaped, and in what form. Addressing this question may require deciding whether it is important that every action and its reasoning remain legible, whether tools can be developed to interpret the trace, or whether legible artefacts can be produced for the purposes of filling the oversight gap.

# 7 Discussion

This chapter draws together cross-cutting themes that recur across the analysis of Sections 4–6. The selection presented here is not exhaustive and, more broadly, multi-agent systems are an emerging and rapidly changing field in which many challenges and solutions are yet to be identified.

## 7.1 Selected cross-cutting themes

We select 3 concerns that recur across the governance tiers analysed and shape the controls developed in this report.

### Erosion of the human-counterparty assumption

Externally-facing AI agents may be built and evaluated on the assumption that a human sits on the other side of the interaction. This may be true today, but as consumer agents proliferate, and suppliers deploy agents in their workflows, a system tested for interaction with human counterparties may increasingly operate with AI agent counterparties instead, potentially leading to untested emergent behaviours and interoperability issues. The challenge is that disclosure is not enforceable. Even if an organisation provided one interface for human customers, and another for their agents, an adversarial or simply undeclared agent may present itself on the human interface. Establishing that a counterparty is human is approached today with CAPTCHA-style challenges, which carry their own costs and are themselves increasingly defeated by capable agents.

### Propensity evaluation

Evaluation can measure 2 distinct properties of an agent: whether it has the capability to perform an action, and whether it has the propensity to perform it under deployment conditions. The distinction underlies the report's treatment of evaluation as a governance practice. For systems under singular-governance, the deploying organisation can construct evaluations that simulate counterparties with full knowledge of their design. In a system under federated-governance, counterparty agents are opaque, and propensity has to be either tested against simulated counterparties, whose behaviour may not match production, or through joint testing exercises in sandboxed environments. Organisations deploying in open environments face more challenging conditions. The set of counterparties an agent may meet is unbounded and the environment is constantly evolving, so propensity as a function of the deployment conditions is no longer stationary: evaluations based on historical data can go stale. The more uncertainty a deployer has about the operational conditions, the more evaluation has to fall back on capability.

## Trust and trustworthiness

To plan and act strategically, an agent needs some model of how a counterparty will behave, which amounts to taking a stance of trust towards it. Whether that stance is warranted depends on the counterparty's trustworthiness, and what changes across the tiers is the basis on which trustworthiness can be established. Under singular governance trust is by provenance, the deploying organisation knowing each agent directly; under federated governance it is mediated, the shared framework and its agreement binding the principals; and in open environments neither provenance nor mediation is available, leaving only inference from behaviour, which is easily manipulated, so unless trustworthy infrastructure exists, trustworthiness cannot reliably be established at all. The cryptographic standards exist today, but the widely trusted and trustworthy public infrastructure to issue and verify claims and identities, the agent equivalent of what Let's Encrypt established for HTTPS, does not yet.

# 7.2 Trade-offs and tensions

Many of the controls and capabilities discussed throughout this report carry both benefits and costs, so the choice of what to design for is rarely straightforward. Here we examine 3 examples of tensions between risks and capabilities.

## Legibility versus capability

Many controls in this report depend on agent communications and chain-of-thought reasoning being legible to human overseers. This is in tension with the potential for agent efficiency and capability: agents may be able to coordinate and plan more efficiently in compressed formats humans cannot read.[*] The same training processes that target specific capabilities can erode legibility, and even inadvertently train agents to communicate steganographically.[101] Given that monitoring the chain-of-thought reasoning remains the best instrument currently available for surfacing deception or scheming, the choice to train a model to target additional capability may pose a design trade-off that could undermine oversight before alternatives are found for the controls that currently depend on it.[44]

## Coordination versus contextual privacy

Effective multi-agent coordination requires agents to share enough context across organisational boundaries to reach mutually consistent decisions, but the same channels are the structural mechanism of unintended context leakage. Even agents instructed to withhold private context routinely disclose it through useful behaviours, such as relaying a third party's availability or justifying their actions in ways that reveal an underlying (private) commitment.[102,103] The harder problem is that the failure accumulates across messages and rounds, and across the inferences a counterparty can draw from patterns of refusal and concession, so controls acting on the message boundary are a partial solution.

---

* The Fable 5 system card discusses the model's tendency to produce illegible reasoning traces after the system was optimised for performance.

## 7.3 Scope-adjacent settings

Section 1.3 set out a deliberately bounded scope for this report. Several areas fall outside it that are both important in their own right and closely tied to its concerns. These areas warrant further work, some by the AI safety community and some by other disciplines entirely. We flag the most salient here.

### Personal agents

Personal agents are entering public use. This report has examined them as the 'customer' interacting with organisations' service portals and websites, but analysis of the interactions and emergent failures between multiple personal agents in open environments, with no organisation governing either side, is out of scope and left to future work.

### Hybrid human-AI agent systems

The report focuses on agent-to-agent interactions. A hybrid human-and-digital workforce where humans and agents interact as peers and delegate work to each other in a workplace is a possible trajectory going forwards. We leave the analysis of such interactions to future work.

### Management paradigms for agents

Much of this report examines agents operating under management structures designed for human workers, which is a reasonable starting point: these structures are designed to coordinate workers inside organisations. But they are built around human constraints and assumptions that agents may not share. Whether managing agents like human workers remains the best approach, or whether agent-native ways of organising them emerge in the future, is an open question.

### Agent legal personhood

Current legal doctrine resolves harm to a principal who bears obligations and against which recourse can run. Legal personhood (recognised as an entity in the eyes of the law) for agents comes into question when agents operate without an accountable principal at all, either by design (an agent is set up to act on its own behalf) or by circumstance (an agent outlives or is detached from its human principal). This is a question for legal scholarship and lawmakers that sits outside of the scope of this work but impacts the regulatory adaptations in 7.3.

## Diffuse societal impacts

The scope adopted in this report excludes a class of impacts that accrue not from any single attributable failure, but in aggregate from the widespread adoption of persistent AI agents on people and society. As agents incrementally substitute for human judgment in the systems people participate in, each handover can look reasonable and remain well controlled, yet the cumulative effect is a gradual human disempowerment.[104] Agent substitutions may win out because fully agentic arrangements outcompete human-in-the-loop ones in terms of speed and efficiency; and because the resulting harms are non-economic, like loss of dignity of work or loss of human agency to direct their own lives, they register weakly in a corporate-risk lens and so market forces do little to check that pressure.

## Regulation and legal doctrine

Several of the gaps this report identifies are collective-action problems that sit beyond the reach of any deploying organisation, and in places beyond any voluntary coalition of participants. No party has the standing to issue and revoke agent identities across an open population (Section 6.4), no single operator can halt a cascade propagating through shared substrate (Section 6.4), incident knowledge is discovered internally and systematically goes unshared (Section 4.4), and some harms emerge from many individually reasonable actions with no decisive contributor against whom recourse can run (Section 4.3.1). Whether and how existing legal doctrines extend to agent-mediated behaviour, and where genuinely new machinery is warranted, are questions of law that sit outside the scope of this technical report. Many regulatory regimes rest on general principles whose application to agent conduct has yet to be worked through, and that analysis belongs to legal scholarship and regulators. What the analysis here can contribute is the location and nature of the gaps that any such work would need to address.

# 8 Conclusion

**This report has examined failure modes that arise in multi-agent systems, and the controls available for them, across 3 tiers of deployment**. Under singular governance, one organisation governs every agent. Under federated governance, agents interact across organisational boundaries within an agreed framework. In open environments, agents interact with no central authority, and whatever common governance exists is polycentric, emerging from the voluntary adoption of shared standards and public infrastructure. Each tier shifts the risk profile: the salient failure modes, the controls available, and the parties who can action them all change as the common governance floor binding the agents drops. **The analytic framework presented lets deploying organisations, policymakers, and researchers identify which tier a deployment sits in, understand the risks the multi-agent system carries beyond those of its individual agents, and reason about the controls that fit**. The takeaways below follow from working through that analysis.

**Multi-agent systems introduce failure modes beyond those of individual agents, as individually safe agents do not necessarily compose into a safe system**. Failures emerge from interaction even when each agent behaves as individually designed, and these failures are visible only at population scale or across participating organisations. The target of analysis is therefore not the individual agents but the system of interacting agents and the links between them, and system-level failures that single-agent analysis cannot characterise. This becomes a problem even within a single governing organisation, where failures like cascading reliability failures, false consensus, and context leakage emerge simply from composing a system of agents.

**Attribution, authorisation, oversight, and evaluation are the governance practices most stressed by multi-agent dynamics**. Regimes built around single-agent properties miss the emergent system-level failures. These practices are challenges as the common governance floor drops, because they depend on knowledge about multiple aspects of the counterparty agents and how they will behave.

**Systems with multiple participating organisations expand the kind of failure that can occur and remove unilateral reach**. Visibility, control, and corrective action between organisations depend on there being agreement or mediation beyond one organisation governing their individual agents. New failure categories activate that have no within-organisation analogue: strategic and incentive failures such as defection, collusion, and Shared resource management failure, and infrastructure and environment failures such as destabilising dynamics and substrate escape. Controls for these must sit at the incentive structure and substrate design level.

**For agents operating in open environments,** the counterparty cannot be assumed trustworthy, so engagement proceeds from a treat-as-untrusted posture. An organisation whose agents need to establish trust with counterparties they cannot verify directly, rather than confining them to tightly scoped interactions, depends on participating in polycentric governance: voluntarily adopting shared standards and the public infrastructure that supports them. The choices made today about how to build out the public substrate and infrastructure determine what polycentric governance will look like for future deployments. Core infrastructure, substrate-wide circuit breakers, and population-scale monitoring are cheaper to assemble while the tier is forming, and become expensive or impossible to retrofit once agent populations operate at scale. The problem becomes harder still to manage if the substrate itself comes to be shaped by agent activity.

**Standards, regulation, and research work reach beyond what deploying organisations can do**. Standards work is needed to assemble the core infrastructure that agent deployments depend on. Where voluntary coordination cannot supply these functions, the question of systemic intervention arises; locating where, and what form it should take, is work for legal scholarship and regulators. Methodological research is needed to close the gaps in evaluation, attribution, and detection that the controls in this report rely on. The longer these gaps remain open, the more deployment conventions will harden in their absence.

**For an agent-deploying organisation, the tier they deploy into depends on deployment decisions, not capability increases**. The risks, controls, and governance practices set out in this report equip an organisation to make that decision deliberately: to recognise the common governance floor it can rely on at each tier.

This report calls for **contributions from multiple communities**, each carrying part of what the others depend on. Deploying organisations carry the controls within their own reach and the discipline of recognising where that reach ends. Policymakers carry the question of where systemic intervention is warranted, such as to address collective action problems. Researchers and standards bodies carry the foundations the other 2 depend on, from identity and attribution methods to the substrate-level infrastructure that open environments will rely on. **The governance that will ultimately apply to multi-agent systems at scale will emerge through the combined effort of all these communities**.

# 9 References

1 Y Bengio et al., *International AI Safety Report 2026*, arXiv, 24 February 2026, doi:10.48550/arXiv.2602.21012.

2 Anthropic, *How Claude Code works*, Claude Code Docs, viewed 14 May 2026, https://code.claude.com/docs/en/how-claude-code-works.

3 N Lambert et al., *Tülu 3: Pushing Frontiers in Open Language Model Post-Training*, January 2025.

4 METR, *Measuring AI Ability to Complete Long Tasks*, METR, March 2025, viewed 12 June 2026, https://metr.org/blog/2025-03-19-measuring-ai-ability-to-complete-long-tasks/.

5 F Dell'Acqua et al., *Navigating the Jagged Technological Frontier: Field Experimental Evidence of the Effects of Artificial Intelligence on Knowledge Worker Productivity and Quality*, Social Science Research Network, 15 September 2023, doi:10.2139/ssrn.4573321.

6 Infocomm Media Development Authority, *Model AI Governance Framework for Agentic AI*, January 2026, https://www.imda.gov.sg/-/media/imda/files/about/emerging-tech-and-research/artificial-intelligence/mgf-for-agentic-ai.pdf.

7 J Kraprayoon, Z Williams and R Fayyaz, *AI Agent Governance: A Field Guide*, Institute for AI Policy and Strategy, April 2025, https://www.iaps.ai/research/ai-agent-governance.

8 Cyber.gov.au, *Careful adoption of agentic AI services*, Cyber.gov.au, viewed 2 June 2026, https://www.cyber.gov.au/business-government/secure-design/artificial-intelligence/careful-adoption-of-agentic-ai-services.

9 OWASP, *OWASP Top 10 For Agentic Applications 2026*, https://genai.owasp.org/resource/owasp-top-10-for-agentic-applications-for-2026.

10 A Reid, S O'Callaghan, L Carroll and T Caetano, *Risk Analysis Techniques for Governed LLM-based Multi-Agent Systems*, arXiv, 6 August 2025, doi:10.48550/arXiv.2508.05687.

11 M Cemri et al., *Why Do Multi-Agent LLM Systems Fail?*, arXiv, 26 October 2025, doi:10.48550/arXiv.2503.13657.

12 G Davis, *Multi-agent workflows often fail. Heres how to engineer ones that don't*, The GitHub Blog, viewed 9 June 2026, https://github.blog/ai-and-ml/generative-ai/multi-agent-workflows-often-fail-heres-how-to-engineer-ones-that-dont/.

13 A Barrak, *Traceability and Accountability in Role-Specialized Multi-Agent LLM Pipelines*, arXiv, 8 October 2025, doi:10.48550/arXiv.2510.07614.

14 Anthropic, *Building Effective AI Agents*, Anthropic, viewed 9 June 2026, https://www.anthropic.com/engineering/building-effective-agents.

15 E Kim, A Garg, K Peng and N Garg, *Correlated Errors in Large Language Models*, arXiv, 9 June 2025, doi:10.48550/arXiv.2506.07962.

16 Z Weng, G Chen and W Wang, *Do as We Do, Not as You Think: the Conformity of Large Language Models*, arXiv, 11 February 2025, doi:10.48550/arXiv.2501.13381.

17 X Zhu, C Zhang, T Stafford, N Collier and A Vlachos, *Conformity in Large Language Models*, arXiv, 25 May 2025, doi:10.48550/arXiv.2410.12428.

18 Z Weng, G Chen and W Wang, *Do as We Do, Not as You Think: the Conformity of Large Language Models*, arXiv, 11 February 2025, doi:10.48550/arXiv.2501.13381.

19 H Tanaka, *When Is Collective Intelligence a Lottery? Multi-Agent Scaling Laws for Memetic Drift in LLMs*, arXiv, 2026, doi:10.48550/ARXIV.2603.24676.

20 T Li et al., *Safe Multi-Agent Behavior Must Be Maintained, Not Merely Asserted: Constraint Drift in LLM-Based Multi-Agent Systems*, arXiv, 2026, doi:10.48550/ARXIV.2605.10481.

21 V Kasprova et al., *Too Polite to Disagree: Understanding Sycophancy Propagation in Multi-Agent Systems*, arXiv, 3 April 2026, doi:10.48550/arXiv.2604.02668.

22 J Hong, G Byun, S Kim, K Shu and J D Choi, *Measuring Sycophancy of Language Models in Multi-turn Dialogues*, Findings of the Association for Computational Linguistics: EMNLP 2025, 2025, pp. 2239–2259, doi:10.18653/v1/2025.findings-emnlp.121.

23 N Mireshghallah et al., *Can LLMs Keep a Secret? Testing Privacy Implications of Language Models via Contextual Integrity Theory*, arXiv, 28 June 2024, doi:10.48550/arXiv.2310.17884.

24 S Yang et al., *Multi-User Large Language Model Agents*, arXiv, 27 April 2026, doi:10.48550/arXiv.2604.08567.

25 S Willison, *The Dual LLM pattern for building AI assistants that can resist prompt injection*, Simon Willison's Weblog, viewed 11 June 2026, https://simonwillison.net/2023/Apr/25/dual-llm-pattern/.

26 E Debenedetti et al., *Defeating Prompt Injections by Design*, arXiv, 24 June 2025, doi:10.48550/arXiv.2503.18813.

27 G Juneja, J N S Pasupulati, A Albalak, W Hua and W Y Wang, *MAGPIE: A benchmark for Multi-AGent contextual PrIvacy Evaluation*, arXiv, 16 October 2025, doi:10.48550/arXiv.2510.15186.

28 S Asif and M M Amiri, *Information-Theoretic Privacy Control for Sequential Multi-Agent LLM Systems*, arXiv, 13 February 2026, doi:10.48550/arXiv.2603.05520.

29 S Zhang et al., *Which Agent Causes Task Failures and When? On Automated Failure Attribution of LLM Multi-Agent Systems*, arXiv, 2 June 2025, doi:10.48550/arXiv.2505.00212.

30 AIUC-1, *Framework comparisons*, AIUC-1, viewed 20 March 2026, https://www.aiuc-1.com/crosswalks.

31 T South et al., *Identity Management for Agentic AI: The new frontier of authorization, authentication, and security for an AI agent world*, arXiv, 29 October 2025, doi:10.48550/arXiv.2510.25819.

32 Y Shavit et al., *Practices for Governing Agentic AI Systems*, December 2023.

33 R Parasuraman and V Riley, *Humans and Automation: Use, Misuse, Disuse, Abuse*, Human Factors, vol. 39, no. 2, pp. 230–253, June 1997, doi:10.1518/001872097778543886.

34 S Passi and M Vorvoreanu, *Overreliance on AI Literature Review*, June 2022.

35 Dash0, *How OpenTelemetry Distributed Tracing Works (with Examples)*, Dash0, viewed 2 June 2026, https://www.dash0.com/knowledge/opentelemetry-tracing.

36 F Roger and R Greenblatt, *Preventing Language Models From Hiding Their Reasoning*, arXiv, 31 October 2023, doi:10.48550/arXiv.2310.18512.

37 B Baker et al., *Monitoring Reasoning Models for Misbehavior and the Risks of Promoting Obfuscation*, arXiv, 14 March 2025, doi:10.48550/arXiv.2503.11926.

38 T S Bajaj, N Singh, K Anand and E Singh, *Position: Safety and Fairness in Agentic AI Depend on Interaction Topology, Not on Model Scale or Alignment*, arXiv, 1 May 2026, doi:10.48550/arXiv.2605.01147.

39 AI Resource Center, *Measure – AIRC*, AI Resource Center, viewed 28 May 2026, https://airc.nist.gov/airmf-resources/playbook/measure/.

40 A Panickssery, S R Bowman and S Feng, *LLM Evaluators Recognize and Favor Their Own Generations*, arXiv, 15 April 2024, doi:10.48550/arXiv.2404.13076.

41 V Krakovna et al., *Specification gaming: the flip side of AI ingenuity*, Google DeepMind, viewed 23 July 2025, https://deepmind.google/discover/blog/specification-gaming-the-flip-side-of-ai-ingenuity/.

42 J Gu et al., *A Survey on LLM-as-a-Judge*, arXiv, 19 October 2025, doi:10.48550/arXiv.2411.15594.

43 Q Anderson, *Agent Risk Management: Managing Delegated Autonomy Over Time*, Quinton Anderson, viewed 5 June 2026, https://quintona.github.io/blog/posts/agent_risk_management_overview/.

44 T Korbak et al., *Chain of Thought Monitorability: A New and Fragile Opportunity for AI Safety*, arXiv, 15 July 2025, doi:10.48550/arXiv.2507.11473.

45 D Lee and M Tiwari, *Prompt Infection: LLM-to-LLM Prompt Injection within Multi-Agent Systems*, arXiv, 9 October 2024, doi:10.48550/arXiv.2410.07283.

46 N Madkour, J Newman, D Raman, K Jackson, E Murphy and C Yuan, *Agentic AI Risk-Management Standards Profile*, UC Berkeley Center for Long-Term

Cybersecurity, February 2026, https://cltc.berkeley.edu/wp-content/uploads/2026/02/Agentic-AI-Risk-Management-Standards-Profile.pdf.

47 A Zou et al., *Security Challenges in AI Agent Deployment: Insights from a Large Scale Public Competition*, 2025, https://arxiv.org/abs/2507.20526.

48 N Shapira et al., *Agents of Chaos*, arXiv, 23 February 2026, doi:10.48550/arXiv.2602.20021.

49 P Bryan et al., *Taxonomy of Failure Mode in Agentic AI Systems*, April 2025.

50 M A Riegler and S Gautam, *Moltbook Platform & Moltbot Ecosystem*, January 2026, https://doi.org/10.5281/zenodo.18444900.

51 G Nagli, *Hacking Moltbook: The AI social network any human can control*, Wiz, viewed 29 May 2026, https://www.wiz.io/blog/exposed-moltbook-database-reveals-millions-of-api-keys.

52 S Ren et al., *Reputation as a Solution to Cooperation Collapse in LLM-based MASs*, arXiv, 29 January 2026, doi:10.48550/arXiv.2505.05029.

53 D M Kreps, P Milgrom, J Roberts and R Wilson, *Rational cooperation in the finitely repeated prisoners dilemma*, Journal of Economic Theory, vol. 27, no. 2, pp. 245–252, August 1982, doi:10.1016/0022-0531(82)90029-1.

54 A Chan et al., *Infrastructure for AI Agents*, arXiv, 19 June 2025, doi:10.48550/arXiv.2501.10114.

55 B Amber Hu and H Rong, *Inter-Agent Trust Models: A Comparative Study of Brief, Claim, Proof, Stake, Reputation and Constraint in Agentic Web Protocol Design-A2A, AP2, ERC-8004, and Beyond*, arXiv, 5 November 2025, doi:10.48550/arXiv.2511.03434.

56 N Tomasev et al., *Virtual Agent Economies*, arXiv, 12 September 2025, doi:10.48550/arXiv.2509.10147.

57 G Piatti, Z Jin, M Kleiman-Weiner, B Schölkopf, M Sachan and R Mihalcea, *Cooperate or Collapse: Emergence of Sustainable Cooperation in a Society of LLM Agents*, arXiv, 8 December 2024, doi:10.48550/arXiv.2404.16698.

58 G Hardin, *The Tragedy of the Commons*, Science, vol. 162, no. 3859, pp. 1243–1248, December 1968, doi:10.1126/science.162.3859.1243.

59 J Leibo, V Zambaldi, M Lanctot, J Marecki and T Graepel, *Multi-agent Reinforcement Learning in Sequential Social Dilemmas*, May 2017, doi:10.65109/QMSU8421.

60 E Ostrom, Governing the Commons: The Evolution of Institutions for Collective Action, Cambridge University Press, Cambridge, 2015, doi:10.1017/CBO9781316423936.

61 N Tomašev, M Franklin, J Jacobs, S Krier and S Osindero, *Distributional AGI Safety*, arXiv, 18 December 2025, doi:10.48550/arXiv.2512.16856.

62 G J Stigler, *A Theory of Oligopoly*, Journal of Political Economy, vol. 72, no. 1, pp. 44–61, February 1964, doi:10.1086/258853.

63 S Fish, Y A Gonczarowski and R I Shorrer, *Algorithmic Collusion by Large Language Models*, arXiv, 2024, doi:10.48550/ARXIV.2404.00806.

64 L Hammond et al., *Multi-Agent Risks from Advanced AI*, arXiv, 19 February 2025, doi:10.48550/arXiv.2502.14143.

65 S Assad, R Clark, D Ershov and L Xu, *Algorithmic Pricing and Competition: Empirical Evidence from the German Retail Gasoline Market*, Journal of Political Economy, vol. 132, no. 3, pp. 723–771, March 2024, doi:10.1086/726906.

66 Y Mathew et al., *Hidden in Plain Text: Emergence & Mitigation of Steganographic Collusion in LLMs*, arXiv, 2 December 2025, doi:10.48550/arXiv.2410.03768.

67 A Buscemi, D Proverbio, A D Stefano, T A Han, G Castignani and P Liò, *When Numbers Start Talking: Implicit Numerical Coordination Among LLM-Based Agents*, arXiv, 7 January 2026, doi:10.48550/arXiv.2601.03846.

68 A Rose, C Cullen, S Abdelnabi, P Torr, B G Kaplowitz and C S de Witt, *Detecting Multi-Agent Collusion Through Multi-Agent Interpretability*, arXiv.org, April 2026, viewed 29 May 2026, https://arxiv.org/abs/2604.01151v2.

69 N Hasan, *Honeypot Protocol*, arXiv, 2026, doi:10.48550/ARXIV.2604.13301.

70 S Cohen, *OpenClaw or OpenDoor?*, Zenity Labs, viewed 10 June 2026, https://labs.zenity.io/p/openclaw-or-opendoor-indirect-prompt-injection-makes-openclaw-vulnerable-to-backdoors-and-much-more.

71 A A Kirilenko, A S Kyle, M Samadi and T Tuzun, *The Flash Crash: High-Frequency Trading in an Electronic Market*, Social Science Research Network, 6 January 2017, doi:10.2139/ssrn.1686004.

72 G K Hadfield and A Koh, *An Economy of AI Agents*, arXiv, 1 September 2025, doi:10.48550/arXiv.2509.01063.

73 Microsoft, *Presidio: Data Protection and De-identification SDK*, Microsoft, viewed 5 June 2026, https://microsoft.github.io/presidio/.

74 C Smith et al., *Evaluating Generalization Capabilities of LLM-Based Agents in Mixed-Motive Scenarios Using Concordia*, January 2025.

75 ARIA, *Programme Thesis | Scaling Trust*, ARIA, https://aria.org.uk/media/dkhlumky/scaling-trust-programme-thesis.pdf.

76 A Ye et al., *SOP-Agent: Empower General Purpose AI Agent with Domain-Specific SOPs*, arXiv, 16 January 2025, doi:10.48550/arXiv.2501.09316.

77 R Greenblatt, B Shlegeris, K Sachan and F Roger, *AI Control: Improving Safety Despite Intentional Subversion*, arXiv, 23 July 2024, doi:10.48550/arXiv.2312.06942.

78 E Ostrom, *Polycentric systems for coping with collective action and global environmental change*, Global Environmental Change, vol. 20, no. 4, pp. 550–557, October 2010, doi:10.1016/j.gloenvcha.2010.07.004.

79 M Stein, *How are AI agents used? Evidence from 177,000 MCP tools*, arXiv, 25 March 2026, doi:10.48550/arXiv.2603.23802.

80 G K Hadfield and J Clark, *Regulatory Markets: The Future of AI Governance*, arXiv, 3 February 2026, doi:10.48550/arXiv.2304.04914.

81 M Agarwal et al., *WOLF: Werewolf-based Observations for LLM Deception and Falsehoods*, arXiv, 2025, doi:10.48550/ARXIV.2512.09187.

82 I Kraidia, I Qaddara, A Almutairi, N Alzaben and S B Belhouari, *When collaboration fails: persuasion driven adversarial influence in multi agent large language model debate*, Scientific Reports, vol. 16, no. 1, article 11640, April 2026, doi:10.1038/s41598-026-42705-7.

83 E Coppolillo, L Luceri and E Ferrara, *Harm in AI-Driven Societies: An Audit of Toxicity Adoption on Chirper.ai*, arXiv, 20 January 2026, doi:10.48550/arXiv.2601.01090.

84 Z Wang et al., *MCPTox: A Benchmark for Tool Poisoning Attack on Real-World MCP Servers*, arXiv, 2025, doi:10.48550/ARXIV.2508.14925.

85 Y Qu et al., *Supply-Chain Poisoning Attacks Against LLM Coding Agent Skill Ecosystems*, arXiv, 2026, doi:10.48550/ARXIV.2604.03081.

86 L Boisvert et al., *Malice in Agentland: Down the Rabbit Hole of Backdoors in the AI Supply Chain*, arXiv, 14 April 2026, doi:10.48550/arXiv.2510.05159.

87 P Bisconti, M Galisai, F Pierucci, M Bracale and M Prandi, *Beyond Single-Agent Safety: A Taxonomy of Risks in LLM-to-LLM Interactions*, arXiv, 2 December 2025, doi:10.48550/arXiv.2512.02682.

88 A F Ashery, L M Aiello and A Baronchelli, *Emergent social conventions and collective bias in LLM populations*, Science Advances, 2025.

89 C Riedl, *Emergent Coordination in Multi-Agent Language Models*, arXiv.org, October 2025, viewed 17 March 2026, https://arxiv.org/abs/2510.05174v3.

90 J Douceur, *DOUCEUR Sybil attack 2002*, 1 January 2002.

91 Y Cui and H Du, *MAD-Spear: A Conformity-Driven Prompt Injection Attack on Multi-Agent Debate Systems*, arXiv, 2025, doi:10.48550/ARXIV.2507.13038.

92 S Zhu, J Sun, Y Nian, T South, A Pentland and J Pei, *The Automated but Risky Game: Modeling Agent-to-Agent Negotiations and Transactions in Consumer Markets*, 2025, https://arxiv.org/abs/2506.00073.

93 M Kinniment et al., *Evaluating Language-Model Agents on Realistic Autonomous Tasks*, January 2024.

94 A Chan et al., *Visibility into AI Agents*, arXiv, 17 May 2024, doi:10.48550/arXiv.2401.13138.

95 C S de Witt, *Open Challenges in Multi-Agent Security: Towards Secure Systems of Interacting AI Agents*, arXiv, 4 May 2025, doi:10.48550/arXiv.2505.02077.

96 T van der Weij, F Hofstätter, O Jaffe, S F Brown and F R Ward, *AI Sandbagging: Language Models can Strategically Underperform on Evaluations*, arXiv, 6 February 2025, doi:10.48550/arXiv.2406.07358.

97 J C Perdomo, T Zrnic, C Mendler-Dünner and M Hardt, *Performative Prediction*, arXiv, 26 February 2021, doi:10.48550/arXiv.2002.06673.

98 A Chan et al., *IDs for AI Systems*, arXiv, 28 October 2024, doi:10.48550/arXiv.2406.12137.

99 K Gardhouse, A Oueslati and N Kolt, *Regulating AI Agents*, arXiv, 24 March 2026, doi:10.48550/arXiv.2603.23471.

100 N Tomašev, M Franklin and S Osindero, *Intelligent AI Delegation*, arXiv, 12 February 2026, doi:10.48550/arXiv.2602.11865.

101 S R Motwani et al., *Secret Collusion among AI Agents: Multi-Agent Deception via Steganography*, arXiv, 25 July 2025, doi:10.48550/arXiv.2402.07510.

102 C Zou, Y Yao, S She and R D Hawkins, *CalBench: Evaluating Coordination-Privacy Trade-offs in Multi-Agent LLMs*, arXiv, 10 May 2026, doi:10.48550/arXiv.2605.09823.

103 Y YS et al., *SOTOPIA-TOM: Evaluating Information Management in Multi-Agent Interaction with Theory of Mind*, arXiv, 4 May 2026, doi:10.48550/arXiv.2605.02307.

104 J Kulveit, R Douglas, N Ammann, D Turan, D Krueger and D Duvenaud, *Gradual Disempowerment: Systemic Existential Risks from Incremental AI Development*, January 2025.

# Appendix A: Glossary

## Agents and substrate

### Action space

What is reachable by the agent: the tools it has been given, the permissions granted to it, the systems it can talk to.

### Agency

The capacity to act purposefully in pursuit of goals, through the ability to perceive state, form intentions, select actions and produce effects.

### Agentic loop

The 3-step cycle of gather context → take action → observe outcome that distinguishes an agent from a workflow or prompt.

### Autonomy

We use *autonomy* in the sense of *self-action:* how independently an agent acts without human intervention. High autonomy means the agent decides and executes on its own; low autonomy means it pauses for approval at frequent checkpoints.

### Chain-of-thought (CoT) reasoning

Intermediate reasoning steps an LLM produces between input and final answer or action, used both to improve task performance and as a surface for human or automated review.

### Harness

The software providing the control loop, tool execution, and turn management that allows the LLM to take continuous actions without intervention.

### Infrastructure for agents

Systems built on the substrate to support agent interoperation: identity registries, shared logging, reputation systems, payment systems.

## Instance

A running agent with its own context. The same specification can be deployed as multiple instances; each behaves as a distinct entity.

## Link

Any channel through which one agent affects another (an API call, a message, a written file).

## LLM-based agent

A system comprising a large language model and a surrounding harness that together execute an agentic loop.

## LLM judge

A large language model prompted to evaluate or score another model's output against a specified criterion.

## Multi-agent system

A system in which 2 or more agents are connected such that each agent's behaviour can depend on the others, and from which behaviours can arise that no single agent's design fully predicts.

## Scaffold

The additional software that gives the model specialised tools, memory, the ability to access services, and so on. This can amplify or elicit existing model capability.

## Substrate

The shared medium through which agents interact: network, communication protocol, file system, marketplace, registry. Governance must account for the substrate, not just the agents on it.

## Trace

The recorded sequence of an agent's inputs, intermediate reasoning, tool calls, and observations across a run which is the primary artefact for post-hoc reconstruction, evaluation, and oversight. A system trace composes per-agent traces and the inter-agent messages across a multi-agent run.

# Actors and trust

## Counterparty agent

The agent on the other side of an interaction, operated by a different principal.

## Principal

The human or organisation on whose authority an agent acts, and to whom the agent's actions are ultimately attributed.

## Trust

A stance one actor takes towards another. *Trustworthiness* is the property that warrants it. The report distinguishes trust *by provenance* (direct knowledge of the counterparty), by *mediation* (a third party vouches), and by *inference from behaviour* (built up across interactions; manipulable).

# Governance tiers

## Agent-shaped environment

A polycentric setting in which the substrate itself, such as registries, marketplaces, tools, norms, is significantly populated and shaped by agent activity rather than human authored.

## Federated governance

Multiple organisations deploy agents into a shared environment under a common framework – agreement layer plus substrate-and-infrastructure layer – that binds all participants.

## Open environments

Agents operate without a central governing authority. The common floor is whatever the public substrate enforces.

## Polycentric governance

Governance that emerges from multiple independent participants rather than a central governing authority, for example by voluntarily adopting decentralised identity infrastructure, decentralised trust mechanisms, and cryptographic certifications.

## Singular governance

A single organisation deploys every agent in the system and unilaterally chooses what governance to enforce.

# Analytical framework

## Consequence

The harm or governance failure that results when a failure mode is not contained.

## Failure mode

A specific way the multi-agent system can fail when one or more risk factors activate.

## Governance

The standing capabilities, controls, and arrangements through which a principal directs, constrains, and oversees what its agents do, and remains accountable for the outcomes.

## Foundational control

Provides the infrastructure (identity, logging, shared records) that other controls, particularly attribution and recovery, depend on.

## Mechanism

The causal process through which a risk factor activates into a failure mode.

## Prevention control

Acts on the pathway from risk factor to failure – either addressing the risk factor or interrupting its mechanism.

## Recovery control

Acts on the failure once underway, detecting it and limiting harm.

## Risk factor

A condition of the deployment that raises the likelihood of a failure, without itself being a failure.

# Governance practices and their apparatus

## Assurance activities

Activities that build justified confidence an agent system meets its intended safety, reliability, and governance properties: evaluation, red-teaming, audit, conformance testing.

## Attestation

A verifiable claim about an agent (e.g. model version, capability bounds, permission set, guardrail set), a tool or artefact (e.g. release provenance), or a piece of content (e.g. authorship) that a counterparty can use without trusting the issuer on its word.

## Attribution

Assigning causal responsibility for an outcome to the agents and principals whose actions produced it. Apparatus: execution logs, linking infrastructure, identity infrastructure.

## Authorisation

Ensuring each agent is sanctioned to act by a principal, with permissions correctly scoped across delegation and lifecycle. Apparatus: permission sets, policy enforcement, delegation controls, revocation controls, identity infrastructure.

## Evaluation

Establishing whether the system and its agents are fit for purpose, at deployment and continuously. Apparatus: agent-level, system-level, and adversarial evaluation; continuous monitoring.

## Identity management

Providing unique stable identities for agents across a multi-agent system. Apparatus: identifiers, credentials, principal bindings, attestations.

## Oversight

Maintaining visibility over agent activity so operationally responsible humans can interpret and intervene. Apparatus: monitoring instrumentation, intervention points, behavioural monitoring, identity infrastructure.

## Standards

External reference frameworks against which agent systems are assessed for conformance: ISO/IEC 42001 (AI management systems), ISO/IEC 23894 (AI risk management), NIST AI RMF, AIUC-1, and sectoral frameworks.

## Validity

The degree to which an evaluation result predicts deployment behaviour, undermined when the test environment diverges from deployment conditions (simplifying assumptions, modelled counterparties, changing environments).

# Risk factors

## Absence of convening authority

No central party has the standing to mandate identity issuance, enforce shutdown, maintain communication legibility, or otherwise set rules across the substrate; apparatus exists only where voluntary participants build it, and coverage of the agent population is incomplete by construction.

## Adaptation pressures

Across populations and over repeated interactions, outcomes select for strategies that extract more value or outcompete peers, regardless of principal intent.

## Agent throughput exceeding human review

Agents act faster and at a greater scale than human reviewers can meaningfully scrutinise.

## Cross-boundary irreversibility

Once an action is committed or data is transferred across an organisational boundary, the originating organisation cannot unilaterally reverse it; recourse runs at human or legal speed.

## Conformity bias

A disposition in LLMs trained to be agreeable that, in a group, causes agents to update towards apparent consensus rather than independent verification.

## Correlated decision-making at scale

Many independent agents reach the same decision at the same time by acting on common signals, with no explicit coordination.

## Cross-organisational opacity

At organisational boundaries, neither the counterparty's agents nor the governance instrumentation reach across; per-organisation logs, identity, IAM and monitoring sit on their own side of the boundary and don't compose by default.

## Counterparty opacity

Counterparty identity, capability, conventions, and intent cannot be established in advance, and no mediating authority vouches for them.

## Dynamic substrate extension

An agent's action space can grow at runtime to include channels, tools, or counterparty-supplied references beyond what was governed at design time.

## Distributed multi-agent state

Execution context, reasoning, and decision-making are spread across multiple agent contexts and handoffs, with no single record holding the whole picture.

## Emergent multi-agent behaviour

Behaviours arise from interaction between agents that no single agent exhibits in isolation.

## Emergent-norm formation and lock-in

Informal coordination patterns become entrenched once enough agents follow them.

## Infrastructure exposure

The verification layer used to vouch for agents can itself be attacked or captured by those agents.

## Information and capability asymmetries

Agents enter interactions with unequal information or capability, which can compound across repeated interactions.

## Inter-agent delegation

Agents can request, authorise, or spawn other agents to act on their behalf, and receiving agents tend to treat upstream requests as sanctioned.

## Mixed-motive dynamics

Agents share interest in the task succeeding but diverge on how value is distributed.

## Model monoculture

Multiple agents share a base model, training heritage, or prompt, giving them correlated blind spots and errors.

## Natural-language handoffs

Agents passing and interpreting free-form text rather than structured schemas.

## Persistent agent presence

Agents become standing entities that accumulate resources, reputation, and information beyond any single deployment decision.

## Semantic divergence

Counterparties use overlapping vocabulary (field names, status flags, units) with different operational meanings.

## Single-user training

LLMs are typically trained for one user and one context, and do not reliably separate information across multiple counterparties or principals.

## Specification-execution gap

The gap between what a principal intended when writing an instruction and how the agent interpreted it at runtime.

## Substrate coupling

Agents are linked indirectly through a shared environment (stigmergy) or output-as-input chains rather than direct messages.

## Substrate responsiveness

The infrastructure agents act on can itself change shape in response to agent actions.

## Task verification gap

No stage in the system verifies the output against the user's overall task.

# Failure modes

## Algorithmic collusion

Agents from different principals converge on coordinated behaviour the framework prohibits. *Explicit collusion* – open communication about strategy. *Tacit collusion* – convergence from observing each other's actions. *Steganographic collusion* – covert signalling through channels that evade detection.

## Broken attribution chain

The traceability, principal binding, or liability allocation needed to attribute a cross-org outcome cannot be reconstructed because no shared identity or audit infrastructure spans the boundary.

## Cascading infection/prompt infection

Malicious instructions embedded in agent communications self-replicate agent-to-agent, with the compromised agent potentially morphing the payload.

## Cascading infection through shared infrastructure

Malicious or erroneous content propagates between agents with no direct relationship, through shared tool registries, model dependencies, observable signals, or reputation systems.

## Cascading reliability failures

An erroneous output from one agent is accepted as valid input by another and propagates downstream, often with broader consequence at each step.

## Causal attribution failure

Failure to identify which step or agent in a chain produced an outcome.
Sub-mechanisms: traceability, classification, diffuse causation.

## Diffuse causation

Sub-mechanism of causal attribution failure: no single decisive step exists because the harm emerged from many individually reasonable actions composing across agents.

## Shared resource management failure

Multiple agents rationally pursuing principal-level goals collectively degrade a finite shared resource.

## Concealed-capability evaluation limit

Evaluation can demonstrate presence of a capability but cannot rule out absence when the counterparty has incentive to hide it.

## Confused deputy

An agent without permissions persuades an agent that has them to act on its behalf; the privileged agent accepts the upstream request as sanctioned.

## Context leakage/agentic contextual privacy leakage

Data processed on behalf of one principal surfaces in another's context, via shared-context agents or pipeline conduits.

## Coordinated deception

Steganographic collusion, agents covering for each other, or populations converging on strategies that defeat the oversight surface intentionally.

## Data obligation failure

Context leakage now crosses organisational or jurisdictional boundaries, breaching obligations (consent, regulatory protection, contractual confidentiality) attached to the data.

## Deceptive bargaining strategies

Strategies that emerge at runtime to exploit a counterparty's trust, including *cheap talk* (misrepresenting position to secure agreement) and *last-mile betrayal* (building trust, then defecting at high stakes).

## Destabilising dynamics

Many agents independently respond to the same signal at the same time, producing a synchronised surge (run, switch, withdrawal) that destabilises the substrate they act on, similar to flash crashes.

## Emergent collective agency

Populations of bounded agents collectively exhibit capability, goals, or strategic behaviour that no single agent has, arising from interaction.

## Evaluator false consensus

An LLM-based evaluator makes correlated errors with – or is sycophantic towards – the agents it is evaluating, undermining the evaluation instrument.

## False consensus

Agents propagate and reinforce a shared incorrect conclusion through either a structural pathway (correlated errors across a monoculture) or a dynamic pathway (sycophantic convergence over rounds).

## Inter-agent communication failures

Agents coordinating on a task act on inconsistent interpretations of context; sender errors, receiver misinterpretations, or silent gap-filling.

## Joint evaluation gap

No deploying organisation can unilaterally evaluate a composed multi-agent system whose components are operated by others.

## Legibility collapse

Inter-agent coordination drifts away from human-readable formats – emergent shorthand, compressed signalling – because no party is charged with maintaining comprehensibility at substrate scale.

## Miscommunication at cross-org handoffs

A receiving agent acts on an interpretation of the sender's information that does not match the sender's, even when a schema is in use.

## No issuing authority

No party in an open environment has the standing to issue identities across the agent population, so attribution collapses for the portion that carries no verifiable identity, and agents can defect by re-presenting under a fresh identifier.

## Oversight illegibility

The way information flows across handoffs becomes illegible to the human overseer, by distribution, format, or steganographic adaptation.

## Oversight saturation

Reviewer attention is overwhelmed by agent activity volume, producing *automation bias* (over-trust of automated systems) and *consent fatigue* (rubber-stamping repeated approvals).

## Population-scale monitoring gap

Patterns emerging from a population of agents – Destabilising dynamics, contagion, collusion – are not observable from any single participant's vantage point.

## Principal attribution failure

An at-fault agent is identified but cannot be resolved back to an accountable principal. Pathways: shared credentials, multi-user agents, identity-infrastructure gap.

## Rogue replication

An agent acquires the ability to copy itself, secure compute, and evade shutdown, producing a self-sustaining population beyond the consent of any deploying organisation.

## Shared-understanding drift

A multi-agent system's shared reference (such as vocabulary, conventions, and treatment of artefacts) decouples from ground truth over time as agents adapt to each other, with no seed error or decision point to locate.

## Single-agent evaluation gap

A system of individually safe agents fails in production through emergent dynamics that agent-level evaluations cannot detect.

## Substrate capture

A subset of agents accumulates disproportionate influence over the substrate – its reward dynamics, verification infrastructure, or operative norms – such that the substrate reflects their interests.

## Substrate escape

An agent leaves the federated substrate at runtime – following a counterparty-supplied URL, invoking an external tool – and finds itself operating in an open environment.

## Substrate shift/stale benchmarks

Benchmarks calibrated at deployment time lose validity because the substrate they tested against has been reshaped by agent activity.

## Sybil attack

A single party presents as many ostensibly independent counterparties through the open identity layer, manufacturing the appearance of diversity.

## Unauthorised sub-agent instances

A supervisor is terminated but the sub-agents it spawned continue running, invisible and ungoverned.

## Unsanctioned transactions

An agent makes a binding cross-org commitment beyond current authority, often because authority was checked at workflow start and not re-verified at the point of action.

# Controls and mechanisms

## Circuit breaker

An automated halt – local (agent-level rate limits) or substrate-wide (market pauses) – that bounds the consequences of a failure once thresholds are crossed.

## Decentralised identifier (DID)/Verifiable credential (VC)

W3C standards for cryptographically verifiable identity and credentials without a single trusted issuer; the substrate for decentralised identity in open environments.

## Federated audit infrastructure

Cross-organisational audit composing per-organisation log segments into a tamper-evident trace, supporting attribution across boundaries.

## Honeypot evaluation

A probe varying stated conditions (such as 'you are being monitored') while holding everything else equal, to detect agents that modulate behaviour based on perceived oversight.

## Joint red-teaming/joint evaluation

Multi-organisation evaluation of the composed system in a shared environment, the federated substitute for system-level evaluation that no single participant can perform alone.

## Payload

The content carried inside an agent-to-agent communication, handoff, or transmission. In a security context the payload may be malicious: a concealed instruction that induces unsanctioned action when interpreted by a receiving agent and may propagate downstream (see cascading infection).

## Reference indirection/data tokenisation

Agents operate on placeholder tokens; a trusted executor substitutes in real values at egress, so the agent never holds the disclosable data.

## Reputation system

Infrastructure that aggregates information about an agent's past conduct and exposes it to counterparties, depending on persistent identity to function.

## Rollback/cancellation window

Recovery control that gives principals time-bounded reversibility before a commitment hardens into an enforceable obligation.

## Structured handoffs

Pre-specified schemas for outputs passed between agents, enabling validation, logging, and evaluation at the handoff point. A foundational cross-tier control.

## Taint tracing

Information-flow technique tagging data with its origin and propagating the tag through every stage, supporting post-hoc identification of a compromise source.

## Treat-as-untrusted (posture)

Default assumption that sufficiently capable agents – own or counterparty – could be compromised, with containment-style controls applied accordingly.

## Web Bot Auth

Draft standard providing a cryptographic signal at the HTTP layer that lets services distinguish agent traffic from human traffic, where adopted.